\documentclass[pra,preprint,showpacs,preprintnumbers,amsmath,amssymb]{revtex4}
\usepackage{graphicx}
\usepackage{dcolumn}
\usepackage{epic}%
\usepackage{eepic}%
\usepackage{bm}

\begin{document}

\preprint{APS/123-QED}

\newcommand{\scscs}{\scriptscriptstyle}

\title{Reference--State One--Particle Density--Matrix Theory}

\author{James P. Finley}

\affiliation{
Department of Physical Sciences, 
Eastern New~Mexico University,
Station \#33, Portales, NM 88130}
\email{james.finley@enmu.edu}

\affiliation{Department of Chemistry, University of Arkansas, Fayetteville,
 Arkansas 72701 }

\affiliation{Department of Applied Chemistry, Graduate School of Engineering,
The University of Tokyo, Tokyo, Japan 113-8656}

\date{\today}

\begin{abstract}

A density-matrix formalism is developed based on the one-particle density-matrix
of a single-determinantal reference-state.  Unlike traditional
density-functional-theory approaches, the $v$-representable problem does {\em not}
appear in the proposed method, nor the need to introduce functionals defined by a
constrained search; furthermore, the necessary conditions for a one-particle
density matrix to come from a single determinant are known, so they can be
expressed as constraints when minimizing variational-energy functionals.  The
correlation-energy functionals are {\em not} universal, in the sense that they
depend on the external potential. Nevertheless, model systems can still be used to
derive universal energy-functionals. For example, the Colle--Salvetti functional
is shown to be compatible with the proposed method. In addition, the
correlation-energy functionals can be partitioned into individual terms that are
-- to a varying degree -- universal; yielding, for example, an {\em electron gas
approximation}, where the gas in {\em no}t assumed to be uniform. Variational and
non-variational energy functionals are introduced that yield the target state when
the reference state -- or its corresponding one-particle density matrix -- is
constructed from Brueckner orbitals.  Using many-body perturbation theory,
diagrammatic expansions are given for the non-variational energy-functionals,
where the individual diagrams {\em explicitly} depend on the one-particle
density-matrix.  Non-variational energy-functionals yield generalized
Hartree--Fock equations involving a non-local correlation-potential and the
Hartree--Fock exchange; these equations are obtained by imposing the
Brillouin--Brueckner condition. The same equations -- for the most part -- are
obtained from variational energy-functionals using functional minimizations,
yielding the (kernel of) correlation potential as the functional derivative of
correlation-energy functionals. Approximations for the correlation-energy
functions are introduced, including a one-particle-density-matrix variant of the
local-density approximation (LDA), a variant of the Lee--Yang--Parr (LYP)
functional, and a variant of the correlation-energy functional used in the
so-called B3LYP approach.  A brief comparison with the {\em exact SCF theory} by
L\"owdin is presented.

\end{abstract}
\maketitle

\section{Introduction} \label{INTRO}

Many variants of density functional theory (DFT)
\cite{Dreizler:90,Parr:89,Springborg:97,Ellis:95,Gross:94,Seminario:95} share
common features with the Hartree--Fock approach. In particular, the Kohn--Sham
method \cite{Kohn:65} uses orbital equations that appear similar, in certain
respects, with the ones from Hartree--Fock theory. On the other hand, unlike the
Hartree--Fock wavefunction, the Kohn-Sham determinantal state shares {\em only} a
common density with the exact wavefunction, and is {\em not} considered an
approximation of the ground-state. Furthermore, in contrast to the exchange
potential from Hartree--Fock theory, the Kohn--Sham exchange-correlation potential
is local.

DFT approaches that use hybrid functionals
\cite{Becke:93,Burke:97,Perdew:96,Ernzerhof:96} introduce a component of exact
exchange-energy, where justification, in part, for this modification comes from
the adiabatic connection
\cite{Becke:93b,Harris:74,Gunnarsson:76,Langreth:77,Harris:84}, yielding an
approach that, again, has more similarities with Hartree--Fock, especially since
its energy functional yields a non-local potential -- the Hartree-Fock exchange --
that depends on the one-particle density matrix of the Kohn-Sham determinantal-
state. The most celebrated hybrid functional, B3LYP \cite{Becke:93,Stephens:94},
contains three parameters, two correlation-energy functionals, the Dirac-exchange
functional (with a correction), and, of course, exact exchange. The LYP density
functional \cite{Lee:88} -- a key component of B3LYP -- is derived from the
Colle--Salvetti correlation-energy functional \cite{Colle:75}, where this
functionals depends on a one-particle density-matrix, say $\tilde{\gamma}$, where
$\tilde{\gamma}$ is from the {\em Hartree-Fock} reference state, and not an {\em
exact} eigenstate, indicating a further evolutionary step of DFT methods towards a
Hartree-Fock generalization with inclusion of electron correlation.

In the Hartree--Fock Kohn--Sham approach \cite{Seidl:96}, the exchange energy is
treated in an exact manner and the non-local, Hartree--Fock exchange-potential
appear in the orbital equations. A generalization of this approach by Lindgren and
Salomonson \cite{Lindgren:02} yields, in addition, a nonlocal correlation
potential and orbitals that, they believe, are very similar to Brueckner orbitals.
Other workers also suggest that Brueckner and Kohn-Sham orbitals are very similar
\cite{Hebelmann}.

Brueckner orbital theory
\cite{Brueckner:54,Nesbet:58,Brenig:61,Lowdin:62,Kutzelnigg:64,Cizek:80,Chiles:81,Stolarczyk:84,Handy:85,Handy:89,Raghavachari:90,Hirao:90,Stanton:92,Hampel:92,Scuseria:94,Lindgren:02}
is a generalization of Hartree--Fock theory that utilizes a single-determinantal
state that has the maximum overlap with an exact eigenfunction
\cite{Kobe:71,Shafer:71}.  Below we use this formalism to develop a density-matrix
theory, in which a variety of variational and non-variational energy-functionals
are introduced that depend on the one-particle density-matrix, say
$\gamma$. Unlike other approaches, where $\gamma$ is the one-particle
density-matrix of an exact eigenfunction
\cite{Gilbert:75,Berrondo:75,Donnelly:78,Levy:79,Valone:80,Ludena:87}, the
introduced method -- called {\em reference-state one-particle density-matrix
theory} -- has $\gamma$ arising from a single-determinantal reference state, where
the energy-functionals yield the exact energy when $\gamma$ is the one from the
Brueckner reference-state.

One advantage that this one-particle, density-matrix approach has over traditional
density-functional formalisms -- or one-particle density-matrix formalisms -- is
that there is no $v$-representable problem \cite{Englisch:83,Parr:89,Dreizler:90}
nor the need to introduce functionals defined by a constrained search
\cite{Levy:79,Levy:82,Levy:85}. Furthermore, the necessary conditions for a
one-particle density matrix $\gamma$ to come from a single determinant are known,
and they can be expressed as constraints when minimizing energy functionals that
depend on $\gamma$ \cite{Lowdin:55b,Blaizot:86}.

Below, generalized Hartree--Fock equations are obtained containing the exact
exchange-potential and a nonlocal correlation-potential, where these equations are
obtained using the Brillouin-Brueckner condition -- using non-variational energy
functionals -- and functional minimization -- using variational functionals. Both
variational and non-variational approached lead to the same correlation potential
and generalized Fock-operator.  (Correlation potentials from either approach are
the same, in the sense that the (occupied) Brueckner-orbitals obtained from the {\em
variational} correlation-potential differs only from the orbitals obtained from the
{\em non-variational} one by unitary transformation.)

Using time-independent many-body perturbation theory
\cite{Goldstone:57,Hugenholtz:57,Sanders:69,Raimes:72,Paldus:75,Lindgren:86},
diagrammatic expansions are given for the non-variational energy-functionals that
are expressed in terms of orbitals and orbital energies.  When restrictions are
placed on the orbital energies, the individual diagrams are shown to {\em
explicitly} depend on the one-particle density-matrix of the reference state. The
diagrammatic expansions for the variational-energy functionals are presented
elsewhere \cite{tobe}.

Kohn--Sham variants of DFT employ a universal exchange-correlation functional,
independent of the external potential; approximations can be derived from model
systems, where, in the vicinity of the model systems, the general form of the
exchange-correlation functional is known. In contrasts, the correlation-energy
functionals introduced below depend on the external potential, and are, therefore,
in this sense, {\em not} universal. Nevertheless, as shown below, the
correlation-energy functionals can be partitioned into individual terms that are
-- to a varying degree -- universal; approximations can be derived from model
systems. For example, the electron-gas correlation-energy can be used in an {\em
electron gas approximation}, where, unlike the local density approximation (LDA)
\cite{Kohn:65}, the gas in not assumed to be uniform.

In addition, even without partitioning of the correlation-energy functionals --
with the external-potential dependence intact -- approximate functionals can still
be derived from model systems. For example, as discussed below, the
Colle--Salvetti functional \cite{Colle:75} -- derived from the helium atom -- is a
valid approximation within the proposed method.

\section{Overview}

Sec.~\ref{TRIAL} introduces four trial wavefunctions -- say
$|\Psi_\Phi^{\scscs(\eta)}\rangle$, where $\eta= \mbox{{\small I, II, III}, and
{\small IV}}$ -- that are defined with respect to a single-determinantal
reference-state, say $|\Phi\rangle$.  The first trial-wavefunction
$|\Psi_\Phi^{\scscs (\mathrm{I})}\rangle$ is simply the target state of interest,
say $|\Psi\rangle$, with the single excitations removed. The second
trial-wavefunction $|\Psi_\Phi^{\scscs (\mathrm{II})}\rangle$ is defined with
respect to the target state expressed by an exponential ansatz: ($|\Psi\rangle=
e^S|\Phi\rangle$), where $|\Psi_\Phi^{\scscs (\mathrm{II})}\rangle$ is generated
by removing the single-excitation amplitudes $S_1$ from the cluster-operator
$S$. All of the trial states $|\Psi_\Phi^{\scscs (\eta)}\rangle$ -- including the
third and fourth ones defined below -- share the property that they contain no
single excitations; furthermore, they generate the target state of interest
$|\Psi\rangle$ when their reference states satisfies
($|\Phi\rangle=|\Theta\rangle$), where $|\Theta\rangle$ is the determinantal state
constructed from occupied Bruckner orbitals. In other words, we have
($|\Psi_\Theta^{\scscs (\eta)}\rangle = |\Psi\rangle$).

Using intermediate normalization, the exact energy of interest, say ${\cal E}$, is just
$\langle\Phi| H |\Psi\rangle$, and it can be partitioned into two terms: the first-order
energy, say $E_1[\Phi]$ -- given by $\langle\Phi| H |\Phi\rangle$ -- and the correlation
energy, say ${\cal E}_{\mathrm{co}}[\Phi]$ -- which, unlike ${\cal E}$, is a functional of the
reference state $|\Phi\rangle$; these concepts are briefly review in Sec.~\ref{basic}.

Four non-variational energy-functionals, say $E_\eta[\Phi]$, are defined in an analogous way
as ${\cal E}$: ($E_\eta[\Phi]= \langle \Phi|H|\Psi_\Phi^{\scscs (\eta)}\rangle$), where they
yield the exact energy for the Brueckner reference-state: (${\cal E}=E_\eta[\Theta]$), and
these functionals can also be partitioned into two terms:
($E_\eta[\Phi]=E_1[\Phi]+E_{\mathrm{co}}^{\scscs (\eta)}[\Phi]$), where
$E_{\mathrm{co}}^{\scscs (\eta)}[\Phi]$ is the {\em correlation-energy functional}, and we have
(${\cal E}_{\mathrm{co}}[\Theta]=E_{\mathrm{co}}^{\scscs(\eta)}[\Theta]$); the details are
presented in Sec.~\ref{TRIAL}.

Because of the one-to-one correspondence between the set of determinant states,
say $\{|\Phi\rangle\}$, and the one-particle density-matrices
\cite{Blaizot:86,Parr:89}, say $\{\gamma\}$, the correlation energy, say ${\cal
E}_{\mathrm{co}}[\gamma]$, and the correlation-energy functionals, say
$E_{\mathrm{co}}^{\scscs (\eta)}[\gamma]$ -- or any other functions and
functionals of $\gamma$ -- can be transfered into ones that depend on the
one-particle density-matrix $\gamma$, as discussed in Sec.~\ref{IMPL}; the
correlation-energy and exact energy satisfy: (${\cal
E}_{\mathrm{co}}[\tau]=E_{\mathrm{co}}^{\scscs(\eta)}[\tau]$) and (${\cal
E}=E_\eta[\tau]$), where $\tau$ is the one-particle density matrix of the
Brueckner reference-state $|\Theta\rangle$.

Unlike the first-two trial wavefunctions, $|\Psi_\gamma^{\scscs (\mathrm{I})}\rangle$ and
$|\Psi_\gamma^{\scscs (\mathrm{II})}\rangle$, that are generated by removing single-excitation
amplitudes from the target state $|\Psi\rangle$, the third and fourth trial wavefunctions,
$|\Psi_\gamma^{\scscs (\mathrm{III})}\rangle$ and $|\Psi_\gamma^{\scscs (\mathrm{IV})}\rangle$,
are obtained by solving the coupled-cluster theory
\cite{Hubbard:57,Coester:58,Cizek:66,Cizek:69,Cizek:71,Lindgren:78,Bartlett:78,
Pople:78,Lindgren:86} and configuration interaction
\cite{Boys:50,Schaefer:72,Roos:77,Shavitt:77,Szabo:82,Harris:92,Dykstra:88} equations,
respectively, in an approximate way -- by neglecting the single-excitation portions.

The coupled cluster formalism is briefly reviewed in Sec.~\ref{tlct}.  A
transparent perturbative treatment of the coupled cluster theory is presented in
Sec.~\ref{LDE} that is useful to obtain a perturbative expansion for the third
correlation-energy functional $E_{\mathrm{co}}^{\scscs (\text{III})}$. (This
approach is compared to Lindgren's variant of the link diagram theorem in
Sec.~\ref{tldt}.) Sec.~\ref{RSPT} reviews Rayleigh-Schr\"{o}dinger perturbation
theory that can be used to generate the fourth correlation-energy functional
$E_{\mathrm{co}}^{\scscs (\text{IV})}$ when the linked diagram theorem is {\em
not} invoked and the single-excitation subspace is neglected.

In Secs.~\ref{EXPL} and \ref{EXPL2}, diagrammatic expansions using many-body
perturbation theory are presented for the correlation energy ${\cal
E}_{\mathrm{co}}[\gamma]$ and the correlation-energy functionals
$E_{\mathrm{co}}^{\scscs (\eta)}[\gamma]$. The individual diagrams depend on the
orbitals -- both occupied and unoccupied -- and the orbital energies -- defined by
the zeroth-order Hamiltonian. By using degenerate sets of occupied and unoccupied
orbitals, and additional methods, it is demonstrated that diagrams can be defined
that {\em explicitly} depend $\gamma$.

In Sec.~\ref{extpot} an approach based on many-body perturbation theory is
introduced, where the perturbation is partitioned into terms that depend on the
external potential, say $v$, and the remaining portion that is $v$ independent;
the perturbation expansions for the correlation energy ${\cal E}_{\mathrm{co}}$
and the correlation-energy functionals $E_{\mathrm{co}}^{\scscs (\eta)}$ mirrors
this partitioning, yielding terms that depend on $v$ and, the remainder, called
the {\em electron gas terms}, that are $v$ independent. The electron gas terms are
the only terms that contribute to ${\cal E}_{\mathrm{co}}$ and
$E_{\mathrm{co}}^{\scscs(\eta)}$ for an electron gas and, in most other cases, are
the dominant portions of ${\cal E}_{\mathrm{co}}$ and
$E_{\mathrm{co}}^{\scscs(\eta)}$; they are also universal functions -- independent
of the external potential. {\em Atomic}, {\em diatomic} and {\em molecular} terms
are defined in an analogous way are are obtained by further partitioning the
perturbation into potential terms from the individual nuclei and selectively
partitioning the perturbation expansions for ${\cal E}_{\mathrm{co}}$ and
$E_{\mathrm{co}}^{\scscs(\eta)}$.

Using the electron gas terms, an {\em electron gas approximation} is proposed in
Sec~\ref{APPROX} that is an alternative to the LDA. Additional approximations are
also considered in this Sec.\ including one that leads to the Colle--Salvetti
functional \cite{Colle:75}.

Generalized Hartree--Fock equations are defined in Sec.~\ref{F}, where the exact
Fock-operators, say ${\cal \hat{F}}_{\!\Theta}^{\scscs (\eta)}$, generate the
Brueckner orbitals, and these operators are defined by the trial wavefunctions
$|\Psi_\Theta^{\scscs(\eta)}\rangle$ and the Brillouin-Brueckner condition, which
is reviewed in Sec.~\ref{BBC}. In addition -- from the one-to-one correspondence
mentioned above -- we can also write ${\cal \hat{F}}_{\tau}^{\scscs (\eta)}$;
where, it is demonstrated that these operators are independent of $\eta$; so, in
addition, we can omit the $\eta$ subscript and write ${\cal \hat{F}}_{\tau}$.
Solving the generalized Hartree--Fock equations permit the determination of the
Brueckner orbitals, and the one-particle density-matrix, $\tau$, that is defined
by these orbitals, permitting the determination of ${\cal E}$ and ${\cal
E}_{\mathrm{co}}[\tau]$, since, as mentioned above, they are given by
$E_\eta[\tau]$ and $E_{\mathrm{co}}^{\scscs(\eta)}[\tau]$, respectively; A
correlation potential, say $v_{\mathrm{co}}^{\scscs \tau}$, is also defined, and
satisfies the following identity: (${\cal \hat{F}}_{\tau} = \hat{F}_\tau +
v_{\mathrm{co}}^{\scscs \tau}$), where $\hat{F}_\tau$ is the Fock operator,
determined by $\tau$; it is constructed from the Brueckner orbitals.

A variational formalism is presented in Sec.~\ref{VARI}, where energy functionals,
say $\bar{E}_\eta[\gamma]$, are defined using the same trial wavefunctions as in
the non-variational case: ($\bar{E}_\eta[\gamma] = \langle\Psi_\gamma^{\scscs
(\eta)}| H|\Psi_\gamma^{\scscs (\eta)}\rangle \left[\langle\Psi_\gamma^{\scscs
(\eta)}|\Psi_\gamma^{\scscs (\eta)}\rangle\right]^{-1}$), where the exact energy
is generated for the Brueckner-state one-particle density-matrix: (${\cal
E}=\bar{E}_\eta[\tau]$).  These functionals are minimized subject to the
constraint that the one-particle density-matrix comes from a single-determinantal
state $\mbox{\large $\gamma$}\mbox{\footnotesize $(|\Phi\rangle)$}$.  The
functional derivative of $\bar{E}_\eta[\gamma]$ -- with respect to the
one-particle density-matrix -- generates ${\cal \zeta}_{\tau}^{\scscs
(\eta)}(\mathbf{x}_1,\mathbf{x}_2)$, where these two-body functions are the
kernels of generalized Fock operators, say $\mathcal{\hat{\zeta}}_{\tau}^{\scscs
(\eta)}$, and it is demonstrated that these operators are independent of $\eta$;
in addition, these operators are -- in a the sense mentioned in the introduction,
Sec.~\ref{INTRO} -- equivalent to the non-variational operators, ${\cal
\hat{F}}_{\tau}$. A correlation potential, say $\hat{\nu}_{\mathrm{co}}^{\scscs
\tau}$, is also obtained that is defined by its kernels which is given by the
functional derivative of variational correlation-energy functionals, say
$\bar{E}_{\mathrm{co}}^{\scscs (\eta)}[\gamma]$. The electron-gas and
Colle--Salvetti functionals, mentioned above, are valid within the variational
approach, permitting the determination of approximate correlation
potentials~$\hat{\nu}_{\mathrm{co}}^{\scscs \tau}$ by functional differentiation.

A brief comparison with the Brueckner-orbital, {\em exact SCF theory} by L\"owdin
\cite{Lowdin:62} and Kobe \cite{Kobe:71} is presented in Appendix~\ref{ESCF}.

\section{Perturbation and Coupled Cluster Theory} \label{THEORY} 

\subsection{The exact and correlation energies} \label{basic} 
 
We seek solutions of the time-independent Schr\"odinger equation,
\begin{equation} \label{SE} 
H|\Psi\rangle ={\cal E}|\Psi\rangle,
\end{equation}
where $|\Psi\rangle$ is an eigenstate of the Hamiltonian operator,
\begin{equation} \label{H} 
H = \sum_{ij} [i|\hat{h}|j]
a_i^\dagger a_j +  
\frac{1}{2}\sum_{ijkl} [ij|kl] 
a_i^\dagger a_k^\dagger a_l a_j,
\end{equation} 
and the integrals are written using chemist's notation \cite{Szabo:82}: 
\begin{eqnarray} \label{h1} 
[i|\hat{h}|j]&=&
[i|({-}\mbox{\small$\frac{1}{2}$}\nabla^2)|j] + [i|v|j], \\
\label{chemist} 
\mbox{} [ij|kl]&=&\sum_{\omega_2\omega_2} \int 
\psi_{i}^*(\mathbf{x}_1)
\psi_{j}(\mathbf{x}_1) 
r_{12}^{-1}
\psi_{k}^*(\mathbf{x}_2) 
\psi_{l}(\mathbf{x}_2) \; d \mathbf{r}_1 d \mathbf{r}_{2},
\end{eqnarray} 
where the spatial and spin coordinates, $\mathbf{r}$ and $\omega$, are denoted
collectively by $\mathbf{x}$.

The wavefunction of interest $|\Psi\rangle$ can be generated by a wave operator
$\Omega_\Phi$:
\begin{equation} \label{WO} 
\Omega_\Phi |\Phi\rangle = |\Psi\rangle,
\end{equation}
where $|\Phi\rangle$ is a determinantal reference-state.

The reference state $|\Phi\rangle$ is completely defined by its {\em o}ccupied
orbitals; we denote these orbitals by $\{\psi_o\mbox{\small $\rightarrow\Phi$}\}$;
the set of {\em u}noccupied orbitals -- the virtual orbital set -- is denoted by
$\{\psi_u\mbox{\small $\rightarrow\Phi$}\}$. The virtual set $\{\psi_u\mbox{\small
$\rightarrow\Phi$}\}$ also determines the occupied set, since the two sets are
orthogonal, and the union of the two sets is a complete set. Hence,
$\{\psi_u\mbox{\small $\rightarrow\Phi$}\}$ also determines $|\Phi\rangle$.
Unless stated otherwise, two sets of either occupied or unoccupied orbitals that
differ by a unitary transformation are considered equivalent.

We use the following orbital convention: Arbitrary orbitals are denoted by $i$ and
$j$; occupied orbitals are denoted by $w$, $x$, and $y$; virtual orbitals are
denoted by $r$, $s$, and $t$:
\begin{subequations}
\label{phi.orbs.index} 
\begin{eqnarray} 
\label{wxy} 
\psi_w,\psi_x,\psi_y&\in&\{\psi_o\rightarrow\Phi\}, \\ 
\label{rst} 
\psi_r,\psi_s,\psi_t&\in&\{\psi_u\rightarrow\Phi\}, \\
\label{ijk} 
\psi_i,\psi_j,\psi_k&\in&\{\psi_o\rightarrow\Phi\} \cup \{\psi_u\rightarrow\Phi\}.
\end{eqnarray}
\end{subequations}

Explicitly, our spin-orbitals $\psi_{i}(\mathbf{x})$ have the following form:
\begin{equation} \label{spinorb} 
\psi_{i}(\mathbf{x})=
\chi_{i\sigma}(\mathbf{r})\sigma(\omega), 
\;\; \sigma=\alpha \mbox{ or } \beta,
\end{equation}
where the spin and spatial portions are given by $\chi_{i\sigma}(\mathbf{r})$ and
$\sigma(\omega)$, respectively, and the spatial functions
$\chi_{i\sigma}(\mathbf{r})$ are permitted to be unrestricted -- two spin orbitals
do not, in general, share the same spatial function.

In principle, $|\Phi\rangle$ can be any determinantal state that overlaps with the
reference state: \mbox{($\langle \Phi |\Psi\rangle \ne 0$)}. However, our interest
is often in cases where the target state $|\Psi\rangle$ is a ground state that is
well described by a closed-shell reference-state $|\Phi\rangle$, and the
Hamiltonian is spin-free -- it contains no spin coordinates.  In these cases,
instead of Eq.~(\ref{spinorb}), we often use a spatially restricted set of
orbitals, given by
\begin{equation} \label{spinorb.res} 
\psi_{i\sigma}(\mathbf{x})=
\chi_i(\bm{r})\sigma(\omega), \;\; \sigma=\alpha,\beta,
\end{equation}
so that $|\Phi\rangle$ is determined by a set of doubly-occupied spatial orbitals,
denoted by $\{\chi_o\mbox{\footnotesize $\rightarrow\Phi$}\}$, where this set also
determines the virtual set $\{\chi_u\mbox{\footnotesize $\rightarrow\Phi$}\}$. Two
sets of orbitals that differ by a unitary transformation are considered, again, as
equivalent.

By multiplying the Schr\"odinger Eq.~(\ref{SE}) from the left by $\langle\Phi|$,
and requiring intermediate normalization to be satisfied,
\begin{equation} \label{IN} 
\langle \Phi |\Psi\rangle = \langle \Phi |\Omega_\Phi |\Phi\rangle=1,
\end{equation}
we get
\begin{equation} \label{E0.a} 
{\cal E} = \langle\Phi| H |\Psi\rangle = E_1[\Phi] + 
{\cal E}_{\mathrm{co}}[\Phi],
\end{equation}
where the first-order energy is
\begin{equation} 
\label{first.energy} 
E_1[\Phi] = \langle\Phi| H |\Phi\rangle 
=
\sum_{w\in \{\psi_o\rightarrow \Phi\}}
[w|({-}
\mbox{\small$\frac{1}{2}$}
\nabla^2) + v + \mbox{\large $\frac12$}
(J_\Phi - K_\Phi)|w], 
\end{equation}
and the Coulomb $J_\Phi(\mathbf{r})$ and exchange
$K_\Phi(\mathbf{x})$ operators have their usual forms:
\begin{eqnarray}
  [i|J_\Phi |j] &=& \sum_{x\in \{\psi_o\rightarrow\Phi\}} [xx|ij],\\
{}[i|K_\Phi |j] &=& \sum_{x\in \{\psi_o\rightarrow\Phi\}} [xi|jx];
\end{eqnarray}
furthermore,  the correlation energy ${\cal E}_{\mathrm{co}}[\Phi]$, given by
\begin{equation} \label{Corr} 
{\cal E}_{\mathrm{co}}[\Phi]=\langle\Phi| H |\Psi_{Q_{\Phi}}\!\rangle,
\end{equation}
is obtained from the correlation function:
\begin{equation} \label{Corr.fun} 
|\Psi_{Q_{\Phi}}\!\rangle= Q_\Phi|\Psi\rangle,
\end{equation}
where the orthogonal-space projector is 
\begin{equation} \label{Q} 
Q_\Phi = 1 - |\Phi\rangle\langle\Phi|.
\end{equation}

The first-order energy can also be written as
\begin{equation} \label{Efirst.cl} 
E_1[\Phi] = \langle\Phi| H |\Phi\rangle=\left(H\right)_{\text{cl}},
\end{equation}
where the cl subscript indicates the closed portion -- the fully contracted terms
that, diagrammatically speaking, have no external free-lines
\cite{Cizek:66,Cizek:69,Lindgren:86,Paldus:75}.  Appendix~\ref{orb.part} presents
partitioning or second-quantized operators into {\em closed} and {\em open}
portions in a slightly different manner than is done by other authors. 

Similar to the first-order energy, for the correlation energy we have
\begin{equation} \label{Ecorr.cl} 
{\cal E}_{\mathrm{co}}[\Phi]=\left(H\chi_\Phi\right)_{\text{cl}},
\end{equation}
where the  correlation operator $\chi_\Phi$, defined by
\begin{equation} \label{corr.op} 
\chi_\Phi= \Omega_\Phi - 1,
\end{equation}
generates the correlation function $|\Psi_{Q_{\Phi}}\!\rangle$ when operating on the
reference state:
\begin{equation}
\chi_\Phi|\Phi\rangle = |\Psi_{Q_{\Phi}}\!\rangle.
\end{equation}
The sum of Eqs.~(\ref{Efirst.cl}) and (\ref{Ecorr.cl}) gives the exact energy:
\begin{equation} \label{exactE} 
{\cal E} =\left(H\Omega_\Phi\right)_{\text{cl}}.
\end{equation}

We also write down the expression for the exchange-correlation energy:
\begin{equation} \label{EXC} 
{\cal E}_{\mathrm{xc}}[\Phi] = {\cal E}_{\mathrm{co}}[\Phi] 
- E_{\mathrm{x}}[\Phi],
\end{equation}
where the exchange energy $E_{\mathrm{x}}[\Phi]$ is the last term on the right side of
Eq.~(\ref{first.energy}):
\begin{equation} \label{EX} 
E_{\mathrm{x}}[\Phi]=
\mbox{\large $\frac12$} \!\!\!\!\!\! \sum_{w\in \{\psi_o\rightarrow \Phi\}}[w|K_\Phi|w].
\end{equation}

\subsection{The linked cluster theorem} \label{tlct} 

The wave operator $\Omega_\Phi$ can be expressed in an exponential form
\cite{Hubbard:57,Coester:58,Cizek:66,Cizek:69,Cizek:71,Lindgren:78,Bartlett:78,Pople:78,Lindgren:86},
\begin{subequations}
\label{OmegaT} 
\begin{equation} \label{OmegaTa} 
\Omega_\Phi=  e^{S_{\Phi}} = 1 + S_{\Phi} + \frac{1}{2!} S^2_{\Phi} 
+ \frac{1}{3!} S^3_{\Phi} + \cdots,
\end{equation}
where the cluster operator $S_{\Phi}$ can be written as a sum of one-, two-, and
higher-body terms,
\begin{equation} \label{T} 
S_{\Phi} = S_1^{\Phi}+S_2^{\Phi}+S_3^{\Phi}+\cdots,
\end{equation}
\end{subequations}
and these amplitudes are defined by the following relations:
\begin{subequations}
\label{T.amp} 
\begin{eqnarray} \label{T1} 
S_1^{\Phi}&=&\sum_{rw} s_{rw}^{\Phi} 
a_r^\dagger a_w,
\\ \label{T2} 
S_2^{\Phi}&=&\frac{1}{2!}\sum_{rwsx} s_{rwsx}^{\Phi} 
a_r^\dagger a_s^\dagger a_x a_w,
\\
\label{T3} 
S_3^{\Phi}&=&\frac{1}{3!}\sum_{rwsxty} s_{rwsxty}^{\Phi} 
a_r^\dagger a_s^\dagger a_t^\dagger a_y a_x a_w,
\\
\vdots\;\;\;\; && \nonumber
\end{eqnarray}
\end{subequations}
which use the orbital convention given by Eqs.~(\ref{phi.orbs.index});
furthermore, the coefficients are required to satisfy exchange symmetry:
\begin{subequations}
\label{exch.symmT} 
\begin{eqnarray}
s_{rwsx}^{\Phi}&=&s_{sxrw}^{\Phi}, \\
s_{rwsxty}^{\Phi}&=&s_{sxrwty}^{\Phi} = s_{rwtysx}^{\Phi},\, \ldots.
\end{eqnarray}
\end{subequations}
The cluster operator $S_\Phi$ and its amplitudes $S_n^{\Phi}$ are invariant
to unitary transformations of the occupied or virtual orbitals
\cite{Stolarczyk:84}.

Since $S_{\Phi}$ -- given by Eqs.~(\ref{T}) and (\ref{T.amp}) -- is open, only
connected (cn) portions contribute to the correlation- and exact-energies, ${\cal
E}_{\mathrm{co}}[\Phi]$ and ${\cal E}$, given by Eqs.~(\ref{Ecorr.cl}) and
(\ref{exactE}). Therefore, we can write
\begin{eqnarray} 
\label{Ecorr.cl.op} 
{\cal E}_{\mathrm{co}}[\Phi] &= &\left(H\chi_\Phi\right)_{\text{cl,cn}},
\\ \label{E.cl.op} 
{\cal E} &=& \left(H\Omega_\Phi\right)_{\text{cl,cn}},
\end{eqnarray}
where the additional cn subscripts indicate that only the connected portions
contribute -- contractions in which all $S_{\Phi}$ amplitudes are connected
together by $H$.  

Eq.~(\ref{E.cl.op}) indicates that the closed part of $(H\Omega_\Phi)_{\text{cn}}$
gives the energy of interest, ${\cal E}$; the open part is the mathematical
statement of the linked-cluster theorem
\cite{Cizek:66,Cizek:69,Cizek:71,Lindgren:78,Bartlett:78,Pople:78,Lindgren:86}:
\begin{equation} \label{lct.op} 
\left(H\Omega_\Phi \right)_{ \text{op,cn}} = 0.
\end{equation}

Using the time-independent form of Wick's theorem
\cite{Bogoliubov:59,Paldus:75,Lindgren:86,Cizek:69}, the operator product
$H\Omega_\Phi$ can be written as a sum of zero-, one-, two- and higher-body
excitations:
\begin{equation} \label{HOm.exp} 
H\Omega_\Phi = {\cal E} + 
\left(H\Omega_\Phi \right)_{1} +
\left(H\Omega_\Phi \right)_{2} +
\left(H\Omega_\Phi \right)_{3} + \cdots,
\end{equation}
where the notation $(\ldots)_{n}$ indicates that the $n$-body term within the
brackets $(\ldots)$ is normal-ordered with respect to the $|\Phi\rangle$ vacuum
state. Substituting this expression into Eq.~(\ref{lct.op}), we get
\begin{equation} \label{lctb} 
\sum_{n=1}^\infty \left[\left(H\Omega_\Phi \right)_{n} \right]_{\text{op,cn}} = 0.
\end{equation}
Since each term is linearly independent, the solution is
\begin{equation} \label{lctc} 
\left[\left(H\Omega_\Phi \right)_{n} \right]_{\text{op,cn}} = 0, \; n\ne 0.
\end{equation}
This relation can be used to obtain the coupled-cluster equations, satisfied by
the $S_{\Phi}$ amplitudes, $S_n^{\Phi}$
\cite{Hubbard:57,Coester:58,Cizek:66,Cizek:69,Cizek:71,Lindgren:78,Bartlett:78,Pople:78,Lindgren:86}.

\subsection{Perturbation treatment of the linked cluster theorem} \label{LDE} 

For Rayleigh-Schr\"{o}dinger perturbation theory
\cite{Goldstone:57,Hugenholtz:57,Sanders:69,Raimes:72,Paldus:75,Lindgren:74,Wilson:85,Lindgren:86,Harris:92},
the wave- and cluster-operators are given by order-by-order expansions:
\begin{subequations}
\label{ordexp} 
\begin{eqnarray} \label{T.ordexp} 
S_{\Phi}&=&S_{\Phi}^{(1)}+S_{\Phi}^{(2)}+S_{\Phi}^{(3)}+\ldots,
\\ \label{wo.ordexp} 
\Omega_{\Phi}&=&\Omega_\Phi^{(0)}+\Omega_{\Phi}^{(1)}+\Omega_{\Phi}^{(2)}+\ldots,
\end{eqnarray}
where
\begin{equation} \label{Omega0} 
\Omega_\Phi^{(0)}=1.
\end{equation}
\end{subequations}
Substituting these relations into Eq.~(\ref{OmegaTa}) and equating each order, we
get the following identities \cite{Lindgren:78,Lindgren:86}:
\begin{subequations}
\label{T=Omega.ord} 
\begin{eqnarray} \label{T1=Omega1.ord} 
\Omega^{(1)}_\Phi&=&S^{(1)}_\Phi, 
\\ \label{T2.Omega2.ord} 
\Omega^{(2)}_\Phi&=&S^{(2)}_\Phi + \frac{1}{2!} \left(S^{(1)}_\Phi\right)^2, \\
\Omega^{(3)}_\Phi&=&S^{(3)}_\Phi + S^{(1)}_\Phi S^{(2)}_\Phi + 
\frac{1}{3!} \left(S^{(1)}_\Phi\right)^3, 
\\
\vdots\;\;\;\;&& \nonumber
\end{eqnarray}
\end{subequations}
where the order from each term is defined as a sum of the superscripts, e.g.,
$S^{\scscs (1)}_\Phi S^{\scscs (2)}_\Phi$ is a third order term.

For a perturbative treatment, we partition the Hamiltonian into a zeroth-order
Hamiltonian $H_0$ and a perturbation~$V$:
\begin{equation} \label{Hpart} 
H = H_0 + V,
\end{equation}
where we require the reference state
$|\Phi\rangle$ to be an eigenfunction of $H_0$, and a one-body operator:
\begin{eqnarray} 
\label{h0.eigen} 
H_0 |\Phi\rangle&=&E_0 |\Phi\rangle, \\
\label{H0.onebod} 
H_0&=&\sum_{ij} \epsilon_{ij} a^\dagger_i a_j.
\end{eqnarray}

The above zeroth-order Hamiltonian is defied by its matrix elements.  We choose
them by requiring the following relation to be satisfied:
\begin{subequations} \label{H0.matel} 
\begin{equation}
\epsilon_{ij}= \epsilon_{ji}= \epsilon_{ij}^{\mbox{\tiny $\Phi$}}, \; 
\end{equation}
where
\begin{eqnarray} 
\label{H0.wr} 
\epsilon_{wr}^{\mbox{\tiny $\Phi$}}&=& 0, \; 
\\ \label{H0.wx} 
\epsilon_{wx}^{\mbox{\tiny $\Phi$}}&=&
\langle \psi_w|\hat{f}_o^{\scscs \Phi}|\psi_x \rangle,
\\ \label{H0.rs} 
\epsilon_{rs}^{\mbox{\tiny $\Phi$}}&=&
\langle \psi_r|\hat{f}_u^{\scscs \Phi}|\psi_s \rangle,
\end{eqnarray}
\end{subequations}
and the one-body operators, $\hat{f}_o^{\scscs \Phi}$ and $\hat{f}_u^{\scscs
\Phi}$, are determined by the reference state $|\Phi\rangle$, but the dependence
of $\hat{f}_o^{\scscs \Phi}$ and $\hat{f}_u^{\scscs \Phi}$ upon $|\Phi\rangle$ is
at our disposal; the orbital subspaces are, again, defined by
Eqs.~(\ref{phi.orbs.index}).

Using the above choice, our zeroth-order Hamiltonian becomes
\begin{equation} \label{H0.nd} 
H_{0}^{\Phi} = \sum_{w,x\in \{\psi_o\rightarrow \Phi\}}
\epsilon_{wx}^{\mbox{\tiny $\Phi$}} a^\dagger_w a_x + \sum_{r,s\in \{\psi_u\rightarrow \Phi\}}
\epsilon_{rs}^{\mbox{\tiny $\Phi$}} a^\dagger_r a_s,
\end{equation}
where the appended ${\Phi}$ superscript indicates that $H_0^{\scscs \Phi}$ now
depends on the reference state~$|\Phi\rangle$.  

A linked diagram expansion for $\Omega_{\Phi}$ and ${\cal E}_{\mathrm{co}} [\Phi]$
is known to exist for a zeroth-order Hamiltonian that is a diagonal, one-body,
operator
\cite{Goldstone:57,Hugenholtz:57,Sanders:69,Raimes:72,Paldus:75,Lindgren:74,Wilson:85,Lindgren:86,Harris:92}. A
diagonal form for our one-body operator, $H_0^\Phi$, is obtained when we choose its
orbital sets -- $\{\psi_o\mbox{\small $\rightarrow\Phi$}\}$ and
$\{\psi_u\mbox{\small $\rightarrow\Phi$}\}$ -- to satisfy the following
conditions:
\begin{subequations} \label{vphi.matel} 
\begin{eqnarray} \label{vphi.wx} 
\langle \psi_w|\hat{f}_o^{\scscs \Phi}|\psi_x \rangle&=& 
\delta_{wx} \epsilon_w^{\mbox{\tiny $\Phi$}},
\\ \label{vphi.rs} 
\langle \psi_r|\hat{f}_u^{\scscs \Phi}|\psi_s \rangle&=&
\delta_{rs} \epsilon_r^{\mbox{\tiny $\Phi$}},
\end{eqnarray}
\end{subequations}
where we denote these particular sets of orbitals by $\{\psi_o \mbox{\small
$\leftarrow\Phi,\hat{f}_o^{\Phi}$}\}$ and $\{\psi_u \mbox{\small
$\leftarrow\Phi,\hat{f}_u^{\Phi}$}\}$, indicating that they are uniquely
determined by $|\Phi\rangle$ and their one-particle operator, $\hat{f}_o^{\scscs
\Phi}$ or $\hat{f}_u^{\scscs \Phi}$.

Using these orbitals, $H_0^{\Phi}$ can be written as
\begin{equation} \label{H0.d} 
H_0^{\Phi} = 
\sum_{w\in \{\psi_o\leftarrow \Phi,\hat{f}_o^{\scscs \Phi}\}} 
\epsilon_{w}^{\mbox{\tiny $\Phi$}} a^\dagger_w a_w + 
\sum_{r\in \{\psi_u\leftarrow \Phi,\hat{f}_u^{\scscs \Phi}\}} 
\epsilon_{r}^{\mbox{\tiny $\Phi$}} a^\dagger_r a_r,
\end{equation}
where our partitioning is
\begin{equation} \label{Hpart.phi} 
H = H_0^{\Phi} + V_\Phi.
\end{equation}

Since $H_0^{\Phi}$ is a second-quantized operator with no projection operators,
it is appropriate to substitute the above partitioning into Eq.~(\ref{lct.op}),
yielding
\begin{equation} \label{lct.op.partb} 
-\left(H_0^{\Phi}S_\Phi \right)_{ \text{op,cn}} 
= \left(V_\Phi\Omega_\Phi \right)_{ \text{op,cn}},
\end{equation}
Substituting Eqs.~(\ref{ordexp}) into (\ref{lct.op.partb}), and equating each
order, we have
\begin{equation} \label{lct.ord} 
-\left(H_0^{\Phi}S_\Phi^{(n)} \right)_{ \text{op,cn}} 
= \left(V_\Phi\Omega_\Phi^{(n-1)} \right)_{ \text{op,cn}},
\end{equation}
where $V_\Phi$ contributes unity to the overall order of the rhs. (In a sense,
we have: $V_\Phi\!=\!V_\Phi^{\scscs(1)}$.)  

Solving Eqs.~(\ref{lct.ord}) and (\ref{T=Omega.ord}) in an iterative and
sequential manner generates a linked diagram expansion for $\Omega_{\Phi}$ and a
connected expansion for $S_{\Phi}$, which is identical to the expansions obtained
by Lindgen \cite{Lindgren:74,Lindgren:78,Lindgren:86}. The procedure begins by
using Eq.~(\ref{Omega0}), which gives $S^{(1)}_\Phi$ and $\Omega^{(1)}_\Phi$ from
Eqs.~(\ref{lct.ord}) and (\ref{T1=Omega1.ord}). Substituting $\Omega^{(1)}_\Phi$
back into Eq.~(\ref{lct.ord}) gives $S^{(2)}_\Phi$, permitting the calculation of
$\Omega^{(2)}_\Phi$ from (\ref{T2.Omega2.ord}), and so on. However, in order to
get the same diagrams as Lindgren, the factorization theorem
\cite{Hugenholtz:57,Frantz:60,Baker:71,Brandow:67,Brandow:77,Sanders:69,Lindgren:74,Lindgren:86}
must by exploited to unfactorize the disconnected products that appear on the right side
on Eq.~(\ref{T=Omega.ord}). The diagrammatic expansions for the correlations and
exact energies, ${\cal E}_{\mathrm{co}}[\Phi]$ and ${\cal E}$, are obtained from
Eqs.~(\ref{Ecorr.cl.op}) and (\ref{E.cl.op}). It is worth mentioning that these
diagrammatic expansions are only obtained when the one-body,
zeroth-order Hamiltonian $H_0^{\Phi}$ is a diagonal operator, permitting the left side
of Eq.~(\ref{lct.ord}) to be evaluated to give terms containing factors of
orbital-energy differences, e.g., ($\epsilon_w^{\mbox{\tiny $\Phi$}} +
\epsilon_x^{\mbox{\tiny $\Phi$}} - \epsilon_r^{\mbox{\tiny $\Phi$}} -
\epsilon_s^{\mbox{\tiny $\Phi$}}$).

Note that Eq.~(\ref{lct.ord}) can also be expressed by
\begin{equation} \label{lct.ord.comm} 
\left[S_\Phi^{(n)},H_0^{\Phi}\right] P_\Phi
= Q_\Phi\left(V_\Phi\Omega_\Phi^{(n-1)} \right)_{ \text{cn}}P_\Phi,
\end{equation}
where the reference state projector is
\begin{equation} \label{P} 
P_\Phi= |\Phi \rangle \langle \Phi |.
\end{equation}
Expression (\ref{lct.ord.comm}) is easily obtained from Eq.~(\ref{lct.ord}) by
using the definition of an open operator, adding
($P_\Phi\left[S_\Phi^{(n)},H_0^{\Phi}\right] P_\Phi = 0$), and using the following
identity:
\begin{equation}
Q_\Phi\left[S_\Phi^{(n)},H_0^{\Phi}\right] P_\Phi = 
-Q_\Phi \left(H_0^{\Phi}S_\Phi^{(n)} \right)_{ \text{cn}} P_\Phi. 
\end{equation}
%

\subsection{Rayleigh-Schr\"{o}dinger perturbation theory and the Bloch Equations}
\label{RSPT} 

Using Eqs.~(\ref{WO}), (\ref{E0.a}), and (\ref{P}), the Schr\"odinger Eq.\
(\ref{SE}) can be written
\begin{equation} \label{Blocha} 
H\Omega_\Phi|\Phi\rangle = \Omega_\Phi  P_\Phi H \Omega_\Phi |\Phi\rangle.
\end{equation}
The operator form of this Eq.\ is the Bloch
Eq.\ \cite{Bloch:58,Kvasnicka:74,Lindgren:74,Lindgren:78}:
\begin{equation} \label{Bloch.gen} 
H\Omega_\Phi P_\Phi = \Omega_\Phi  P_\Phi H \Omega_\Phi P_\Phi,
\end{equation}
which yields Eq.~(\ref{Blocha}) when multiplied by $|\Phi\rangle$ from the right
side.  Substituting Eq.~(\ref{Hpart}) into (\ref{Bloch.gen}), and using
(\ref{IN}) and (\ref{h0.eigen}), where $H_0$ is Hermitian, gives a variant of the
Bloch equation \cite{Bloch:58,Lindgren:74,Kvasnicka:74,Lindgren:78}:
\begin{equation} \label{Bloch} 
\left(E_0 - H_0 \right) \Omega_\Phi P_\Phi = Q_\Phi \left(V \Omega_\Phi - 
\Omega_\Phi P_\Phi V\Omega_\Phi \right) P_\Phi,
\end{equation}
where we have used the following identity:
\begin{equation}
P_\Phi\left(V \Omega_\Phi - \Omega_\Phi P_\Phi V\Omega_\Phi \right) = 0,
\end{equation}
and this identity follows from intermediate normalization, given by
Eq.~(\ref{IN}), i.e., ($P_\Phi \Omega_\Phi P_\Phi = P_\Phi$).

For Rayleigh-Schr\"{o}dinger perturbation theory, the wave-operator
$\Omega_{\Phi}$ is given by the order-by-order expansion (\ref{wo.ordexp}).
Substituting this expression into Eq.~(\ref{Bloch}) and equating the individual
orders, gives \cite{Lindgren:74,Lindgren:86}
\begin{eqnarray} \label{Bloch.order} 
\left(E_0 - H_0 \right) \Omega_\Phi^{(n)} P_\Phi =
Q_\Phi \left[V \Omega_\Phi^{(n-1)} - 
\sum_{m=1}^{n-1}\Omega_\Phi^{(n-m)} P_\Phi V\Omega_\Phi^{(m-1)} \right] P_\Phi,
\;\;\;\;
\end{eqnarray}
where the second term on the right side does not appear for ($n=1$).

\subsection{The Linked diagram theorem} \label{tldt} 

When the zeroth-order Hamiltonian is in the diagonal, one-body form, given by Eq.\
(\ref{H0.d}), it can be shown that the wave-operator $\Omega_\Phi$ satisfies a
{\em linked diagram theorem} \cite{Lindgren:78,Lindgren:86}:
\begin{subequations}
\begin{equation} \label{ldt.ho2} 
\left(E_0 - H_0^{\Phi}\right)  \Omega_\Phi P_\Phi= 
Q_\Phi\left( V_\Phi \Omega_\Phi \right)_{\text{l}}P_\Phi,
\end{equation} 
where the individual orders, defined by Eq.~(\ref{wo.ordexp}), satisfy
\begin{equation} \label{ldt.order} 
\left(E_0 - H_0^{\Phi}\right)  \Omega_\Phi^{(n)} P_\Phi=
Q_\Phi\left( V_\Phi \Omega_\Phi^{(n-1)} \right)_{\text{l}}P_\Phi,
\end{equation} 
\end{subequations}
and the additional $l$ subscripts indicate that only the linked portions
contribute -- all disconnected terms are open.  

In order to solve either of the above two Eqs, the wave operator $\Omega_\Phi$ is
written as a sum of one-, two-, and higher-body excitations,
\begin{equation} \label{Omega} 
\Omega_\Phi=1+\Omega_1^{\Phi}+\Omega_2^{\Phi}+\Omega_3^{\Phi}+\cdots,
\end{equation}
where
\begin{subequations}
\label{Omega.comp} 
\begin{eqnarray} \label{Omega1} 
\Omega_1^{\Phi}&=&\sum_{rw} x_{rw}^{\Phi}
a_r^\dagger a_w,
\\ \label{Omega2} 
\Omega_2^{\Phi}&=&\frac{1}{2!} \sum_{rwsx} x_{rwsx}^{\Phi} 
a_r^\dagger a_s^\dagger a_x a_w, 
\\ 
\label{Omega3} 
\Omega_3^{\Phi}&=&\frac{1}{3!}\sum_{rwsxty} x_{rwsxty}^{\Phi} 
a_r^\dagger a_s^\dagger a_t^\dagger a_y a_x a_w,
\\
\vdots\;\;\;\;&& \nonumber
\end{eqnarray}
\end{subequations}
and Eqs.~(\ref{phi.orbs.index}) remain valid.  In addition, we require our
coefficients to have exchange symmetry:
\begin{subequations}
\label{exch.symm} 
\begin{eqnarray}
x_{rwsx}^{\Phi}&=&x_{sxrw}^{\Phi}, \\
x_{rwsxty}^{\Phi}&=&x_{sxrwty}^{\Phi} = x_{rwtysx}^{\Phi}, \ldots . 
\end{eqnarray}
\end{subequations}
The wave operator $\Omega_\Phi$ and its $n$-body operators $\Omega_n^{\Phi}$ are
also invariant to a unitary transformation of either its occupied orbitals
$\{\psi_o\mbox{\footnotesize $\rightarrow\Phi$}\}$ or its virtual orbitals
$\{\psi_u\mbox{\footnotesize $\rightarrow\Phi$}\}$.  

The linked diagram theorem provides an exponential form for the wave operator
$\Omega_\Phi$; comparing the two forms of the wave operator, Eqs.~(\ref{OmegaT})
and (\ref{Omega}), the following identities are easily proven:
\begin{subequations}
\label{T=Omega} 
\begin{eqnarray} \label{T1=Omega1} 
\Omega_1^{\Phi}&=&S_1^{\Phi}, 
\\ \label{T2.Omega2} 
\Omega_2^{\Phi}&=&S_2^{\Phi} + \frac12 \left(S_1^{\Phi}\right)^2, \\
\Omega_3^{\Phi}&=&S_3^{\Phi} + S_1^{\Phi} S_2^{\Phi} + 
\frac{1}{3!} \left(S_1^{\Phi}\right)^3, 
\\
\vdots\;\;\;\;&& \nonumber
\end{eqnarray}
\end{subequations}

\section{Brillouin-Brueckner condition} \label{BBC} 

Consider the Slater determinantal state, say $|\Theta\rangle$, that satisfies the
Brillouin-Brueckner condition
\cite{Brenig:57,Lowdin:62,Nesbet:58,Lowdin:62,Kobe:71,Schafer:71},
\begin{equation} \label{BB-cond} 
\langle \Theta_{w}^{r}|H|\Psi\rangle = 0,
\end{equation}
for any single excitation from $|\Theta\rangle$:
\begin{equation}
|\Theta_{w}^{r}\rangle= a^\dagger_{r} a_{w} |\Theta\rangle,
\end{equation}
where both the occupied and virtual orbitals determine the Brueckner
determinantal-state~$|\Theta\rangle$:
\begin{subequations}
\label{theta.orbs} 
\begin{eqnarray} \label{theta.orbw} 
\psi_w &\in& \{\psi_o \rightarrow \Theta \}, \\
\psi_r &\in& \{\psi_u \rightarrow \Theta \}.
\end{eqnarray}
\end{subequations}

Using Eqs.~(\ref{SE}) and (\ref{BB-cond}), it is easily demonstrated that the
wavefunction $|\Psi\rangle$ contains no single excitations from $|\Theta\rangle$:
\begin{equation} \label{no.sing} 
\frac{1}{{\cal E}}\langle \Theta_{w}^{r}|H|\Psi\rangle = 
\langle \Theta_{w}^{r}|\Psi\rangle = 0.
\end{equation}
Since the states $|\Theta_{w}^{r}\rangle$ are linearly independent, the
wavefunction satisfies the following condition:
\begin{equation}\label{beta-BB} 
P_{11}^{\Theta}|\Psi\rangle = 0.
\end{equation}
where the projector for the singly-excited states is
\begin{equation} \label{P11} 
P_{11}^{\Theta} = 
\sum_{w\in \{\psi_o\rightarrow \Theta\}} 
\sum_{r\in \{\psi_u\rightarrow \Theta\}}
|\Theta_{w}^{r}\rangle\langle\Theta_{w}^{r}|,
\end{equation}
and this subspace is completely determined by $|\Theta\rangle$; $P_{11}^{\Theta}$
is also invariant to a unitary transformation of occupied, or virtual, orbitals
\cite{Stolarczyk:84}.

Using Eqs.~(\ref{no.sing}), (\ref{beta-BB}) and (\ref{P11}), Eq.~(\ref{BB-cond})
can be generalized:
\begin{equation}\label{BB2} 
P_{11}^{\Theta}H\left(1-P_{11}^{\Theta}\right)|\Psi\rangle = 0.
\end{equation}
The occupied set of orbitals $\{\psi_o \mbox{\footnotesize $\rightarrow\Theta$}\}$
that satisfy Eq.~(\ref{BB2}) are called Brueckner orbitals. However, since these
orbitals are invariant to a unitary transformation, Eq.~(\ref{BB2}) actually defines
the Brueckner-determinantal state $|\Theta\rangle$, since $|\Theta\rangle$ determines
$P_{11}^{\scriptscriptstyle \Theta}$.

Note also the following identities:
\begin{subequations}
\label{Om1=T1=0} 
\begin{eqnarray} \label{Omega1=0} 
\Omega_1^{\Theta}&=&0,
\\ \label{T1=0} 
S_1^{\Theta}&=&0. 
\end{eqnarray}
\end{subequations}
The first identity is obtained by substituting Eq.~(\ref{WO}) into (\ref{beta-BB})
and using (\ref{Omega}) and (\ref{Omega.comp}) for ($\Phi=\Theta$). The second identity
uses either Eq.~(\ref{T1=Omega1}) or Eqs.~(\ref{OmegaT}) and (\ref{T.amp}).

Since $S_1^{\Theta}$ is zero, we have
\begin{equation}
\Omega_\Theta = e^{-S_1^{\Theta}} \Omega_\Theta.
\end{equation}
Multiplying this equation from the right by $|\Theta\rangle$ and using Eq.\
(\ref{WO}) gives
\begin{equation} \label{BB2.ccA} 
|\Psi\rangle = e^{-S_1^{\Theta}}|\Psi\rangle. 
\end{equation}
Substituting this equation into Eq.~(\ref{BB-cond}), and using Eq.~(\ref{P11}),
gives
\begin{equation}\label{BB2.cc} 
P_{11}^{\Theta}He^{-S_1^{\Theta}}|\Psi\rangle = 0.
\end{equation}
This equation is the Brillouin-Brueckner condition for coupled cluster
theory \cite{Stolarczyk:84,Finley:ehf}. As in Eq.~(\ref{BB2}), the
Brueckner orbitals that satisfy Eq.~(\ref{BB2.cc}) are invariant to a
unitary transformation, so Eq.~(\ref{BB2.cc}) defines the
determinantal state $|\Theta\rangle$, since $|\Theta\rangle$
determines $P_{11}^{\scriptscriptstyle \Theta}$ and $S_1^{\scriptscriptstyle \Theta}$.

\section{Trial Wavefunctions and energy functionals} \label{TRIAL} 

\subsection{General Requirements} \label{GTRIAL} 

Consider four trial wavefunctions, denoted by $|\Psi_\Phi^{\scscs (\eta)}\rangle$,
where $\eta= \mbox{{\small I, II, III}, and {\small IV}}$. Each of these four
states depends on the reference state $|\Phi\rangle$, 
satisfies intermediate normalization,
\begin{equation}
\langle \Phi|\Psi_\Phi^{\scscs (\eta)}\rangle = 1,
\end{equation}
has no components within the singly-excited subspace,
\begin{equation} \label{trial.wf} 
|\Psi_\Phi^{\scscs (\eta)}\rangle= \left(1 - P_{11}^{\Phi}\right) 
|\Psi_\Phi^{\scscs (\eta)}\rangle,
\end{equation}
and yields the exact state of
interest when $|\Phi\rangle$ is the Brueckner determinantal-state:
\begin{equation} \label{trwf=exact} 
|\Psi_\Theta^{\scscs (\eta)}\rangle = |\Psi\rangle.
\end{equation}

From these trial wavefunctions $|\Psi_\Phi^{\scscs (\eta)}\rangle$, we can
construct energy functionals:
\begin{equation} \label{Efunct} 
E_\eta[\Phi]= \langle \Phi|H|\Psi_\Phi^{\scscs (\eta)}\rangle= E_1[\Phi] + 
E_{\mathrm{co}}^{\scscs (\eta)}[\Phi],
\end{equation}
where the correlation (co) energy-functionals are given by
\begin{equation} \label{Ecorr} 
E_{\mathrm{co}}^{\scscs (\eta)}[\Phi] = 
\langle\Phi|H|\Psi_{Q_{\Phi}}^{\scscs (\eta)}\hspace{-0.1ex}\rangle,
\end{equation}
the trial correlation-functions are
\begin{equation} \label{Corr.funal} 
Q_{\Phi}|\Psi_\Phi^{\scscs (\eta)}\rangle=
|\Psi_{Q_{\Phi}}^{\scscs (\eta)}\hspace{-0.1ex}\rangle,
\end{equation}
and $E_1[\Phi]$ is given by Eq.~(\ref{first.energy}).  Operating on Eq.\
(\ref{trwf=exact}) by $Q_{\Theta}$ and using Eqs.~(\ref{Corr.fun}) and
(\ref{Corr.funal}) we have
\begin{equation} \label{Ecorr.fun.br} 
|\Psi_{Q_{\Theta}}^{\scscs (\eta)}\hspace{-0.1ex}\rangle =
|\Psi_{Q_{\Theta}}\rangle.
\end{equation}

Let us also define exchange-correlation (xc) energy-functionals:
\begin{equation} \label{Excorr} 
E_{\mathrm{xc}}^{\scscs (\eta)}[\Phi] = E_{\mathrm{co}}^{\scscs (\eta)}[\Phi] - 
E_{\mathrm{x}}[\Phi],
\end{equation}
where the exchange energy $E_{\mathrm{x}}[\Phi]$ is given by Eq.~(\ref{EX}).

Eqs.~(\ref{E0.a}), (\ref{trwf=exact}), and (\ref{Efunct}) indicate that the energy
functionals $E_\eta[\Phi]$ yield the exact energy ${\cal E}$ when the reference
state $|\Phi\rangle$ is the Brueckner determinantal state $|\Theta\rangle$:
\begin{equation} \label{fy.exact} 
{\cal E}= E_\eta[\Theta], 
\end{equation}
and from Eqs.~(\ref{Corr}), (\ref{EXC}), (\ref{trwf=exact}), (\ref{Ecorr}), and
(\ref{Excorr}), the following identities are obtained for the correlation
and exchange-correlation energies, ${\cal E}_{\mathrm{co}}[\Theta]$ and ${\cal
E}_{\mathrm{xc}}[\Theta]$:
\begin{eqnarray} \label{Eco.ident} 
{\cal E}_{\mathrm{co}}[\Theta]&=&E_{\mathrm{co}}^{\scscs (\eta)}[\Theta],\\
\label{Exc.ident} 
{\cal E}_{\mathrm{xc}}[\Theta]&=&E_{\mathrm{xc}}^{\scscs (\eta)}[\Theta].
\end{eqnarray}

Substituting Eq.~(\ref{trwf=exact}) into Eq.~(\ref{BB2}), and using Eq.\
(\ref{trial.wf}) for ($|\Phi\rangle = |\Theta\rangle$), gives the
Brillouin-Brueckner condition for our trial wavefunctions:
\begin{equation} \label{trial.BBcond} 
P_{11}^{\Theta}H|\Psi_\Theta^{\scscs (\eta)}\rangle = 0.
\end{equation}

We now define the explicit forms of these trial wavefunctions and give expressions
for their correlation-energy functionals. Additional expressions for the
correlation-functionals in terms of their wave-operator, or cluster-operator,
amplitudes, e.g., $t_{rwsx}^{\Phi}$, are given in Sec.~\ref{Cef}; diagrammatic
expressions are presented in Sec.~\ref{EXPL2}.

\subsection{The first trial wavefunction} \label{Case1} 

The first trial-wavefunction is given by
\begin{equation} \label{trial.I} 
|\Psi_\Phi^{\scscs (\mathrm{I})}\rangle = \left(1-P_{11}^{\Phi}\right)|\Psi\rangle.
\end{equation}
It follows from Eqs.~(\ref{Omega}) and (\ref{Omega.comp}) that
$\Omega_1^{\Phi}$ exclusively generates the singly-excited portion of the
orthogonal space:
\begin{subequations}
\begin{eqnarray}
P_{11}^{\Theta}\Omega_1^{\Phi}|\Phi\rangle&=&\Omega_1^{\Phi}|\Phi\rangle,\\
P_{11}^{\Theta}\Omega_n^{\Phi}|\Phi\rangle&=& 0, \; n \ne 1.
\end{eqnarray}
\end{subequations}
Therefore, this trial wavefunction can be written
\begin{equation} \label{trial.Ib} 
|\Psi_\Phi^{\scscs (\mathrm{I})}\rangle = 
\left(\Omega_\Phi - \Omega_1^{\Phi} \right)|\Phi\rangle.
\end{equation}
Using this expression and Eq.~(\ref{corr.op}), after substituting
Eq.~(\ref{Corr.funal}) into (\ref{Ecorr}), yields the first correlation-energy
functional:
\begin{equation} \label{Ecorr.I} 
E_{\mathrm{co}}^{\scscs (\mathrm{I})}[\Phi] = 
\left[H\left( \chi_\Phi - \Omega_1^{\Phi} \right)\right]_{\text{cl}},
\end{equation}
where we also used ($P_{\Phi}(\chi_\Phi - \Omega_1^{\Phi})|\Phi\rangle=0$).

\subsection{The second trial wavefunction} \label{Case2} 

The second trial-wavefunction is given by
\begin{equation} \label{trial.IIa} 
|\Psi_\Phi^{\scscs (\mathrm{II})}\rangle=e^{-S_1^{\Phi}} |\Psi\rangle.
\end{equation}
Using Eqs.~(\ref{WO}) and (\ref{OmegaTa}), this Eq.\ becomes 
\begin{equation} \label{trial.IIb} 
|\Psi_\Phi^{\scscs (\mathrm{II})}\rangle=e^{(S_{\Phi}-S_1^{\Phi})}|\Phi\rangle,
\end{equation}
where we used the identity, given by
\begin{equation} \label{ident.II} 
e^{-S_1^{\Phi}}\Omega_\Phi = e^{(S_{\Phi}-S_1^{\Phi})}, 
\end{equation}
and this relation follows from Eqs.~(\ref{OmegaT}), since $S_{\Phi}$ and
$S_1^{\Phi}$ commute \cite{Mattuck:76}.

Using Eq.~(\ref{trial.IIb}) after substituting Eq.~(\ref{Corr.funal}) into
(\ref{Ecorr}), gives the second correlation-energy functional:
\begin{equation} \label{Ecorr.II} 
E_{\mathrm{co}}^{\scscs (\mathrm{II})}[\Phi] = 
\left[H\left(e^{(S_{\Phi}-S_1^{\Phi})} - 1 
\right)\right]_{\text{cl}},
\end{equation}
where we also used the following:
\begin{equation}
\langle \Phi |HP_{\Phi} e^{(S_{\Phi}-S_1^{\Phi})} |\Phi\rangle -  
\langle \Phi |H|\Phi\rangle = 0.
\end{equation}

\subsection{The third trial wavefunction} \label{Case3} 

The third trial-wavefunction $|\Psi_\Phi^{\scscs (\mathrm{III})}\rangle$ can be
generated by its wave-operator:
\begin{equation} \label{ccf.wo} 
\hat{\Omega}_\Phi|\Phi\rangle = |\Psi_\Phi^{\scscs (\mathrm{III})}\rangle,
\end{equation}
that can be expressed in an exponential form,
\begin{subequations}
\label{OmegaT.ccf} 
\begin{equation} \label{OmegaTa.ccf} 
\hat{\Omega}_\Phi= 
e^{\hat{S}_{\Phi}} = 1 + \hat{S}_{\Phi} + \frac{1}{2!}
\hat{S}^2_{\Phi} + \frac{1}{3!} \hat{S}^3_{\Phi} + \cdots,
\end{equation}
where $\hat{S}_{\Phi}$ can be written as of sum $n$-body excitations, with the
exclusion of a one-body operator:
\begin{equation} \label{T.ccf} 
\hat{S}_{\Phi}= \hat{S}_2^{\Phi}+\hat{S}_3^{\Phi}+\cdots.
\end{equation}
\end{subequations}
The individual amplitudes are defined by the following equations:
\begin{subequations} \label{T.ccf.amp} 
\begin{eqnarray} \label{T2.ccf} 
\hat{S}_2^{\Phi}&=&\frac{1}{2!}\sum_{rwsx} \hat{s}_{rwsx}^{\Phi} 
a_r^\dagger a_s^\dagger a_x a_w,
\\
\label{T3.ccf} 
\hat{S}_3^{\Phi}&=&\frac{1}{3!}\sum_{rwsxty} \hat{s}_{rwsxty}^{\Phi} 
a_r^\dagger a_s^\dagger a_t^\dagger a_y a_x a_w,
\\
\vdots\;\;\;\;&& \nonumber
\end{eqnarray}
\end{subequations}
Let us also mention that the orbital convention, Eqs.~(\ref{phi.orbs.index}),
remains valid; we also require the coefficients to possess exchange symmetry, as in
Eqs.~(\ref{exch.symmT}).

Using Eqs.~(\ref{ccf.wo}) and (\ref{OmegaTa.ccf}), after substituting
Eq.~(\ref{Corr.funal}) into (\ref{Ecorr}), gives the third correlation-energy
functional:
\begin{equation} \label{Ecorr.III} 
E_{\mathrm{co}}^{\scscs (\mathrm{III})}[\Phi] = 
\left[H\left(e^{\hat{S}_{\Phi}} - 1 
\right)\right]_{\text{cl}},
\end{equation}
where we used the following:
\begin{equation}
\langle \Phi |HP_{\Phi} e^{\hat{S}_{\Phi}}
|\Phi\rangle - 
\langle \Phi |H|\Phi\rangle
= 0.
\end{equation}

We define $\hat{\Omega}_{\Phi}$ as a solution to the following variant of Eq.\
(\ref{lct.op}):
\begin{equation} \label{lct.op.iv} 
\left(1-P_{11}^{\Phi}\right) \left(H\hat{\Omega}_\Phi \right)_{\text{op,cn}} = 0,
\end{equation}
which defines the trial functional $|\Psi_\Phi^{\scscs (\mathrm{III})}\rangle$
using Eq.~(\ref{ccf.wo}). 

As in Eq.~(\ref{HOm.exp}), the operator $H\hat{\Omega}_\Phi$ can also be written
as a sum of zero-, one-, two- and higher-body excitations:
\begin{equation} \label{HOm.expIII} 
H\hat{\Omega}_\Phi = E_{\scscs \mathrm{III}}[\Phi] + 
\left(H\hat{\Omega}_\Phi \right)_{1} +
\left(H\hat{\Omega}_\Phi \right)_{2} +
\left(H\hat{\Omega}_\Phi \right)_{3} + \cdots,
\end{equation}
where we use the identity, given by
\begin{equation}
E_{\scscs \mathrm{III}}[\Phi] =
\left(H\hat{\Omega}_\Phi\right)_{\text{cl}} =
\left(H\hat{\Omega}_\Phi\right)_{0},
\end{equation}
and this relation follow from Eqs.~(\ref{Efunct}) and (\ref{ccf.wo}).  (See also
Appendix \ref{orb.part}.)

Substituting Eq.~(\ref{HOm.expIII}) into Eq.~(\ref{lct.op.iv}) yields
\begin{equation} \label{lct.op.ivb} 
\left(H\hat{\Omega}_\Phi \right)_{\text{op,cn}} -
\left[\left(H\hat{\Omega}_\Phi \right)_{1} \right]_{\text{op,cn}}
= 0.
\end{equation}
Substituting expansion (\ref{HOm.expIII}) into Eq.~(\ref{lct.op.ivb}) and noting
that each term is linearly independent, we have
\begin{equation} \label{lctc.op} 
\left[\left(H\hat{\Omega}_\Phi \right)_{n} \right]_{\text{op,cn}} = 0, 
\; n \ge 2,
\end{equation}
and this relation can be used to obtain the coupled cluster equations for the
$\hat{S}_n^{\Phi}$ amplitudes.

We now demonstrate that $|\Psi_\Phi^{\scscs (\mathrm{III})}\rangle$ is a valid
trial wavefunction: Eq.~(\ref{trwf=exact}) and the other relations for Sec.\
(\ref{GTRIAL}) are  satisfied.

Consider a determinantal state, say $|\Theta^\prime\rangle$, that we require to
satisfy the following condition:
\begin{subequations}
\label{bbc.ccf.op.1} 
\begin{equation} \label{bbc.ccf.op.n} 
P_{11}^{\Theta^\prime}\left(H\hat{\Omega}_{\Theta^\prime} \right)_{\text{op,cn}} = 0.
\end{equation}
Using Eq.~(\ref{HOm.expIII}), we can easily verify that the following conditions
causes Eq.~(\ref{bbc.ccf.op.n}) to be satisfied:
\begin{equation} \label{bbc.ccf.op} 
\left[\left(H\hat{\Omega}_{\Theta^\prime} \right)_{1} \right]_{\text{op,cn}} = 0.
\end{equation}
\end{subequations}
Adding Eqs.~(\ref{lct.op.iv}) and (\ref{bbc.ccf.op.n}), for ($|\Phi\rangle=
|\Theta^\prime\rangle$), and comparing the result with Eq.~(\ref{lct.op}),
indicates that 
\begin{equation} 
\hat{\Omega}_{\Theta^\prime} = \Omega_{\Theta^\prime},
\end{equation}
and therefore we have
\begin{equation} \label{psi.thetap.op} 
|\Psi_{\Theta^\prime}^{\scscs (\mathrm{III})}\rangle = |\Psi\rangle.
\end{equation}
(Combining Eqs.~(\ref{lctc.op}) and (\ref{bbc.ccf.op}), and comparing the result
with Eq.~(\ref{lctc}), also yields the above two Eqs.)

Multiplying Eq.~(\ref{psi.thetap.op}) by $P_{11}^{\Theta^\prime}$ and using
Eqs.~(\ref{ccf.wo}) and (\ref{OmegaT.ccf}) gives
\begin{equation} \label{psi.thpri.op} 
P_{11}^{\Theta^\prime}|\Psi\rangle = 0.
\end{equation}
Comparing this Eq.\ with Eq.~(\ref{beta-BB}) indicate that $|\Theta^\prime
\rangle$ is the Brueckner state:
\begin{equation} \label{bst.thpri.op} 
|\Theta^\prime \rangle=|\Theta\rangle.
\end{equation}
Substituting this result into Eq.~(\ref{psi.thetap.op}) indicates that Eq.\
(\ref{trwf=exact}) is satisfies. All other relations from Sec.\ (\ref{GTRIAL}) are
easily verified. For example, Eqs.~(\ref{fy.exact}), (\ref{Eco.ident}), and
(\ref{Exc.ident}) follow from Eqs.~(\ref{Efunct}) through (\ref{Excorr}).

Substituting Eq.~(\ref{bst.thpri.op}) into Eqs.~(\ref{bbc.ccf.op.1}) yields the
following identities:
\begin{subequations}
\begin{eqnarray} 
\label{bbc.alt.op} 
P_{11}^{\Theta}\left(H\hat{\Omega}_{\Theta} \right)_{\text{op,cn}}&=&0,
\\ \label{bbc.alt.op.n} 
\left[\left(H\hat{\Omega}_{\Theta} \right)_{1} \right]_{\text{op,cn}}&=&0.
\end{eqnarray}
\end{subequations}
These equivalent relations are alternative representations of the
Brillouin-Brueckner condition, since if they satisfied, then Eq.\
(\ref{trial.BBcond}) is also satisfied. Note that $\hat{\Omega}_{\Phi}$ does not
possess a single-excitation operator. i.e., $\hat{S}_1^{\Phi}$ is absent in Eq.\
(\ref{T.ccf}).

Using the definition of an open operator, Eqs.~(\ref{lct.op.iv}) and
(\ref{bbc.alt.op}) can be represented by the following:
\begin{eqnarray} \label{lct.iv} 
\tilde{Q}_\Phi \left(H\hat{\Omega}_\Phi \right)_{\text{cn}} P_\Phi&=&0,
\\ \label{bbc.alt} 
P_{11}^{\Theta} \left( H \hat{\Omega}_{\Theta} \right)_{\text{cn}} P_\Theta&=&0,
\end{eqnarray}
where
\begin{equation} \label{Qb} 
\tilde{Q}_\Phi= Q_\Phi- P_{11}^{\Phi}.
\end{equation}

As in Eqs.~(\ref{ordexp}), the wave- and cluster-operators of interest
are given by order-by-order expansions:
\begin{subequations}
\label{ordexp.III} 
\begin{eqnarray} \label{T.ordexp.III} 
\hat{S}_{\Phi}&=&\hat{S}_{\Phi}^{(1)}+\hat{S}_{\Phi}^{(2)}+\hat{S}_{\Phi}^{(3)}+\ldots,
\\ \label{wo.ordexp.III} 
\hat{\Omega}_{\Phi}&=&\hat{\Omega}_\Phi^{(0)}+\hat{\Omega}_{\Phi}^{(1)}+
\hat{\Omega}_{\Phi}^{(2)}+\ldots,
\end{eqnarray}
where ($\hat{\Omega}_\Phi^{(0)}=1.$)
\end{subequations}
Substituting these relations into Eq.~(\ref{OmegaTa.ccf}) and equating each order, we
get similar identities as in Eqs.~(\ref{T=Omega.ord}):
\begin{subequations}
\label{T=Omega.ord.III} 
\begin{eqnarray} \label{T1=Omega1.ord.III} 
\hat{\Omega}^{(1)}_\Phi&=&\hat{S}^{(1)}_\Phi, 
\\ \label{T2.Omega2.ord.III} 
\hat{\Omega}^{(2)}_\Phi&=&\hat{S}^{(2)}_\Phi + \frac12
\left(\hat{S}^{(1)}_\Phi\right)^2, 
\end{eqnarray}
\end{subequations}
and so so. 

From Eq.~(\ref{lct.op.iv}), and using Eqs.~(\ref{Hpart.phi}) and
(\ref{ordexp.III}) -- as in the derivation of Eq.~(\ref{lct.ord}) -- we have
\begin{equation} \label{lct.ord.III} 
- \left(1-P_{11}^{\Phi}\right)
\left(H_0^{\Phi}\hat{S}_\Phi^{(n)} \right)_{\text{op,cn}} 
= 
\left(1-P_{11}^{\Phi}\right)
\left(V_\Phi\hat{\Omega}_\Phi^{(n-1)} \right)_{\text{op,cn}}.
\end{equation}
Solving Eqs.~(\ref{lct.ord.III}) and (\ref{T=Omega.ord.III}) in an iterative and
sequential manner generates a linked diagram expansion for $\hat{\Omega}_{\Phi}$
and a connected expansion for $\hat{S}_{\Phi}$.  The diagrams representing
$E_{\mathrm{co}}^{\scscs (\mathrm{III})}[\Phi]$ and $\hat{S}_\Phi^{(n)}$ are
discussed in Sec.\ \ref{EXPL} and elsewhere \cite{tobe}, respectively.

Of the four trial wavefunctions $|\Psi_\Phi^{\scscs (\eta)}\rangle$, the third one
$|\Psi_\Phi^{\scscs (\mathrm{III})}\rangle$, we believe, is the most applicable;
the fourth one is presented below for completeness.

\subsection{The fourth trial wavefunction} \label{Case4} 

The fourth trial wavefunction is a solution of the Sch\"odinger Eq.\ within the
subspace that neglects the single-excited states:
\begin{equation} \label{trial.IVa} 
\left(1 - P_{11}^{\Phi}\right)H|\Psi_\Phi^{\scscs (\mathrm{IV})}\rangle =
E_{\scscs \mathrm{IV}}[\Phi] |\Psi_\Phi^{\scscs (\mathrm{IV})}\rangle.
\end{equation}
From the variational theorem, it follows that the above energy functional provides an upper
bound to the exact energy:
\begin{equation} \label{var.IV} 
E_{\scscs \mathrm{IV}}[\Phi] \ge {\mathcal E}.
\end{equation}
We now prove that the exact wavefunction and energy satisfy Eqs.\
(\ref{trwf=exact}) and (\ref{fy.exact}), where \mbox{($\eta=\mbox{IV}$)}.

The proof uses the Schr\"odinger Eq.~(\ref{SE}), which can be written
\begin{equation}
P_{11}^{\Theta}H|\Psi\rangle + \left(1 - P_{11}^{\Theta}\right)H|\Psi\rangle =
{\cal E}P_{11}^{\Theta}|\Psi\rangle+
{\cal E} \left(1 - P_{11}^{\Theta}\right)|\Psi\rangle, 
\end{equation}
where he have added and subtracted $P_{11}^{\Theta}$ terms. Eqs.~(\ref{BB-cond}),
(\ref{beta-BB}), and (\ref{P11}) indicate that the first terms on the right and
left sides vanish, so we have
\begin{equation} \label{SE.equiv} 
\left(1 - P_{11}^{\Theta}\right)H|\Psi\rangle =
{\cal E} \left(1 - P_{11}^{\Theta}\right)|\Psi\rangle.
\end{equation}
Eqs.~(\ref{trwf=exact}) and (\ref{fy.exact}), for ($\eta=\mbox{IV}$), are
obtained by comparing Eqs.~(\ref{trial.IVa}) and (\ref{SE.equiv}), and using
(\ref{trial.wf}). All other relations from Sec.\ (\ref{GTRIAL}) are easily
verified.

As in the exact wavefunction of interest $|\Psi\rangle$, the trial wavefunction
$|\Psi_\Phi^{\scscs (\mathrm{IV})}\rangle$ can be generated by a wave operator
$\tilde{\Omega}_\Phi$:
\begin{equation} \label{WO.var} 
\tilde{\Omega}_\Phi |\Phi\rangle = |\Psi_\Phi^{\scscs (\mathrm{IV})}\rangle,
\end{equation}
where $\tilde{\Omega}_\Phi$ is similar to $\Omega_\Phi$ -- defined by Eqs.\
(\ref{Omega}) and (\ref{Omega.comp}) -- except that there is no excitation
operator into $P_{11}^{\Phi}$:
\begin{equation} \label{Omega.var} 
\tilde{\Omega}_\Phi=1+\tilde{\Omega}_2^{\Phi}+\tilde{\Omega}_3^{\Phi}+\cdots,
\end{equation}
where
\begin{subequations}
\label{Omega.comp.var} 
\begin{eqnarray} \label{Omega2.var} 
\tilde{\Omega}_2^{\Phi}&=&\frac{1}{2!}\sum_{rwsx} \tilde{x}_{rwsx}^{\Phi} 
a_r^\dagger a_s^\dagger a_x a_w, 
\\ 
\label{Omega3.var} 
\tilde{\Omega}_3^{\Phi}&=&\frac{1}{3!}\sum_{rwsxty} \tilde{x}_{rwsxty}^{\Phi} 
a_r^\dagger a_s^\dagger a_t^\dagger a_y a_x a_w,
\\
\vdots\;\;\;\;&& \nonumber
\end{eqnarray}
\end{subequations}

Using Eq.~(\ref{WO.var}), after substituting Eq.~(\ref{Corr.funal}) into
(\ref{Ecorr}) gives the fourth correlation-energy functional:
\begin{equation} \label{Ecorr.IV} 
E_{\mathrm{co}}^{\scscs (\mathrm{IV})}[\Phi] = 
\left[H \left(\tilde{\Omega}_\Phi -1\right) \right]_{\text{cl}},
\end{equation}
where we added ($\langle \Phi|HP_\Phi (\tilde{\Omega}_\Phi -1)|\Phi\rangle=0$).

It is easily proven that $\tilde{\Omega}_\Phi$ is a solution to the following
variants of Eqs.~(\ref{wo.ordexp}) and (\ref{Bloch.order}):
\begin{eqnarray} \label{wo.exp.var} 
\tilde{\Omega}_{\Phi}&=&1+\tilde{\Omega}_{\Phi}^{(1)}+\tilde{\Omega}_{\Phi}^{(2)}+\ldots,
\\ \label{Bloch.var} 
\left(E_0 - H_0 \right) \tilde{\Omega}_\Phi^{(n)} P_\Phi&=&
\tilde{Q}_\Phi \left[V \tilde{\Omega}_\Phi^{(n)} - 
\sum_{m=1}^{n-1}\tilde{\Omega}_\Phi^{(n-m)} P_\Phi V\tilde{\Omega}_\Phi^{(m-1)} \right] P_\Phi,
\end{eqnarray}
where $\tilde{Q}_\Phi$ is defined by Eq.~(\ref{Qb}), and we require $H_0$ to
satisfy
\begin{equation}
P_{11}^{\Phi}H_0\left(1-P_{11}^{\Phi}\right) = 0.
\end{equation}
%

\section{Expressions for the correlation-energy functionals:} \label{Cef} 

Using Wick's theorem \cite{Bogoliubov:59,Paldus:75,Lindgren:86,Cizek:69}, the
Hamiltonian can be separated in to zero-, one-, and two-body parts:
\begin{equation} \label{H.norm} 
H= {H}^{\Phi}_c + {H}^{\Phi}_1 + {H}^{\Phi}_2, 
\end{equation}
where
\begin{subequations}
\label{H.comp} 
\begin{eqnarray} \label{Hc} 
{H}^{\Phi}_c &=& \langle \Phi | H | \Phi\rangle = E_1[\Phi]=(H)_{\text{cl}}, 
\\  \label{H1} 
{H}^{\Phi}_1 &=& \sum_{ij} [i|\hat{F}_\Phi|j] \{a_i^\dagger a_j\},
\\ \label{H2} 
{H}^{\Phi}_2 &=& \frac{1}{2}\sum_{ijkl} [ij|kl]
\{a_i^\dagger a_k^\dagger a_l a_j\},
\end{eqnarray}
\end{subequations}
where the Fock operator is given by
\begin{equation} \label{Fock} 
\hat{F}_\Phi
= - \mbox{$\frac{1}{2}$}\nabla^2+
v +  J_\Phi - K_\Phi,
\end{equation}
and the identities within Eqs.~(\ref{Hc}) follow from Eqs.~(\ref{first.energy}) and
(\ref{Efirst.cl}).

Substituting Eqs.~(\ref{corr.op})  and (\ref{H.norm}) into (\ref{Ecorr.cl}),
and using (\ref{Omega}), yield an expression for the correlation energy:
\begin{equation} \label{Corrd} 
{\cal E}_{\mathrm{co}}[\Phi] = \left({H}^{\Phi}_1\Omega_1^{\Phi} + 
{H}^{\Phi}_2\Omega_2^{\Phi} 
\right)_{\text{cl}},
\end{equation}
and from Eqs.~(\ref{H.comp}) and (\ref{Omega.comp}) we have
\begin{eqnarray}  
\label{Ec.eval} 
{\cal E}_{\mathrm{co}}[\Phi]=
\sum_{rw}x_{rw}^{\Phi}[w|\hat{F}_\Phi|r] + 
\frac12 \sum_{rwsx}x_{rwsx}^{\Phi}\left([wr|xs] - [ws|xr]\right),
\end{eqnarray}
where the coefficients are are assumed to have exchange symmetry, defined by
Eq.~(\ref{exch.symm}).

As is the correlation-energy expression (\ref{Ecorr.cl}), the correlation-energy
functionals, given by Eq.~(\ref{Ecorr}), can be written
\begin{equation} \label{Ecorr.cl.eta} 
E_{\mathrm{co}}^{\scscs (\eta)}[\Phi] = \left(H\chi_\Phi^{\scscs \eta}\right)_{\text{cl}},
\end{equation} 
where the trial correlation-operators $\chi_\Phi^{\scscs \eta}$ generate the
trial correlation-functions:
\begin{equation} \label{chi.all} 
\chi_\Phi^{\scscs \eta}|\Phi\rangle = 
|\Psi_{Q_{\Phi}}^{\scscs (\eta)}\hspace{-0.1ex}\rangle,
\end{equation} 
and $|\Psi_{Q_{\Phi}}^{\scscs (\eta)}\hspace{-0.1ex}\rangle$ is given by
Eq.~(\ref{Corr.funal}). Comparing this definition with Eqs.~(\ref{Efunct}) and
(\ref{Ecorr}), and using Eq.~(\ref{Hc}), we have
\begin{equation} \label{Efunct.b} 
E_\eta[\Phi]= \langle \Phi|H\left(1+\chi_\Phi^{\scscs
\eta}\right)|\Phi\rangle = \left[H\left(1+\chi_\Phi^{\scscs
\eta}\right)\right]_{\text{cl}}.
\end{equation}

Comparing Eqs.~(\ref{Ecorr.I}), (\ref{Ecorr.II}), (\ref{Ecorr.III}), and
(\ref{Ecorr.IV}) with (\ref{Ecorr.cl.eta}), yields the following relations:
\begin{subequations} \label{chi.alls} 
\begin{eqnarray}    
\label{chi.I} 
\chi_\Phi^{\scscs \mathrm{I}} &=& \chi_\Phi - \Omega_1^{\Phi}, 
\\ \label{chi.II} 
\chi_\Phi^{\scscs \mathrm{II}}&=& e^{(S_{\Phi}-S_1^{\Phi})} - 1,
\\ \label{chi.III} 
\chi_\Phi^{\scscs \mathrm{III}}&=& 
e^{\hat{S}_{\Phi}} - 1,
\\ \label{chi.IV} 
\chi_\Phi^{\scscs \mathrm{IV}}&=& \tilde{\Omega}_\Phi - 1.
\end{eqnarray}
\end{subequations}
Substituting Eqs.~(\ref{Ecorr.cl}) and (\ref{Ecorr.cl.eta}) into
(\ref{Eco.ident}) for ($|\Phi\rangle = |\Theta\rangle$), we get
\begin{equation} \label{chi.ident} 
\chi_\Theta^{\scscs \eta} = \chi_\Theta, 
\end{equation} 
and this expression indicates that any of the trial correlation-functions --
$\chi_\Theta^{\scscs \text{I}}$, $\chi_\Theta^{\scscs \text{I}}$,
$\chi_\Theta^{\scscs \text{III}}$, and $\chi_\Theta^{\scscs \text{IV}}$ -- can be
used to obtain the Brueckner one, $\chi_\Theta$.

Substituting Eqs.~(\ref{chi.alls}) into Eq.~(\ref{Ecorr.cl.eta}), and using
Eqs.~(\ref{corr.op}), (\ref{Omega}), (\ref{OmegaT}), (\ref{OmegaT.ccf}),
(\ref{Omega.var}), and (\ref{H.norm}), give expressions for the correlation energy
functionals:
\begin{subequations} \label{Eco} 
\begin{eqnarray}  
\label{Eco.I} 
E_{\mathrm{co}}^{\scscs (\mathrm{I})}[\Phi]&=& 
\left({H}_2^{\Phi}\Omega_2^{\Phi}\right)_{\text{cl}},
\\ \label{Eco.II} 
E_{\mathrm{co}}^{\scscs (\mathrm{II})}[\Phi]&=&
\left( {H}^{\Phi}_2 S_2^{\Phi} \right)_{\text{cl}},
\\ \label{Eco.III} 
E_{\mathrm{co}}^{\scscs (\mathrm{III})}[\Phi]&=&
\left({H}^{\Phi}_2 \hat{S}_2^{\Phi} \right)_{\text{cl}},
\\ \label{Eco.IV} 
E_{\mathrm{co}}^{\scscs (\mathrm{IV})}[\Phi]&=& 
\left({H}^{\Phi}_2 \tilde{\Omega}_2^{\Phi} \right)_{\text{cl}}.
\end{eqnarray}
\end{subequations} 
Using Eqs.~(\ref{Omega2}), (\ref{T2}), (\ref{T2.ccf}), (\ref{Omega2.var}), and
(\ref{H2}), the above energy functionals above can be evaluated, giving the
following relations:
\begin{subequations} \label{Ecoev} 
\begin{eqnarray} \label{Eco.Iev} 
E_{\mathrm{co}}^{\scscs (\mathrm{I})}[\Phi]
&=&\frac12 \sum_{rwsx}x_{rwsx}^{\Phi}\left([wr|xs] - [ws|xr]\right),
\;\;\;\;\\\label{Eco.IIev} 
E_{\mathrm{co}}^{\scscs (\mathrm{II})}[\Phi]
&=&\frac12 \sum_{rwsx}s_{rwsx}^{\Phi}\left([wr|xs] - [ws|xr]\right),
\\ \label{Eco.IIIev} 
E_{\mathrm{co}}^{\scscs (\mathrm{III})}[\Phi]
&=&\frac12 \sum_{rwsx} \hat{s}_{rwsx}^{\Phi}\left([wr|xs] - [ws|xr]\right), 
\\ \label{Eco.IVev} 
E_{\mathrm{co}}^{\scscs (\mathrm{IV})}[\Phi]
&=&\frac12 \sum_{rwsx}\tilde{x}_{rwsx}^{\Phi}\left([wr|xs] - [ws|xr]\right), 
\end{eqnarray}
\end{subequations}
where the coefficients are assumed to satisfy exchange symmetry, e.g.,
Eqs.~(\ref{exch.symmT}) and (\ref{exch.symm}).

For later use, we also mention that the correlation energy and the first
correlation-energy functional, given by Eqs.~(\ref{Corrd}) and (\ref{Eco.I}), can
be written using the $S_{\Phi}$ amplitudes:
\begin{eqnarray}  
{\cal E}_{\mathrm{co}}[\Phi]&=& \left({H}^{\Phi}_1 S_1^{\Phi} + 
\frac12 {H}^{\Phi}_2 S_1^{\Phi}S_1^{\Phi} + {H}^{\Phi}_2 
S_2^{\Phi} \right)_{\text{cl}},
\nonumber \\ \label{Ecorr.cc} 
\\ \label{E.corrbb} 
E_{\mathrm{co}}^{\scscs (\mathrm{I})}[\Phi]&=&
\left(\frac12 {H}^{\Phi}_2 S_1^{\Phi}S_1^{\Phi} +
{H}^{\Phi}_2 S_2^{\Phi} \right)_{\text{cl}},
\end{eqnarray}
where we have used Eqs.~(\ref{T1=Omega1}) and (\ref{T2.Omega2}).

\section{Exact Fock operators}\label{F} 

Consider generalized, or exact, Fock operators ${\cal \hat{F}}_{\!\Phi}^{\scscs
(\eta)}$, that are defined, in part, by the following matrix elements:
\begin{eqnarray} \label{EFock} 
\langle\psi_r| {\cal \hat{F}}_{\!\Phi}^{\scscs (\eta)}|\psi_w\rangle 
= \langle\Phi_{w}^{r}|H
|\Psi_\Phi^{\scscs (\eta)}\rangle; \; \;\;
\eta= \mbox{{\footnotesize I, II, III, IV,}}
\end{eqnarray}
where the $w$ and $r$ orbitals are occupied and unoccupied within $|\Phi\rangle$,
respectively, as noted by Eqs.~(\ref{phi.orbs.index}). By multiplying
Eq.~(\ref{trial.BBcond}) from the left by $\langle\Theta_{w}^{r}|$, using
Eq.~(\ref{P11}), and comparing the resulting relation to the above Eq., we have
\begin{equation}
\label{Fock.bbc} 
\langle\psi_r| {\cal \hat{F}}_{\!\Theta}^{\scscs (\eta)}|\psi_w\rangle = 0,
\end{equation}
where the orbitals are defined by Eq.~(\ref{theta.orbs}). When satisfied by all
orbitals, this expression is equivalent to the Brillouin-Brueckner condition,
given by Eq.~(\ref{trial.BBcond}). The operator form of Eq.~(\ref{Fock.bbc}) is
\begin{equation} \label{FBB-cond} 
\mbox{\large $\hat{\kappa}$}_\Theta \hspace{0.15ex} 
{\cal \hat{F}}_{\!\Theta}^{\scscs (\eta)}  
\mbox{\large $\hat{\gamma}$}_\Theta = 0,
\end{equation}
where $\hat{\gamma}_\Phi$ is the one-particle, density-matrix operator for the
determinantal-state $|\Phi\rangle$ \cite{Dirac:30,Dirac:31,Lowdin:55a,McWeeny:60}:
\begin{equation} \label{F.dmo} 
\hat{\gamma}_\Phi= \sum_{x\in \{\psi_o\rightarrow\Phi\}} |\psi_x\rangle\langle \psi_x|;
\end{equation}
$\hat{\kappa}_\Phi$ is the projector into the virtual-orbital subspace:
\begin{equation} \label{virt.proj} 
\hat{\kappa}_\Phi= \sum_{r\in \{\psi_u\rightarrow\Phi\}} |\psi_r\rangle\langle \psi_r|,
\end{equation}
and the identity operator \mbox{\small $\hat{I}$} can be expressed by
\begin{equation} \label{ident.oper} 
\mbox{\small $\hat{I}$} = \hat{\gamma}_\Phi + \hat{\kappa}_\Phi.
\end{equation}
Multiplying Eq.~(\ref{FBB-cond}) from the left and right by $\langle\psi_r|$ and
$|\psi_w\rangle$ gives Eq.~(\ref{Fock.bbc}).

Since {\em all} of our generalized Fock operators -- ${\cal
\hat{F}}_{\!\Theta}^{\scscs (\text{I})}$, ${\cal \hat{F}}_{\!\Theta}^{\scscs
(\text{II})}$, ${\cal \hat{F}}_{\!\Theta}^{\scscs (\text{III})}$, and ${\cal
\hat{F}}_{\!\Theta}^{\scscs (\text{IV})}$ -- satisfy Eq.~(\ref{FBB-cond}), any one
can be used to define an exact Fock operator ${\cal \hat{F}}_{\!\Theta}$:
\begin{equation} \label{FBB-cond.ident} 
\mbox{\large $\hat{\kappa}$}_\Theta \hspace{0.15ex}
{\cal \hat{F}}_{\!\Theta}
\mbox{\large $\hat{\gamma}$}_\Theta = 
\mbox{\large $\hat{\kappa}$}_\Theta \hspace{0.15ex} 
{\cal \hat{F}}_{\!\Theta}^{\scscs (\eta)}
\mbox{\large $\hat{\gamma}$}_\Theta
=
\mbox{\large $\hat{\kappa}$}_\Theta \hspace{0.15ex} 
{\cal \hat{F}}_{\!\Theta}^{\scscs (\eta\prime)}  
\mbox{\large $\hat{\gamma}$}_\Theta,
\end{equation}
and the Brillouin-Brueckner condition, Eq.~(\ref{FBB-cond}), becomes
\begin{equation} \label{FBB-cond.ex} 
\mbox{\large $\hat{\kappa}$}_\Theta \hspace{0.15ex} 
{\cal \hat{F}}_{\!\Theta}
\mbox{\large $\hat{\gamma}$}_\Theta = 0.
\end{equation}
Using Eq.~(\ref{ident.oper}), this Eq.\ can be written as
\begin{equation} \label{FBB-cond.ufactor} 
\left( \mbox{\small $\hat{I}$} - \hat{\gamma}_\Theta \right) 
{\cal \hat{F}}_{\!\Theta}  \hat{\gamma}_\Theta=0.
\end{equation}
Since $\hat{\gamma}_\Theta$ is idempotent,
\begin{equation} \label{indepotent} 
\hat{\gamma}_\Theta \hat{\gamma}_\Theta = \hat{\gamma}_\Theta,
\end{equation}
Eq.~(\ref{FBB-cond.ufactor}) can be written as
\begin{equation} \label{ident.g} 
\left({\cal \hat{F}}_{\!\Theta} \hat{\gamma}_\Theta 
-\hat{\gamma}_\Theta {\cal \hat{F}}_{\!\Theta}\right) \hat{\gamma}_\Theta
= 0.
\end{equation}
By requiring ${\cal \hat{F}}_{\!\Theta} $ to be, at least in part,
Hermitian:
\begin{equation} \label{Herm.F} 
\mbox{\large $\hat{\gamma}$}_{\!\Theta} \hspace{0.15ex} 
{\cal \hat{F}}_{\!\Theta} 
\mbox{\large $\hat{\kappa}$}_\Theta
= 0,
\end{equation}
yields the following identity:
\begin{equation} \label{ident.k} 
\left({\cal \hat{F}}_{\!\Theta} \hat{\gamma}_\Theta 
-\hat{\gamma}_\Theta {\cal \hat{F}}_{\!\Theta}\right) \hat{\kappa}_\Theta
= 0.
\end{equation}
Adding together Eqs.~(\ref{ident.g}) and (\ref{ident.k}), and using
Eq.~(\ref{ident.oper}), indicates that $\hat{\gamma}_\Theta$ and ${\cal
\hat{F}}_{\!\Theta}$ commute:
\begin{equation} \label{commute} 
\left[ {\cal \hat{F}}_{\!\Theta},\hat{\gamma}_\Theta \right] = 0.
\end{equation}
Eq.~(\ref{commute}) is a generalization of the one obtained for
Hartree-Fock theory \cite{Parr:89,Blaizot:86,Lowdin:55b}.

Note that for any reference state, say $|\Phi^\prime\rangle$, we can find a
corresponding state, $|\Phi\rangle$, in which the following relation is satisfied:
\begin{equation} \label{exact.scf} 
\mbox{\large $\hat{\kappa}$}_{\Phi} \vspace{0.1ex}
{\cal \hat{F}}_{\!\Phi^\prime}^{\scscs (\eta)}
\mbox{\large $\hat{\gamma}$}_\Phi = 0.
\end{equation}
Solving this expression in an iterative and self-consistent-field manner leads to
the Brillouin-Brueckner condition, Eq.~(\ref{FBB-cond.ex}), being satisfied, since
when ($|\Phi\rangle=|\Phi^\prime\rangle$), we have ($|\Phi\rangle=|\Theta\rangle$).

%

Consider now the following application of the identity operator:
\begin{equation}
\mbox{\small $\hat{I}$} {\cal \hat{F}}_{\!\Phi}|\psi_w\rangle =
{\cal \hat{F}}_{\!\Phi}|\psi_w\rangle =
\sum_{x\in \{\psi_o\rightarrow\Phi\}} 
\varepsilon_{xw}^{\mbox{\tiny $\Phi$}}
|\psi_x\rangle
+
\sum_{r\in \{\psi_u\rightarrow\Phi\}} 
\varepsilon_{rw}^{\mbox{\tiny $\Phi$}}
|\psi_r\rangle,
\end{equation}
where
\begin{equation}
\varepsilon_{ij}^{\mbox{\tiny $\Phi$}} = 
\langle \psi_i |{\cal \hat{F}}_{\!\Phi}| \psi_j\rangle.
\end{equation}
Setting ($\Phi=\Theta$), and using Eq.~(\ref{Fock.bbc}), gives exact Hartree--Fock Eqs:
\begin{equation}\label{eF.eqs} 
{\cal \hat{F}}_{\!\Theta}|\psi_w\rangle =
\sum_{x\in \{\psi_o\rightarrow\Theta\}} 
\varepsilon_{xw}^{\mbox{\tiny $\Theta$}}
|\psi_x\rangle, 
\end{equation}
where the orbital $|\psi_w\rangle$ is from the set $\{\psi_o\rightarrow\Theta\}$. 

Returning to Eq.~(\ref{EFock}), inserting the identity operator -- defined by
Eq.~(\ref{Q}) -- we have
\begin{eqnarray}  \label{F.mat} 
\langle\psi_r| {\cal \hat{F}}_{\!\Phi}^{\scscs (\eta)}|\psi_w\rangle 
= \langle\Phi_{w}^{r}|H|\Phi\rangle +
\langle\Phi_{w}^{r}|H|\Psi_{Q_{\Phi}}^{\scscs (\eta)}\hspace{-0.1ex}\rangle,
\end{eqnarray}
where the trial-correlation functions $|\Psi_{Q_{\Phi}}^{\scscs
(\eta)}\hspace{-0.1ex}\rangle$ are given by Eq.~(\ref{Corr.funal}); the first term
on the right side of Eq.~(\ref{F.mat}) is the off-diagonal block of the
Fock-operator $\hat{F}_\Phi$:
\begin{eqnarray} \label{F.ident} 
\langle\psi_r| \hat{F}_\Phi
|\psi_w\rangle 
= \langle\Phi_{w}^{r}|H|\Phi\rangle,
\end{eqnarray}
where is $\hat{F}_\Phi$ defined by Eq.~(\ref{Fock}).  Substituting
Eqs.~(\ref{F.ident}) and (\ref{chi.all}) into (\ref{F.mat}), we have
\begin{eqnarray}  \label{F.mat3} 
\langle\psi_r| {\cal \hat{F}}_{\!\Phi}^{\scscs (\eta)}|\psi_w\rangle
&=&\langle\psi_r| \hat{F}_\Phi |\psi_w\rangle +
\langle\Phi_{w}^{r}| H \chi_\Phi^{\scscs \eta} |\Phi\rangle
\\
&=&\langle\psi_r| (\hat{F}_\Phi)_{\text{op}} |\psi_w\rangle +
\langle\psi_r|
\left[ \left(H \chi_\Phi^{\scscs \eta}
\right)_1\right]_{\text{op}} 
|\psi_w\rangle,
\nonumber 
\end{eqnarray}
where we use the more restrictive definition of an open (op) operator, presented
in Appendix~\ref{orb.part}.

Setting ($\Phi = \Theta$), and using Eqs.~(\ref{chi.ident}) and (\ref{Fock.bbc}),
gives another variant of the Brillouin-Brueckner condition:
\begin{equation} \label{bbc.op} 
\left(\hat{F}_\Theta + 
\left( H \chi_\Theta \right)_1 \right)_{\mbox{$\!$}\text{op}} = 0,
\end{equation}
where this expression acts within the one-body sector of the Hilbert space, even
though the subscript op indicates the open portion -- defined by the $n$-body
sector.

Now let the second term on the right side of Eq.~({\ref{F.mat}) define the
off-diagonal block of correlation potentials $v_{\mathrm{co}}^{\scscs
\Phi\eta}(\mathbf{x})$, given by
\begin{equation} \label{corr.pot} 
\langle\psi_r|v_{\mathrm{co}}^{\scscs \Phi\eta}|\psi_w\rangle = 
\langle\Phi_{w}^{r}|H|\Psi_{Q_{\Phi}}^{\scscs (\eta)}\hspace{-0.1ex}\rangle.
\end{equation}
Similarly, exchange-correlation potentials $v_{\mathrm{xc}}^{\scscs
\Phi\eta}(\mathbf{x})$ are defined, in part, by
\begin{equation} \label{Br.exc} 
\langle\psi_r|v_{\mathrm{xc}}^{\scscs \Phi\eta}|\psi_w\rangle=
\langle\psi_r|v_{\mathrm{co}}^{\scscs \Phi\eta}|\psi_w\rangle - 
\langle\psi_r|K_\Phi|\psi_w\rangle.
\end{equation}
Using Eqs.~(\ref{F.mat}), (\ref{F.ident}), and (\ref{corr.pot}), and with no loss
of generality, our exact Fock operators ${\cal \hat{F}}_{\!\Phi}^{\scscs (\eta)}$
can be written
\begin{equation} \label{Falpha} 
{\cal \hat{F}}_{\!\Phi}^{\scscs (\eta)} = \hat{F}_\Phi +
v_{\mathrm{co}}^{\scscs \Phi\eta}.
\end{equation}

Multiplying Eq.~(\ref{Falpha}) from the left and right by $\langle\psi_r|$ and
$|\psi_w\rangle$, and using the one-body partitioning method of
Eqs.~(\ref{one.part}), gives
\begin{eqnarray}  \label{F.mat2} 
\langle\psi_r| {\cal \hat{F}}_{\!\Phi}^{\scscs (\eta)}|\psi_w\rangle 
= \langle\psi_r| (\hat{F}_\Phi)_{\text{ex}} |\psi_w\rangle
+
\langle\psi_r|\left(v_{\mathrm{co}}^{\scscs \Phi\eta}\right)_{\text{ex}}|\psi_w\rangle.
\end{eqnarray}
Comparing Eq.~(\ref{F.mat3}) with (\ref{F.mat2}), and using (\ref{ex=opa}),  we have
\begin{equation} \label{co.ex} 
\left(v_{\mathrm{co}}^{\scscs \Phi\eta}\right)_{\text{ex}} = 
\left[ \left( H \chi_\Phi^{\scscs \eta} \right)_1\right]_{\text{op}},
\end{equation}
and the Brillouin-Brueckner condition (\ref{bbc.op}) becomes
\begin{equation} \label{bbc.op2} 
\left({\cal \hat{F}}_{\!\Theta}\right)_{\text{ex}} = 0,
\end{equation}
where 
\begin{equation} \label{exactF} 
{\cal \hat{F}}_{\!\Theta}=
\hat{F}_\Theta + v_{\mathrm{co}}^{\scscs \Theta},
\end{equation}
and the $\eta$ superscript is suppressed, since, in general we have
\begin{equation}
\left(v_{\mathrm{co}}^{\scscs \Theta}\right)_{\text{ex}} =
\left(v_{\mathrm{co}}^{\scscs \Theta\eta}\right)_{\text{ex}} = 
\left(v_{\mathrm{co}}^{\scscs \Theta\eta^\prime}\right)_{\text{ex}}. 
\end{equation}

The remaining matrix elements of $v_{\mathrm{co}}^{\scscs \Phi \eta}$ --
$[\psi_w|v_{\mathrm{co}}^{\scscs \Phi\eta}|\psi_x]$ and
$\langle\psi_r|v_{\mathrm{co}}^{\scscs \Phi\eta}|\psi_s]$ -- are at our disposal.
By defining these matrix elements in a manner that is independent of $\eta$, but
dependent on $|\Phi\rangle$, $v_{\mathrm{co}}^{\scscs \Theta}(\mathbf{x})$ and ${\cal
\hat{F}}_{\!\Theta}$ are completely, and unambiguously determined; our exact Fock
operator can be diagonalized:
\begin{equation} \label{F.eigen} 
{\cal \hat{F}}_{\!\Theta} \psi_i^{\Theta}(\mathbf{x})
=\varepsilon_{i}^{\mbox{\tiny $\Theta$}}
\psi_i^{\Theta}(\mathbf{x}),
\end{equation}
where orbital energies $\varepsilon_{i}^{\mbox{\tiny $\Theta$}}$ can be defined to
give exact ionization potentials and electron affinities -- exact Koopman's
theorems \cite{Lindgren:02,tobe}.  In addition, since the operators,
$\hat{f}_o^{\scscs \Phi}$ and $\hat{f}_u^{\scscs \Phi}$, that give the
zeroth-order Hamiltonian $H_{0}^{\Phi}$, and the exact Fock operators, ${\cal
\hat{F}}_{\!\Phi}^{\scscs (\eta)}$, are {\em not} mutually exclusive, one tempting
choice is
\begin{equation}
\hat{f}_o^{\scscs \Phi} = \hat{f}_u^{\scscs \Phi} = {\cal \hat{F}}_{\!\Phi}^{\scscs (\eta)}.
\end{equation}

By using the diagrammatic expansion for $\chi_\Phi^{\scscs \eta}$, and
Eq.~(\ref{co.ex}), a diagrammatic expansion for $(v_{\mathrm{co}}^{\scscs
\Phi\eta})_{\text{ex}}$ can be obtained that is a subset of the open one-body
diagrams of $H\!\chi_\Phi$ \cite{tobe}.  

\section{Reference--State One--Particle Density--Matrix Theory} \label{BDMT} 

\subsection{Functionals of the one-particle density-matrix $\gamma$} \label{IMPL} 

There is a one-to-one correspondence between the set of determinant states,
$\{|\Phi\rangle\}$, and their one-particle density-matrices
\cite{Blaizot:86,Parr:89}, $\{\gamma\}$, where these density-matrices are given by
\cite{Dirac:30,Dirac:31,Lowdin:55a,Lowdin:55b,McWeeny:60}
\begin{equation} \label{dm} 
\gamma(\mbox{$\mathbf{x}$},\mbox{$\mathbf{x^\prime}$}) 
= \sum_{w\in \{\psi_o\rightarrow \Phi\}} 
\psi_{w}(\mbox{$\mathbf{x}$}) \psi_{w}^*(\mbox{$\mathbf{x^\prime}$}).
\end{equation} 
Because of this correspondence, determinantal states are uniquely determined by their
one-particle density-matrix: $|\Phi(\gamma)\rangle$; functionals, or functions,
that depend on $|\Phi\rangle$, can be written as ones depending on $\gamma$. For
example, the total energy ${\mathcal E}$, Eq.~(\ref{E0.a}), and our energy
functionals $E_\eta[\Phi]$, Eq~(\ref{Efunct}), can be written
\begin{eqnarray} \label{Ec.dm} 
{\mathcal E}&=& E_1[\gamma] + {\cal E}_{\mathrm{co}}[\gamma], \\
\label{Ecorr.dm} 
E_\eta[\gamma]&=& E_1[\gamma] + E_{\mathrm{co}}^{\scscs (\eta)}[\gamma],
\end{eqnarray}
where, in addition, our trial wavefunctions $|\Psi_{\Phi(\gamma)}^{\scscs (\eta)}\rangle$ can be
denoted by $|\Psi_\gamma^{\scscs (\eta)}\rangle$.

For simplicity, we require the external potential $v(\mbox{$\mathbf{r}$})$ to be a
spin-free operator, so the first-order energy can be written as
\begin{eqnarray} \label{first.eb} 
E_1[\gamma]
=
\int \left[{-}\mbox{\small$\frac{1}{2}$}\nabla_{\mathbf{r}}^2\,
\gamma(\mbox{$\mathbf{x}$},\mbox{$\mathbf{x^\prime}$}) 
\right]_{\mbox{\tiny $\mathbf{x^\prime}\!\!=\!\!\mathbf{x}$}}
\!d\mbox{$\mathbf{x}$} + 
\int v(\mbox{$\mathbf{r}$}) 
\gamma(\mbox{$\mathbf{x}$},\mbox{$\mathbf{x}$})
\,d\mbox{$\mathbf{x}$}
+ E_{\mathrm{J}}[\gamma] - E_{\mathrm{x}}[\gamma],
\end{eqnarray}
where the Coulomb and exchange energies are
\begin{eqnarray} \label{coul.dm} 
E_{\mathrm{J}}[\gamma]&=&
\frac12 \int \!\! \int 
r_{12}^{-1}
\gamma(\mbox{$\mathbf{x_1}$},\mbox{$\mathbf{x_1}$}) 
\gamma(\mbox{$\mathbf{x_2}$},\mbox{$\mathbf{x_2}$})
\,d\mbox{$\mathbf{x_1}$}
\,d\mbox{$\mathbf{x_2}$},
\\ \label{exch.dm} 
E_{\mathrm{x}}[\gamma] &=&
\frac12 \int\!\!\int 
r_{12}^{-1}
\gamma(\mbox{$\mathbf{x_1}$},\mbox{$\mathbf{x_2}$}) 
\gamma(\mbox{$\mathbf{x_2}$},\mbox{$\mathbf{x_1}$})
\,d\mbox{$\mathbf{x_1}$}
\,d\mbox{$\mathbf{x_2}$}.
\end{eqnarray}
(As in Eq.~(\ref{chemist}), an integration over $\mathbf{x}_i$ actually implies a
summation over the spin variable $\omega_i$ and an integration over the spatial
portion $\mathbf{r}_i$.)

Similarly, the one-body operators ${\cal \hat{F}}_{\!\Phi}^{\scscs (\eta)}$,
Eq.~(\ref{Falpha}), can be written
\begin{equation} \label{efock.dm} 
{\cal \hat{F}}_{\gamma}^{\scscs (\eta)}
= \hat{F}_\gamma +
v_{\mathrm{co}}^{\scscs \gamma\eta},
\end{equation}
where, instead of Eq.~(\ref{Fock}), the Fock operator is given
by
\begin{equation} \label{fock.dm} 
\hat{F}_\gamma=
{-}\mbox{\small$\frac{1}{2}$}\nabla^2 + v  + J_\gamma - K_\gamma,
\end{equation}
and the Coulomb $J_\gamma$ and exchange $K_\gamma$ operators satisfy:
\begin{equation} \label{Coulomb} 
J_\gamma\phi({\mathbf x}_1)
= \int r_{12}^{-1}\gamma({\mathbf x}_2,{\mathbf x}_2) \, 
\phi({\mathbf x}_1) \, 
d{\mathbf x}_2,
\end{equation}
\begin{equation} \label{Exchange} 
K_\gamma \phi({\mathbf x}_1) = 
\int r_{12}^{-1}
\gamma({\mathbf x}_1,{\mathbf x}_2)
\phi({\mathbf x}_2) \, d{\mathbf x}_2.
\end{equation}

In addition, the identity operator, given by Eq.~(\ref{ident.oper}), can be
written
\begin{equation}\label{ident.op} 
\mbox{\small $\hat{I}$} = \hat{\gamma} + \hat{\kappa}_\gamma,
\end{equation}
where the density-matrix operator $\hat{\gamma}$ is defined by its kernel,
$\gamma(\mbox{$\mathbf{x}$},\mbox{$\mathbf{x^\prime}$})$ \cite{McWeeny:60}:
\begin{equation} \label{denmat.op} 
\hat{\gamma} \phi(\mbox{$\mathbf{x}$}) 
= 
\int 
\gamma(\mbox{$\mathbf{x}$},\mbox{$\mathbf{x^\prime}$})
\phi(\mbox{$\mathbf{x^\prime}$})
\,d\mbox{$\mathbf{x^\prime}$},
\end{equation}
and $\kappa_\gamma(\mbox{$\mathbf{x}$},\mbox{$\mathbf{x^\prime}$})$ is the kernel
of the virtual-space projector:
\begin{equation} \label{virt.op} 
\hat{\kappa}_\gamma \phi(\mbox{$\mathbf{x}$}) 
= 
\int 
\kappa_\gamma(\mbox{$\mathbf{x}$},\mbox{$\mathbf{x^\prime}$})
\phi(\mbox{$\mathbf{x^\prime}$})
\,d\mbox{$\mathbf{x^\prime}$},
\end{equation}
where 
\begin{equation}
\label{virt.dm} 
\kappa_\gamma(\mbox{$\mathbf{x}$},\mbox{$\mathbf{x^\prime}$}) 
=\sum_{r\in \{\psi_u\rightarrow \gamma\}} 
\psi_{r}(\mbox{$\mathbf{x}$}) \psi_{r}^*(\mbox{$\mathbf{x^\prime}$}),
\end{equation}
and $\hat{\kappa}_\Phi$ is given by Eq.~(\ref{virt.proj}). Similarly, the
one-particle density-matrix for the Brueckner state, say $\mbox{\large
$\tau$}(\mbox{$\mathbf{x}$},\mbox{$\mathbf{x^\prime}$})$, and its density-matrix
operator, say $\hat{\tau}$, are given by the following expressions:
\begin{eqnarray} 
\label{denmat.bru} 
\mbox{ \large $\tau$}(\mbox{$\mathbf{x}$},\mbox{$\mathbf{x^\prime}$}) 
&=&
\sum_{w\in \{\psi_o\rightarrow \Theta\}}
\psi_{w}(\mbox{$\mathbf{x}$}) \psi_{w}^\dagger(\mbox{$\mathbf{x^\prime}$}), \\
\label{denmat.op.bru} 
\hat{\tau} \phi(\mbox{$\mathbf{x}$}) 
&=& \int 
\tau(\mbox{$\mathbf{x}$},\mbox{$\mathbf{x^\prime}$})
\phi(\mbox{$\mathbf{x^\prime}$})
\,d\mbox{$\mathbf{x^\prime}$},
\end{eqnarray}
where we have
\begin{equation}
|\Theta\rangle=|\Phi(\tau)\rangle.
\end{equation}

Since the one-particle density-matrix, $\hat{\tau}$, also satisfies
\begin{equation} 
\hat{\mbox{\large $\tau$}} = \hat{\gamma}_\Theta,
\end{equation}
where $\hat{\gamma}_\Theta$ is given by Eq.~(\ref{F.dmo}), the
Brillouin-Brueckner condition, given by Eq.~(\ref{FBB-cond.ex}), and its complex
conjugate, given by Eq.~(\ref{Herm.F}), become
\begin{subequations} 
\label{FBB-conds.dm} 
\begin{eqnarray} 
\label{FBB-cond.dm} 
\mbox{\large $\hat{\kappa}_\tau$} \,
{\cal \hat{F}}_{\tau}
\mbox{\large $\hat{\tau}$}
&=&0,\\
\label{Herm.F.dm} 
\mbox{\large $\hat{\tau}$} \,
{\cal \hat{F}}_{\tau}
\mbox{\large $\hat{\kappa}_\tau$}
&=&0;
\end{eqnarray}
\end{subequations} 
the commutation condition, given by Eq.~(\ref{commute}), can be written
\begin{equation} \label{commute.dm} 
\left[{\cal \hat{F}}_{\tau},\hat{\mbox{\large $\tau$}}\right] = 0;
\end{equation}
the exact Hartree--Fock Eq.~(\ref{eF.eqs}) is
\begin{equation} \label{eF.eqs.dm} 
{\cal \hat{F}}_{\tau}|\psi_w\rangle =
\sum_{x\in \{\psi_o\rightarrow\tau\}} 
\varepsilon_{xw}^{\mbox{\tiny $\tau$}}
|\psi_x\rangle, 
\end{equation}
where the occupied orbital, $|\psi_w\rangle$, is from $\{\psi_o\rightarrow\tau\}$; furthermore,
the other Brillouin-Brueckner condition, Eq.~(\ref{bbc.op2}), can be written as
\begin{equation} \label{bbc.op.dm} 
\left({\cal \hat{F}}_{\tau}\right)_{\text{ex}} = 0,
\end{equation}
where Eq.~(\ref{exactF}) becomes
\begin{equation} \label{exactF.dm} 
{\cal \hat{F}}_{\tau}=
\hat{F}_\tau + v_{\mathrm{co}}^{\scscs \tau}.
\end{equation}

In the following subsections, we illustrate how the correlation energy and
correlation-energy functionals, ${\cal E}_{\mathrm{co}}[\gamma]$ and
$E_{\mathrm{co}}^{\scscs (\eta)}[\gamma]$, can be obtained from perturbation theory, in
which all terms (or diagrams) explicitly depend on $\gamma$. (An explicit expression for
the first-order energy, $E_1[\gamma]$, is given by Eq.~(\ref{first.eb}).)

\subsection{The correlation energy as a functional of the one-particle density
matrix:~${\cal E}_{\mathrm{co}}[\gamma]$} 
\label{EXPL} 

Consider the normal-ordered form of the Hamiltonian $H$, given by
Eqs.~(\ref{H.norm}) and (\ref{H.comp}), where we choose to represent this operator
using the following orbital sets: $\{\psi_o \mbox{\footnotesize
$\leftarrow\gamma,\hat{f}_o^\gamma$}\}$ and $\{\psi_u \mbox{\footnotesize
$\leftarrow\gamma,\hat{f}_u^\gamma$}\}$, and both sets are introduced in
Sec.~\ref{LDE}.  Since the one-body portion of the Hamiltonian $H_1^\gamma$ is
determined by the Fock operator $\hat{F}_\gamma$, it seems appropriate to use the
following notation:
\begin{subequations}
\label{H.F} 
\begin{equation} \label{H1.F} 
{H}_1^{\gamma} =\{\hat{F}_\gamma\},
\end{equation}
or, equivalently, the expression $\{\hat{F}_\gamma\}$ is defined by the following
procedure: Obtain the Fock operator $\hat{F}_\gamma$ for the one-particle Hilbert
space, given by Eq.~(\ref{fock.dm}); write this operator in its second quantized
form using the true vacuum state $|\;\rangle$; re-write $\hat{F}_\gamma$ using
normal ordering with respect to the new vacuum state, $|\Phi(\gamma)\rangle$:
($\hat{F}_\gamma= (\hat{F}_\gamma)_{\text{cl}} + \{\hat{F}_\gamma\}$); the
uncontracted term is $\{\hat{F}_\gamma\}$. In other words, $\{\hat{F}_\gamma\}$ is
the uncontracted term when the operator $\hat{F}_\gamma$ is written in normal
ordered form using Wick's theorem
\cite{Bogoliubov:59,Paldus:75,Lindgren:86,Cizek:69}.

Using this notation, we also write
\begin{equation} \label{H2.r12} 
{H}_2^{\gamma} =\{r_{12}^{-1}\}_\gamma,
\end{equation}
\end{subequations}
where, again, $r_{12}^{-1}$ is written in second quantization using the true
vacuum state $|\;\rangle$ -- given by the second term on the right side of
Eq.~(\ref{H}) -- and then it is re-written using normal-ordering with respect to
the new vacuum state; the uncontracted term is $\{r_{12}^{-1}\}_\gamma$, where the
additional $\gamma$-subscript from $\{\ldots\}_\gamma$, serves to remind us that
the vacuum state is $|\Phi(\gamma)\rangle$. (This $\gamma$ subscript is
suppressed in Eq.~(\ref{H1.F}): The vacuum state is understood, since $\gamma$
determines $\hat{F}_\gamma$.)

The correlation energy ${\mathcal E}_{\mathrm{co}}$ is determined by the one- and
two-body parts of the Hamiltonian, $\{\hat{F}_\gamma\}$ and
$\{r_{12}^{-1}\}_\gamma$. However, the individual orders of the perturbation
expansion for ${\mathcal E}_{\mathrm{co}}$, also depends, in addition, on the
zeroth-order Hamiltonian, given by Eq.~(\ref{H0.d}); this operator can be written
as
\begin{equation} \label{h0.part} 
H_0^\gamma=\hat{\mbox{\sc \Large $o$}}_\gamma + \hat{\mbox{\sc \large $u$}}_\gamma,
\end{equation}
where these terms -- $\hat{\mbox{\sc \Large $o$}}_\gamma$ and $\hat{\mbox{\sc \large $u$}}_\gamma$ --
are the occupied and unoccupied portions of $H_0^\gamma$ --
$(H_0^\gamma)_{\text{oc}}$ and $(H_0^\gamma)_{\text{un}}$ -- and are given by the
following:
\begin{subequations}
\begin{eqnarray}
\label{H01o} 
\hat{\mbox{\sc \Large $o$}}_\gamma &=& 
\!\!\!\!\!\!\!\!
\sum_{w\in \{\psi_o\leftarrow \gamma,\hat{f}_o^{\gamma}\}} 
\epsilon_w^{\mbox{\tiny $\gamma$}} \, a_w^\dagger a_w,
\\  \label{H01u} 
\hat{\mbox{\sc \large $u$}}_\gamma &=& 
\sum_{r\in \{\psi_u\leftarrow \gamma,\hat{f}_u^{\gamma}\}} 
\epsilon_r^{\mbox{\tiny $\gamma$}} \, a_r^\dagger a_r.
\end{eqnarray}
\end{subequations}
Using this notation, our zeroth-order Hamiltonian in normal-ordered form can be
written as
\begin{equation} \label{h0.norm} 
H_0^\gamma=E_0[\gamma] + 
\{\hat{\mbox{\sc \Large $o$}}_\gamma\} + \hat{\mbox{\sc \large $u$}}_\gamma,
\end{equation}
where $\hat{\mbox{\sc \large $u$}}_\gamma$ is already normal-ordered; the constant term
$E_0[\gamma]$ is the zeroth-order energy of $|\Phi(\gamma)\rangle$:
\begin{equation}
H_0^\gamma|\Phi(\gamma)\rangle= E_0[\gamma]\,|\Phi(\gamma)\rangle,
\end{equation}
and is given by
\begin{equation}
E_0[\gamma]\;=\sum_{w\in \{\psi_o\leftarrow \gamma,\hat{f}_o^{\gamma}\}} 
\epsilon_w^{\mbox{\tiny $\gamma$}}.
\end{equation}
Note that the first-order and the correlation energies, $E_1[\gamma]$ and
${\mathcal E}_{\mathrm{co}}$, do {\em not} depend the zeroth-order energy
$E_0[\gamma]$.

The perturbation $V_\gamma$, defined by Eqs.~(\ref{Hpart.phi}), can also be written
in normal-ordered form:
\begin{equation}
V_\gamma=V_c^{\gamma} + V_1^{\gamma} + V_2^{\gamma},
\end{equation}
where, from Eqs.~(\ref{H.norm}), (\ref{Hc}), (\ref{H.F}), and (\ref{h0.norm}), the
individual terms are given by the following expressions:
\begin{subequations}
\begin{eqnarray} \label{Vc} 
V_c^{\gamma}&=& E_1[\gamma] - E_0[\gamma],
\\  \label{V1} 
V_1^{\gamma}&=& \{\hat{F}_\gamma\} - \{\hat{\mbox{\sc \Large $o$}}_\gamma\} - \hat{\mbox{\sc \large $u$}}_\gamma,
\\ \label{V2} 
V_2^{\gamma}&=&\{r_{12}^{-1}\}_\gamma.
\end{eqnarray}
\end{subequations}

The one- and two-body parts of $H$ are given by Eqs.~(\ref{H1}) and (\ref{H2}),
and are equal to $\{\hat{F}_\gamma\}$ and $\{r_{12}^{-1}\}_\gamma$, as indicated
by Eqs.~(\ref{H.F}). The Goldstone diagrammatic representation of these operators
can be written in the following manner
\cite{Goldstone:57,Hugenholtz:57,Sanders:69,Raimes:72,Paldus:75,Lindgren:74,Wilson:85,Lindgren:86}:
\begin{subequations}  
\label{h.diag} 
\begin{eqnarray}  \label{h1.diag} 
\raisebox{3.6ex}{$\{\hat{F}_\gamma\}\,={H}_1^{\gamma}$} &\raisebox{3.6ex}{$\;=\;\;$}& 
\setlength{\unitlength}{0.00062500in}
\begingroup\makeatletter\ifx\SetFigFont\undefined%
\gdef\SetFigFont#1#2#3#4#5{%
  \reset@font\fontsize{#1}{#2pt}%
  \fontfamily{#3}\fontseries{#4}\fontshape{#5}%
  \selectfont}%
\fi\endgroup%
{\renewcommand{\dashlinestretch}{30}
\begin{picture}(470,939)(0,-10)
\path(458,912)(458,12)
\put(83,462){\blacken\ellipse{150}{80}}
\put(83,462){\ellipse{150}{80}}
\dottedline{45}(158,462)(428,462)
\end{picture}
}
\raisebox{4.0ex}{\, ,} \\ \label{h2.diag} 
\raisebox{3.6ex}{$\{r_{12}^{-1}\}_\gamma\,={H}_2^{\gamma}$} &\raisebox{3.6ex}{$\;=\;\;$}& 
\setlength{\unitlength}{0.00062500in}
\begingroup\makeatletter\ifx\SetFigFont\undefined%
\gdef\SetFigFont#1#2#3#4#5{%
  \reset@font\fontsize{#1}{#2pt}%
  \fontfamily{#3}\fontseries{#4}\fontshape{#5}%
  \selectfont}%
\fi\endgroup%
{\renewcommand{\dashlinestretch}{30}
\begin{picture}(474,939)(0,-10)
\path(462,912)(462,12)
\path(12,912)(12,12)
\dottedline{45}(12,462)(462,462)
\end{picture}
}
\raisebox{4.0ex}{\, .}
\end{eqnarray}
\end{subequations} 

The one-body part of the perturbation $V_1^{\gamma}$ is usually represented by a
{\em single} diagrammatic operator. However, for our purposes, it is convenient to
use separate diagrammatic operators for the three terms on the right side of
Eq.~(\ref{V1}), where $\{\hat{F}_\gamma\}$ is presented by
Eq.~(\ref{h1.diag}). Since the other two terms are diagonal, it is appropriate is
simply represent them as (unfilled) arrows:
\begin{subequations}
\label{H01.diag} 
\begin{eqnarray}
\label{H01o.diag} 
\raisebox{3.6ex}{$-\{\hat{\mbox{\sc \Large $o$}}_\gamma\}$} &\raisebox{3.6ex}{$\;=\;\;\;$}& 
\setlength{\unitlength}{0.00062500in}
\begingroup\makeatletter\ifx\SetFigFont\undefined%
\gdef\SetFigFont#1#2#3#4#5{%
  \reset@font\fontsize{#1}{#2pt}%
  \fontfamily{#3}\fontseries{#4}\fontshape{#5}%
  \selectfont}%
\fi\endgroup%
{\renewcommand{\dashlinestretch}{30}
\begin{picture}(159,939)(0,-10)
\path(80,387)(80,386)
\whiten\path(35.000,521.000)(80.000,386.000)(125.000,521.000)(80.000,480.500)(35.000,521.000)
\path(80,387)(80,386)
\whiten\path(12.500,551.000)(80.000,386.000)(147.500,551.000)(80.000,501.500)(12.500,551.000)
\path(80,912)(80,12)
\end{picture}
}
\raisebox{4.0ex}{\, ,} \\
\label{H01u.diag} 
\raisebox{3.6ex}{$- \hat{\mbox{\sc \large $u$}}_\gamma$} &\raisebox{3.6ex}{$\;=\;\;\;$}& 
\setlength{\unitlength}{0.00062500in}
\begingroup\makeatletter\ifx\SetFigFont\undefined%
\gdef\SetFigFont#1#2#3#4#5{%
  \reset@font\fontsize{#1}{#2pt}%
  \fontfamily{#3}\fontseries{#4}\fontshape{#5}%
  \selectfont}%
\fi\endgroup%
{\renewcommand{\dashlinestretch}{30}
\begin{picture}(159,939)(0,-10)
\path(80,612)(80,613)
\whiten\path(125.000,478.000)(80.000,613.000)(35.000,478.000)(80.000,518.500)(125.000,478.000)
\path(80,612)(80,613)
\whiten\path(147.500,448.000)(80.000,613.000)(12.500,448.000)(80.000,497.500)(147.500,448.000)
\path(80,12)(80,912)
\end{picture}
}
\raisebox{4.0ex}{\, .} 
\end{eqnarray}
\end{subequations}
In contrast, hole- and particle-lines, by themselves, are represented by
filled arrows: 
\put(-4,-1){
\setlength{\unitlength}{0.00062500in}
\begingroup\makeatletter\ifx\SetFigFont\undefined%
\gdef\SetFigFont#1#2#3#4#5{%
  \reset@font\fontsize{#1}{#2pt}%
  \fontfamily{#3}\fontseries{#4}\fontshape{#5}%
  \selectfont}%
\fi\endgroup%
{\renewcommand{\dashlinestretch}{30}
\begin{picture}(114,223)(0,-10)
\path(57,13)(57,12)
\whiten\path(12.000,147.000)(57.000,12.000)(102.000,147.000)(57.000,106.500)(12.000,147.000)
\path(57,13)(57,12)
\blacken\path(12.000,147.000)(57.000,12.000)(102.000,147.000)(57.000,106.500)(12.000,147.000)
\end{picture}
}
} \hspace{1.5ex} and 
\put(-3,-3){
\setlength{\unitlength}{0.00062500in}
\begingroup\makeatletter\ifx\SetFigFont\undefined%
\gdef\SetFigFont#1#2#3#4#5{%
  \reset@font\fontsize{#1}{#2pt}%
  \fontfamily{#3}\fontseries{#4}\fontshape{#5}%
  \selectfont}%
\fi\endgroup%
{\renewcommand{\dashlinestretch}{30}
\begin{picture}(114,222)(0,-10)
\path(57,194)(57,195)
\blacken\path(102.000,60.000)(57.000,195.000)(12.000,60.000)(57.000,100.500)(102.000,60.000)
\end{picture}
}
} \,\,\, .

As a slight alternative to the usual approach to evaluate the diagrams of the
correlation energy ${\mathcal E}_{\mathrm{co}}$
\cite{Goldstone:57,Hugenholtz:57,Sanders:69,Raimes:72,Paldus:75,Lindgren:74,Szabo:82,Wilson:85,Lindgren:86},
we associate a hole line corresponding to a $w$-occupied orbital with a
$\psi_w({\mathbf x}_1)\psi_w^*({\mathbf x}_2)$ factor; we associate a particle
line corresponding to an $r$-unoccupied orbital with a $\psi_r({\mathbf
x}_2)\psi_r^*({\mathbf x}_1)$ factor, where ${\mathbf x}_1$ and ${\mathbf x}_2$
denote the dummy integration variables that arise from the vertices.  Using this
convention, the sole diagram involving the Fock operator $\hat{F}_\gamma$ from
second-order perturbation theory can be evaluated in the following manner:
\begin{widetext}
\begin{equation} \label{secondfa} 
\raisebox{-4ex}{
\setlength{\unitlength}{0.00062500in}
\begingroup\makeatletter\ifx\SetFigFont\undefined%
\gdef\SetFigFont#1#2#3#4#5{%
  \reset@font\fontsize{#1}{#2pt}%
  \fontfamily{#3}\fontseries{#4}\fontshape{#5}%
  \selectfont}%
\fi\endgroup%
{\renewcommand{\dashlinestretch}{30}
\begin{picture}(890,1009)(0,-10)
\path(833,422)(833,421)
\blacken\path(788.000,556.000)(833.000,421.000)(878.000,556.000)(833.000,515.500)(788.000,556.000)
\path(533,572)(533,573)
\blacken\path(578.000,438.000)(533.000,573.000)(488.000,438.000)(533.000,478.500)(578.000,438.000)
\put(83.000,497.000){\arc{1500.000}{5.6397}{6.9267}}
\put(1283.000,497.000){\arc{1500.000}{2.4981}{3.7851}}
\put(83,47){\blacken\ellipse{150}{80}}
\put(83,47){\ellipse{150}{80}}
\put(83,947){\blacken\ellipse{150}{80}}
\put(83,947){\ellipse{150}{80}}
\dottedline{60}(158,947)(633,947)
\dottedline{60}(158,47)(633,47)
\end{picture}
}
} \raisebox{0.5ex}{$ \displaystyle  
\;\;\; = \; 
(\varepsilon_{rw}^{\mbox{\tiny $\gamma$}})^{-1}
\int 
d\,{\mathbf x}_1 \,d\,{\mathbf x}_2 \,
\left(\hat{F}_{\gamma \mbox{\tiny $1$}}
\gamma_w({\mathbf x}_1,{\mathbf x}_2)
\right) \cdot
\hat{F}_{\gamma \mbox{\tiny $2$}}
\gamma_r({\mathbf x}_2,{\mathbf x}_1),$}
\end{equation}
\end{widetext}
where 
\begin{equation}
\varepsilon_{rw}^{\mbox{\tiny $\gamma$}}=
\epsilon_w^{\mbox{\tiny $\gamma$}} - \epsilon_r^{\mbox{\tiny $\gamma$}},
\end{equation}
and the repeated indices -- $r$ and $w$ -- are summed over;
$\hat{F}_{\gamma \mbox{\tiny $i$}}$ denotes the Fock operator $\hat{F}_\gamma$ --
given by Eq.~(\ref{fock.dm}) -- acting upon $({\mathbf x}_i)$; the term
$(\hat{F}_{\gamma \mbox{\tiny $i$}}\cdots)\cdot$ indicates that $\hat{F}_{\gamma
\mbox{\tiny $i$}}$ {\em exclusively} acts within the brackets; furthermore, the
$w$th component of the (one-particle) density-matrix $\gamma$ is denoted by
\begin{subequations}
\label{gammawr} 
\begin{equation} \label{gammaw} 
\gamma_w({\mathbf x}_1,{\mathbf x}_2)=\psi_w({\mathbf x}_1)\psi_w^*({\mathbf x}_2);
\end{equation}
the $r$th orthogonal-component of $\gamma$ is denoted by
\begin{equation} \label{gammar} 
\gamma_r({\mathbf x}_1,{\mathbf x}_2)=\psi_r({\mathbf x}_1)\psi_r^*({\mathbf
x}_2),
\end{equation}
\end{subequations}
where, for a complete set of orbital states, we have \cite{Raimes:72}
\begin{equation} \label{Comp} 
\delta({\mathbf x}_1 - {\mathbf x}_2) = \sum_w\gamma_w({\mathbf x}_1,{\mathbf x}_2) 
+ \sum_r \gamma_r({\mathbf x}_1,{\mathbf x}_2),
\end{equation}
which is a shorthand notations for
\begin{equation} 
\delta({\mathbf x}_1 - {\mathbf x}_2) = 
\delta({\mathbf r}_1 - {\mathbf r}_2)\delta_{\omega_1\omega_2}.
\end{equation}

In order to further compress our notation, we use the convention that all repeated
dummy indices are integrated over and restrict the Fock operator $\hat{F}_{\gamma
\mbox{\tiny $i$}}$ to exclusively act upon the first index of any two-body
function, i.e., ($\hat{F}_{\gamma \mbox{\tiny $i$}} \alpha^\prime({\mathbf
x}_j,{\mathbf x}_i) \alpha({\mathbf x}_i,{\mathbf x}_j) = \alpha^\prime({\mathbf
x}_j,{\mathbf x}_i) \hat{F}_{\gamma \mbox{\tiny $i$}} \alpha({\mathbf
x}_i,{\mathbf x}_j) $); Eq.~(\ref{secondfa}) can then be written as
\begin{subequations}
\begin{equation} \label{secondfb} 
\raisebox{-4ex}{
\setlength{\unitlength}{0.00062500in}
\begingroup\makeatletter\ifx\SetFigFont\undefined%
\gdef\SetFigFont#1#2#3#4#5{%
  \reset@font\fontsize{#1}{#2pt}%
  \fontfamily{#3}\fontseries{#4}\fontshape{#5}%
  \selectfont}%
\fi\endgroup%
{\renewcommand{\dashlinestretch}{30}
\begin{picture}(890,1009)(0,-10)
\path(833,422)(833,421)
\blacken\path(788.000,556.000)(833.000,421.000)(878.000,556.000)(833.000,515.500)(788.000,556.000)
\path(533,572)(533,573)
\blacken\path(578.000,438.000)(533.000,573.000)(488.000,438.000)(533.000,478.500)(578.000,438.000)
\put(83.000,497.000){\arc{1500.000}{5.6397}{6.9267}}
\put(1283.000,497.000){\arc{1500.000}{2.4981}{3.7851}}
\put(83,47){\blacken\ellipse{150}{80}}
\put(83,47){\ellipse{150}{80}}
\put(83,947){\blacken\ellipse{150}{80}}
\put(83,947){\ellipse{150}{80}}
\dottedline{60}(158,947)(633,947)
\dottedline{60}(158,47)(633,47)
\end{picture}
}
} \raisebox{0.5ex}{$ \displaystyle  
\;\;\; = \; 
(\varepsilon_{rw}^{\mbox{\tiny $\gamma$}})^{-1}
\hat{F}_{\gamma \mbox{\tiny $1$}}
\gamma_w({\mathbf x}_1,{\mathbf x}_2)
\hat{F}_{\gamma \mbox{\tiny $2$}}
\gamma_r({\mathbf x}_2,{\mathbf x}_1) ,$}
\end{equation}
and the other two diagrams from second-order perturbation theory
have the following forms:
\begin{widetext}
\begin{equation} \label{secondj.d} 
\raisebox{-4ex}{
\setlength{\unitlength}{0.00062500in}
\begingroup\makeatletter\ifx\SetFigFont\undefined%
\gdef\SetFigFont#1#2#3#4#5{%
  \reset@font\fontsize{#1}{#2pt}%
  \fontfamily{#3}\fontseries{#4}\fontshape{#5}%
  \selectfont}%
\fi\endgroup%
{\renewcommand{\dashlinestretch}{30}
\begin{picture}(1164,939)(0,-10)
\path(57,537)(57,538)
\blacken\path(102.000,403.000)(57.000,538.000)(12.000,403.000)(57.000,443.500)(102.000,403.000)
\path(357,387)(357,386)
\blacken\path(312.000,521.000)(357.000,386.000)(402.000,521.000)(357.000,480.500)(312.000,521.000)
\path(807,537)(807,538)
\blacken\path(852.000,403.000)(807.000,538.000)(762.000,403.000)(807.000,443.500)(852.000,403.000)
\path(1107,387)(1107,386)
\blacken\path(1062.000,521.000)(1107.000,386.000)(1152.000,521.000)(1107.000,480.500)(1062.000,521.000)
\put(-393.000,462.000){\arc{1500.000}{5.6397}{6.9267}}
\put(807.000,462.000){\arc{1500.000}{2.4981}{3.7851}}
\put(357.000,462.000){\arc{1500.000}{5.6397}{6.9267}}
\put(1557.000,462.000){\arc{1500.000}{2.4981}{3.7851}}
\dottedline{60}(207,912)(957,912)
\dottedline{60}(207,12)(957,12)
\end{picture}
}
}
\raisebox{0.3ex}{$\displaystyle  
\;\;\; 
= \frac12 
(\varepsilon_{rwsx}^{\mbox{\tiny $\gamma$}})^{-1}
r_{12}^{-1}r_{34}^{-1}
\gamma_w({\mathbf x}_1,{\mathbf x}_3)
\gamma_r({\mathbf x}_3,{\mathbf x}_1)
\gamma_x({\mathbf x}_2,{\mathbf x}_4)
\gamma_s({\mathbf x}_4,{\mathbf x}_2),$} 
\end{equation}
\begin{equation} \label{secondk.d} 
\raisebox{-4ex}{
\setlength{\unitlength}{0.00062500in}
\begingroup\makeatletter\ifx\SetFigFont\undefined%
\gdef\SetFigFont#1#2#3#4#5{%
  \reset@font\fontsize{#1}{#2pt}%
  \fontfamily{#3}\fontseries{#4}\fontshape{#5}%
  \selectfont}%
\fi\endgroup%
{\renewcommand{\dashlinestretch}{30}
\begin{picture}(1314,943)(0,-10)
\path(57,389)(57,388)
\blacken\path(12.000,523.000)(57.000,388.000)(102.000,523.000)(57.000,482.500)(12.000,523.000)
\path(319,127)(531,339)
\blacken\path(467.360,211.721)(531.000,339.000)(403.721,275.360)(464.178,272.178)(467.360,211.721)
\path(1019,102)(807,314)
\blacken\path(934.279,250.360)(807.000,314.000)(870.640,186.721)(873.822,247.178)(934.279,250.360)
\path(1257,389)(1257,388)
\blacken\path(1212.000,523.000)(1257.000,388.000)(1302.000,523.000)(1257.000,482.500)(1212.000,523.000)
\put(807.000,464.000){\arc{1500.000}{2.4981}{3.7851}}
\put(507.000,464.000){\arc{1500.000}{5.6397}{6.9267}}
\path(207,914)(1109,12)
\path(207,14)(1109,916)
\dottedline{60}(207,914)(1107,914)
\dottedline{60}(207,14)(1107,14)
\end{picture}
}
}
\raisebox{0.3ex}{$ \displaystyle  \;\;\;
=-\frac12 
(\varepsilon_{rwsx}^{\mbox{\tiny $\gamma$}})^{-1}
r_{12}^{-1}r_{34}^{-1}
\gamma_w({\mathbf x}_1,{\mathbf x}_3)
\gamma_r({\mathbf x}_3,{\mathbf x}_2)
\gamma_x({\mathbf x}_2,{\mathbf x}_4)
\gamma_s({\mathbf x}_4,{\mathbf x}_1),$}
\end{equation}
\end{widetext}
\end{subequations}
where
\begin{equation}
\varepsilon_{rwsx}^{\mbox{\tiny $\gamma$}}=
\varepsilon_{rw}^{\mbox{\tiny $\gamma$}}+
\varepsilon_{sx}^{\mbox{\tiny $\gamma$}}.
\end{equation}

The diagonal terms arising from the zeroth-order Hamiltonian, given by
$-\{\hat{\mbox{\sc \Large $o$}}_\gamma\}$ and $-\hat{\mbox{\sc \large $u$}}_\gamma$, and represented by Eqs.~(\ref{H01.diag}),
first appear in third order. For example, the following two diagrams can be
obtained by inserting $- \{\hat{\mbox{\sc \Large $o$}}_\gamma\}$ and $- \hat{\mbox{\sc \large $u$}}_\gamma$ into the diagram
on the left side of Eq.~(\ref{secondfb}):
\begin{subequations}
\begin{eqnarray} \label{thirdo} 
\raisebox{-4ex}{
\setlength{\unitlength}{0.00062500in}
\begingroup\makeatletter\ifx\SetFigFont\undefined%
\gdef\SetFigFont#1#2#3#4#5{%
  \reset@font\fontsize{#1}{#2pt}%
  \fontfamily{#3}\fontseries{#4}\fontshape{#5}%
  \selectfont}%
\fi\endgroup%
{\renewcommand{\dashlinestretch}{30}
\begin{picture}(912,1009)(0,-10)
\path(533,572)(533,573)
\blacken\path(578.000,438.000)(533.000,573.000)(488.000,438.000)(533.000,478.500)(578.000,438.000)
\path(833,422)(833,421)
\whiten\path(765.500,586.000)(833.000,421.000)(900.500,586.000)(833.000,536.500)(765.500,586.000)
\put(83.000,497.000){\arc{1500.000}{5.6397}{6.9267}}
\put(1283.000,497.000){\arc{1500.000}{2.4981}{3.7851}}
\put(83,47){\blacken\ellipse{150}{80}}
\put(83,47){\ellipse{150}{80}}
\put(83,947){\blacken\ellipse{150}{80}}
\put(83,947){\ellipse{150}{80}}
\dottedline{60}(158,947)(633,947)
\dottedline{60}(158,47)(633,47)
\end{picture}
}
} \hspace{-4ex} &&\raisebox{0.5ex}{$ \displaystyle  
\;\;\; = \; 
-\frac{(-\epsilon_w)}{(\varepsilon_{rw}^{\mbox{\tiny $\gamma$}})^{2}}
\hat{F}_{\gamma \mbox{\tiny $1$}}
\gamma_w({\mathbf x}_1,{\mathbf x}_2)
\hat{F}_{\gamma \mbox{\tiny $2$}}
\gamma_r({\mathbf x}_2,{\mathbf x}_1) ,$} 
\nonumber \\ \\
\label{thirdu} 
\raisebox{-4ex}{
\setlength{\unitlength}{0.00062500in}
\begingroup\makeatletter\ifx\SetFigFont\undefined%
\gdef\SetFigFont#1#2#3#4#5{%
  \reset@font\fontsize{#1}{#2pt}%
  \fontfamily{#3}\fontseries{#4}\fontshape{#5}%
  \selectfont}%
\fi\endgroup%
{\renewcommand{\dashlinestretch}{30}
\begin{picture}(890,1009)(0,-10)
\path(833,422)(833,421)
\blacken\path(788.000,556.000)(833.000,421.000)(878.000,556.000)(833.000,515.500)(788.000,556.000)
\path(533,572)(533,573)
\whiten\path(600.500,408.000)(533.000,573.000)(465.500,408.000)(533.000,457.500)(600.500,408.000)
\put(83.000,497.000){\arc{1500.000}{5.6397}{6.9267}}
\put(1283.000,497.000){\arc{1500.000}{2.4981}{3.7851}}
\put(83,47){\blacken\ellipse{150}{80}}
\put(83,47){\ellipse{150}{80}}
\put(83,947){\blacken\ellipse{150}{80}}
\put(83,947){\ellipse{150}{80}}
\dottedline{60}(158,947)(633,947)
\dottedline{60}(158,47)(633,47)
\end{picture}
}
} \hspace{-4ex} &&\raisebox{0.5ex}{$ \displaystyle  
\;\;\; = \; 
\frac{(-\epsilon_r)}{(\varepsilon_{rw}^{\mbox{\tiny $\gamma$}})^{2}}
\hat{F}_{\gamma \mbox{\tiny $1$}}
\gamma_w({\mathbf x}_1,{\mathbf x}_2)
\hat{F}_{\gamma \mbox{\tiny $2$}}
\gamma_r({\mathbf x}_2,{\mathbf x}_1).$}
\nonumber \\ 
\end{eqnarray}
\end{subequations}
The hole-line operator $\{\hat{\mbox{\sc \Large $o$}}_\gamma\}$ generates an additional hole line when
inserted into a diagram and, therefore, a factor of $-1$ is included when diagram
(\ref{thirdo}) is evaluated, where this factor cancels the $-1$ factor from
$-\epsilon_w$. Since this type of cancellation always occurs, as an alternative,
we associate a factor of $\epsilon_w$ for $\{\hat{\mbox{\sc \Large $o$}}_\gamma\}$ insertions, and treat
$\{\hat{\mbox{\sc \Large $o$}}_\gamma\}$ vertices as ones that do not generate additional hole lines;
$\hat{\mbox{\sc \large $u$}}_\gamma$ is associated with a $-\epsilon_r$ factor.  Keep in mind, also,
that these operators generate an additional energy-denominator factor, e.g.,
$\varepsilon_{rw}^{\mbox{\tiny $\gamma$}}$, when inserted into a diagram.

One advantage of partitioning the one-body part of the perturbation $V_1^{\gamma}$
into individual components, as indicated by Eq.~(\ref{V1}), is that it yields
correlation-energy diagrams that explicitly depend on the Fock operator
$\hat{F}_\gamma$.  Furthermore, an overall dependence of the correlation energy
${\mathcal E}_{\mathrm{co}}$ on the one-particle density matrix $\gamma$ becomes, to a
certain extent, transparent, by using Eqs.~(\ref{gammawr}), (\ref{dm}), and
(\ref{virt.dm}), yielding the following identities:
\begin{eqnarray} \label{onepart} 
\gamma({\mathbf x}_1,{\mathbf x}_2)
&=&
\sum_w \gamma_w({\mathbf x}_1,{\mathbf x}_2),\\
\label{onepart.virt} 
\kappa_\gamma({\mathbf x}_1,{\mathbf x}_2)
&=&
\sum_r \gamma_r({\mathbf x}_1,{\mathbf x}_2),
\end{eqnarray}
and note that $\kappa_\gamma$ depends, explicitly, on $\gamma$:
\begin{equation} \label{delta} 
\delta({\mathbf x}_1 - {\mathbf x}_2) =
\gamma({\mathbf x}_1,{\mathbf x}_2) + \kappa_\gamma({\mathbf x}_1,{\mathbf x}_2),
\end{equation}
where the operator form of this relation is Eq.~(\ref{ident.op}).

The individual diagrams depend, in part, on each of the $\gamma_w$ components,
given by Eq.~(\ref{gammaw}), and the orthogonal components $\gamma_r$, given by
Eq.~(\ref{gammar}).  In addition, each diagram depends on the set of orbital
energies $\{\epsilon^{\mbox{\tiny $\gamma$}}\}$, which are at our disposal. In
order to make each diagram an explicit functional of the one-particle density
matrix $\gamma$, we choose all occupied orbitals to be degenerate, with energy
$\epsilon_o^\gamma$; also, we choose all unoccupied orbitals to be degenerate,
with energy $\epsilon_u^\gamma$. With these choices, the zeroth-order Hamiltonian,
given by Eq.~(\ref{H0.d}), becomes
\begin{equation} \label{H0.deg} 
H_0^{\gamma} = 
\epsilon_o^\gamma \!\!\!\!\!\!\!\!
\sum_{w\in \{\psi_o\leftarrow \gamma,\hat{f}_o^{\gamma}\}} 
a^\dagger_w a_w 
+ \epsilon_u^\gamma \!\!\!\!\!\!\!\!
\sum_{r\in \{\psi_u\leftarrow \gamma,\hat{f}_u^{\gamma}\}} 
a^\dagger_r a_r,
\end{equation}
and since this operator is invariant to a unitary transformation of occupied or
unoccupied orbitals, it no longer depends on $\hat{f}_o^{\gamma}$ and
$\hat{f}_u^{\gamma}$ -- any set of orbitals defining $\gamma$ is appropriate -- so
we can write
\begin{equation} \label{H0.degb} 
H_0^\gamma =
\epsilon_o^\gamma \!\!\!\!\!\!
\sum_{w\in \{\psi_o\rightarrow \gamma\}} 
a^\dagger_w a_w 
+ \epsilon_u^\gamma \!\!\!\!\!\!
\sum_{r\in \{\psi_u\rightarrow \gamma\}}
a^\dagger_r a_r.
\end{equation}
It is easily proven that all perturbative orders, except for the zeroth-order,
depend on the orbital-energy difference $\varepsilon^\gamma$, given by
\begin{equation} \label{veps.gamma} 
\varepsilon_\gamma=
\epsilon_o^\gamma - \epsilon_u^\gamma,
\end{equation}
and not on the individual orbital-energies, $\epsilon_o^\gamma$ and
$\epsilon_u^\gamma$.  Therefore, we can choose ($\epsilon_u^\gamma=0$), and so our
only parameter is $\varepsilon_\gamma$. With this choice we have
\begin{equation} \label{H0.gamma} 
H_0^\gamma = \varepsilon_\gamma \hat{N}_\gamma,
\end{equation}
where $\hat{N}_\gamma$ is the number operator for the occupied orbitals,
\begin{equation} \label{Nocc} 
\hat{N}_\gamma = \sum_{w\in \{\psi_o\rightarrow \gamma\}} a^\dagger_w a_w,
\end{equation}
and it gives the total number of occupied orbitals when acting on a single
determinant. In the {\em one-particle} Hilbert space, this operator is the
projector for the occupied subspace -- spanned by $\{\psi_o\mbox{\small
$\rightarrow\gamma$}\}$ -- or, the one-particle density-matrix operator:
\begin{equation} \label{Nocc.dm} 
\hat{N}_\gamma =
\sum_{w\in \{\psi_o\rightarrow \gamma\}} |\psi_w\rangle\langle\psi_w| = 
\hat{\gamma}.
\end{equation}
Using the above two expressions, let us generalize the definition of $\hat{\gamma}$:
\begin{equation} \label{gen.gam} 
\hat{\gamma}= \sum_{w\in \{\psi_o\rightarrow \gamma\}} a^\dagger_w a_w,
\end{equation}
and write the zeroth-order Hamiltonian in a simplified form, given by
\begin{equation} \label{H0.gammab} 
H_0^\gamma = 
\varepsilon_\gamma\,\hat{\gamma}.
\end{equation}
By normal-ordering this expression, we have
\begin{equation} \label{H0.gamma.n} 
H_0^\gamma = \varepsilon_\gamma N_\gamma + 
\varepsilon_\gamma\{\hat{\gamma}\},
\end{equation}
where $N_\gamma$ is the number of particles within $|\Phi(\gamma)\rangle$, and
from Eq.~(\ref{h0.norm}), we get the following identities:
\begin{eqnarray} 
E_0[\gamma] &=&\varepsilon_\gamma N_\gamma, \\
\label{ident.oc} 
\{\hat{\mbox{\sc \Large $o$}}_\gamma\}&=&\varepsilon_\gamma\{\hat{\gamma}\},\\
\label{ident.un} 
\hat{\mbox{\sc \large $u$}}_\gamma &=&0;
\end{eqnarray}
furthermore, our zero- and one-body portion of the perturbation, Eqs.~(\ref{Vc})
and (\ref{V1}), have the following modified forms:
\begin{subequations}
\begin{eqnarray} \label{Vcb} 
V_c^{\gamma}&=& E_1[\gamma] - \varepsilon_\gamma N_\gamma,
\\  \label{V1b} 
V_1^{\gamma}&=& \{\hat{F}_\gamma\} - \varepsilon_\gamma\{\hat{\gamma}\}.
\end{eqnarray}
\end{subequations}

Eq.~(\ref{ident.un}) indicates that the unoccupied operator, $\hat{\mbox{\sc \large
$u$}}_\gamma$, represented by Eq.~(\ref{H01u.diag}), does not appear in the
expansion of the correlation-energy ${\mathcal E}_{\mathrm{co}}$;
$\{\hat{\mbox{\sc \Large $o$}}_\gamma\}$, represented by Eq.~(\ref{H01o.diag}) and
given by $\varepsilon_\gamma\{\hat{\gamma}\}$, is associated with a factor of
$\varepsilon_\gamma$.  Each diagram now becomes an explicit functional of $\gamma$
and $\kappa_\gamma$. For example, the second-order diagrams can be written in the
following manner:
\begin{widetext}
\begin{equation}
\label{secondf} 
\setlength{\unitlength}{0.00062500in}
\begingroup\makeatletter\ifx\SetFigFont\undefined%
\gdef\SetFigFont#1#2#3#4#5{%
  \reset@font\fontsize{#1}{#2pt}%
  \fontfamily{#3}\fontseries{#4}\fontshape{#5}%
  \selectfont}%
\fi\endgroup%
{\renewcommand{\dashlinestretch}{30}
\begin{picture}(890,1009)(0,-10)
\path(833,422)(833,421)
\blacken\path(788.000,556.000)(833.000,421.000)(878.000,556.000)(833.000,515.500)(788.000,556.000)
\path(533,572)(533,573)
\blacken\path(578.000,438.000)(533.000,573.000)(488.000,438.000)(533.000,478.500)(578.000,438.000)
\put(83.000,497.000){\arc{1500.000}{5.6397}{6.9267}}
\put(1283.000,497.000){\arc{1500.000}{2.4981}{3.7851}}
\put(83,47){\blacken\ellipse{150}{80}}
\put(83,47){\ellipse{150}{80}}
\put(83,947){\blacken\ellipse{150}{80}}
\put(83,947){\ellipse{150}{80}}
\dottedline{60}(158,947)(633,947)
\dottedline{60}(158,47)(633,47)
\end{picture}
}
\raisebox{3.7ex}{$ \displaystyle  
\;\;\; = \varepsilon_\gamma^{-1}
\hat{F}_{\gamma \mbox{\tiny $1$}}
\gamma({\mathbf x}_1,{\mathbf x}_2)
\hat{F}_{\gamma \mbox{\tiny $2$}}
\kappa_\gamma({\mathbf x}_2,{\mathbf x}_1), $}
\end{equation}
\begin{equation} \label{secondj} 
\setlength{\unitlength}{0.00062500in}
\begingroup\makeatletter\ifx\SetFigFont\undefined%
\gdef\SetFigFont#1#2#3#4#5{%
  \reset@font\fontsize{#1}{#2pt}%
  \fontfamily{#3}\fontseries{#4}\fontshape{#5}%
  \selectfont}%
\fi\endgroup%
{\renewcommand{\dashlinestretch}{30}
\begin{picture}(1164,939)(0,-10)
\path(57,537)(57,538)
\blacken\path(102.000,403.000)(57.000,538.000)(12.000,403.000)(57.000,443.500)(102.000,403.000)
\path(357,387)(357,386)
\blacken\path(312.000,521.000)(357.000,386.000)(402.000,521.000)(357.000,480.500)(312.000,521.000)
\path(807,537)(807,538)
\blacken\path(852.000,403.000)(807.000,538.000)(762.000,403.000)(807.000,443.500)(852.000,403.000)
\path(1107,387)(1107,386)
\blacken\path(1062.000,521.000)(1107.000,386.000)(1152.000,521.000)(1107.000,480.500)(1062.000,521.000)
\put(-393.000,462.000){\arc{1500.000}{5.6397}{6.9267}}
\put(807.000,462.000){\arc{1500.000}{2.4981}{3.7851}}
\put(357.000,462.000){\arc{1500.000}{5.6397}{6.9267}}
\put(1557.000,462.000){\arc{1500.000}{2.4981}{3.7851}}
\dottedline{60}(207,912)(957,912)
\dottedline{60}(207,12)(957,12)
\end{picture}
}
\raisebox{3.7ex}{$ \displaystyle  
\;\;\; = \frac{1}{4} \varepsilon_\gamma^{-1}
r_{12}^{-1}r_{34}^{-1}
\gamma({\mathbf x}_1,{\mathbf x}_3)
\kappa_\gamma({\mathbf x}_3,{\mathbf x}_1)
\gamma({\mathbf x}_2,{\mathbf x}_4)
\kappa_\gamma({\mathbf x}_4,{\mathbf x}_2),$}
\end{equation}
\begin{equation} \label{secondk} 
\setlength{\unitlength}{0.00062500in}
\begingroup\makeatletter\ifx\SetFigFont\undefined%
\gdef\SetFigFont#1#2#3#4#5{%
  \reset@font\fontsize{#1}{#2pt}%
  \fontfamily{#3}\fontseries{#4}\fontshape{#5}%
  \selectfont}%
\fi\endgroup%
{\renewcommand{\dashlinestretch}{30}
\begin{picture}(1314,943)(0,-10)
\path(57,389)(57,388)
\blacken\path(12.000,523.000)(57.000,388.000)(102.000,523.000)(57.000,482.500)(12.000,523.000)
\path(319,127)(531,339)
\blacken\path(467.360,211.721)(531.000,339.000)(403.721,275.360)(464.178,272.178)(467.360,211.721)
\path(1019,102)(807,314)
\blacken\path(934.279,250.360)(807.000,314.000)(870.640,186.721)(873.822,247.178)(934.279,250.360)
\path(1257,389)(1257,388)
\blacken\path(1212.000,523.000)(1257.000,388.000)(1302.000,523.000)(1257.000,482.500)(1212.000,523.000)
\put(807.000,464.000){\arc{1500.000}{2.4981}{3.7851}}
\put(507.000,464.000){\arc{1500.000}{5.6397}{6.9267}}
\path(207,914)(1109,12)
\path(207,14)(1109,916)
\dottedline{60}(207,914)(1107,914)
\dottedline{60}(207,14)(1107,14)
\end{picture}
}
\raisebox{3.7ex}{$ \displaystyle  
\;\;\; = - \frac{1}{4} 
\varepsilon_\gamma^{-1}
r_{12}^{-1}r_{34}^{-1}
\gamma({\mathbf x}_1,{\mathbf x}_3)
\kappa_\gamma({\mathbf x}_3,{\mathbf x}_2)
\gamma({\mathbf x}_2,{\mathbf x}_4)
\kappa_\gamma({\mathbf x}_4,{\mathbf x}_1),$}
\end{equation}
\end{widetext}
where $\kappa_\gamma$ is given by Eq.~(\ref{delta}).  

In order to remove the explicit dependence on $\kappa_\gamma$, we first note that
the kernel of the Fock operator is given by \cite{Parr:89}:
\begin{equation} \label{kernfock} 
F_\gamma(\mathbf{x}_1,\mathbf{x}_2)
=
\delta(\mathbf{x}_2-\mathbf{x}_1) 
\left[
-\mbox{\small$\frac{1}{2}$}
\nabla_{\!\mbox{\tiny $2$}}^2
+ v(\mathbf{r}_2)\right]
+
\delta(\mathbf{x}_2-\mathbf{x}_1)
\int 
r_{23}^{-1}
\gamma(\mathbf{x}_3,\mathbf{x}_3) 
\,d\mathbf{x}_3
-
r_{12}^{-1}
\gamma(\mathbf{x}_1,\mathbf{x}_2),
\end{equation}
or we can use its matrix representation from a complete basis set:
\begin{equation} \label{kernfock.mat} 
F_\gamma(\mathbf{x}_1,\mathbf{x}_2) =
\sum_{ij} \psi_i(\mathbf{x}_1) 
\langle\psi_i|
\hat{F}_\gamma |\psi_j\rangle
\psi_j^*(\mathbf{x}_2).
\end{equation}
In any case, we use the following identity:
\begin{equation}
\hat{F}_{\gamma \mbox{\tiny $1$}}
\alpha({\mathbf x}_1,{\mathbf x}_2)
= \int d\,{\mathbf x}_3 \,
F_\gamma(\mathbf{x}_1,\mathbf{x}_3)
\alpha({\mathbf x}_3,{\mathbf x}_2),
\end{equation}
to modify Eq.~(\ref{secondf}), yielding
\begin{equation}
\label{secondf2} 
\setlength{\unitlength}{0.00062500in}
\begingroup\makeatletter\ifx\SetFigFont\undefined%
\gdef\SetFigFont#1#2#3#4#5{%
  \reset@font\fontsize{#1}{#2pt}%
  \fontfamily{#3}\fontseries{#4}\fontshape{#5}%
  \selectfont}%
\fi\endgroup%
{\renewcommand{\dashlinestretch}{30}
\begin{picture}(890,1009)(0,-10)
\path(833,422)(833,421)
\blacken\path(788.000,556.000)(833.000,421.000)(878.000,556.000)(833.000,515.500)(788.000,556.000)
\path(533,572)(533,573)
\blacken\path(578.000,438.000)(533.000,573.000)(488.000,438.000)(533.000,478.500)(578.000,438.000)
\put(83.000,497.000){\arc{1500.000}{5.6397}{6.9267}}
\put(1283.000,497.000){\arc{1500.000}{2.4981}{3.7851}}
\put(83,47){\blacken\ellipse{150}{80}}
\put(83,47){\ellipse{150}{80}}
\put(83,947){\blacken\ellipse{150}{80}}
\put(83,947){\ellipse{150}{80}}
\dottedline{60}(158,947)(633,947)
\dottedline{60}(158,47)(633,47)
\end{picture}
}
\raisebox{3.7ex}{$ \displaystyle  
\;\;\; = \varepsilon_\gamma^{-1} 
F_\gamma(\mathbf{x}_1,\mathbf{x}_3)
\gamma({\mathbf x}_3,{\mathbf x}_2)
F_\gamma(\mathbf{x}_2,\mathbf{x}_4)
\kappa_\gamma({\mathbf x}_4,{\mathbf x}_1) 
.$}
\end{equation}
Using Eq.~(\ref{delta}), this term becomes
\begin{eqnarray}
\label{secondf3} 
\varepsilon_\gamma^{-1} 
F_\gamma(\mathbf{x}_1,\mathbf{x}_3)
\gamma({\mathbf x}_3,{\mathbf x}_2)
F_\gamma(\mathbf{x}_2,\mathbf{x}_4)
\kappa_\gamma({\mathbf x}_4,{\mathbf x}_1) 
= \hspace{43ex}
\\ \nonumber \hspace{2ex}
\varepsilon_\gamma^{-1} 
\left[
F_\gamma(\mathbf{x}_1,\mathbf{x}_3)
\gamma({\mathbf x}_3,{\mathbf x}_2)
F_\gamma(\mathbf{x}_2,\mathbf{x}_4)
\delta({\mathbf x}_4-{\mathbf x}_1) 
-
F_\gamma(\mathbf{x}_1,\mathbf{x}_3)
\gamma({\mathbf x}_3,{\mathbf x}_2)
F_\gamma(\mathbf{x}_2,\mathbf{x}_4)
\gamma({\mathbf x}_4,{\mathbf x}_1)
\right].
\end{eqnarray}
The first term on the right side reduces to 
\begin{equation} \label{f.delta} 
\varepsilon_\gamma^{-1} 
F_\gamma(\mathbf{x}_1,\mathbf{x}_3)
\gamma({\mathbf x}_3,{\mathbf x}_2)
F_\gamma(\mathbf{x}_2,\mathbf{x}_4)
\delta({\mathbf x}_4-{\mathbf x}_1) 
=
\varepsilon_\gamma^{-1}
F_\gamma(\mathbf{x}_1,\mathbf{x}_3) 
\gamma({\mathbf x}_3,{\mathbf x}_2)
F_\gamma(\mathbf{x}_2,\mathbf{x}_1),
\end{equation}
where $F_\gamma(\mathbf{x}_2,\mathbf{x}_1)$ is given by Eq.~(\ref{kernfock.mat}),
and {\em not} by Eq.~(\ref{kernfock}), since this form of the Fock kernel contains
the laplacian $\nabla^2$, which is only defined when acting upon a function. As a
possible alternative, the term on the left side of Eq~(\ref{f.delta}) can be
treated using the Fock kernel, given by Eq.~(\ref{kernfock}), and a representation
of delta function that is convenient to differentiate. This approach is probably
more efficient, since it avoids the sums over the one-particle basis that appears
in the matrix representation of the Fock kernel, Eq.~(\ref{kernfock.mat}).

The substitution for $\kappa_\gamma({\mathbf x}_4,{\mathbf x}_1)$ in
Eq.~(\ref{secondf3}) is applicable for any diagram.  For example, consider the
second order (correction to the) energy, denoted by $E_2[\gamma]$, that is given
by a sum of the right sides of Eqs.~(\ref{secondf2}), (\ref{secondj}), and
(\ref{secondk}); where, using Eq.~(\ref{delta}), it can be written as a explicit
functional of $\gamma$: \vspace{-5ex} {\footnotesize
\begin{widetext} 
\begin{eqnarray} \label{seconds} 
& & 
E_2[\gamma]= 
\varepsilon_\gamma^{-1}
\left[
\mbox{\rule{0ex}{2.15ex}}
F_\gamma(\mathbf{x}_1,\mathbf{x}_2)
\gamma({\mathbf x}_2,{\mathbf x}_3)
F_\gamma(\mathbf{x}_3,\mathbf{x}_1)
-
F_\gamma(\mathbf{x}_1,\mathbf{x}_2)
\gamma({\mathbf x}_2,{\mathbf x}_3)
F_\gamma(\mathbf{x}_3,\mathbf{x}_4)
\gamma({\mathbf x}_4,{\mathbf x}_1)\right]
\\ \nonumber 
&& \mbox{} \hspace{2ex} + 
\mbox{$\frac{1}{4}$}
\varepsilon_\gamma^{-1}
\left[\mbox{\rule{0ex}{2.15ex}} \mbox{\vspace{10ex}} 
r_{12}^{-2}
\gamma({\mathbf x}_1,{\mathbf x}_1)
\gamma({\mathbf x}_2,{\mathbf x}_2) - 
2 r_{12}^{-1} r_{13}^{-1}
\gamma({\mathbf x}_1,{\mathbf x}_1)
|\gamma({\mathbf x}_2,{\mathbf x}_3)|^2
+ 
r_{12}^{-1} r_{34}^{-1}
|\gamma({\mathbf x}_1,{\mathbf x}_3)|^2
|\gamma({\mathbf x}_2,{\mathbf x}_4)|^2 \right.
\\ && \!\!\!
\hspace{2ex} \left. \mbox{} -  
r_{12}^{-2} |\gamma({\mathbf x}_1,{\mathbf x}_2)|^2 
+ 2 r_{31}^{-1} r_{12}^{-1}
\gamma({\mathbf x}_1,{\mathbf x}_2)
\gamma({\mathbf x}_2,{\mathbf x}_3)
\gamma({\mathbf x}_3,{\mathbf x}_1)
-
r_{13}^{-1} r_{24}^{-1}
\gamma({\mathbf x}_1,{\mathbf x}_2)
\gamma({\mathbf x}_2,{\mathbf x}_3)
\gamma({\mathbf x}_3,{\mathbf x}_4)
\gamma({\mathbf x}_4,{\mathbf x}_1)\mbox{\rule{0ex}{2.15ex}}\right].
\nonumber 
\end{eqnarray}
\end{widetext}}

The individual terms from Eq.~(\ref{seconds}), as well as higher-order terms, can
be represented by diagrams, where we write the Fock kernel in the following
manner:
\begin{equation}
\label{fockkern.d} 
\raisebox{3.6ex}{$F_\gamma(\mathbf{x}_1,\mathbf{x}_2)$} \raisebox{3.6ex}{$\;=\;\;\;$}
\setlength{\unitlength}{0.00062500in}
\begingroup\makeatletter\ifx\SetFigFont\undefined%
\gdef\SetFigFont#1#2#3#4#5{%
  \reset@font\fontsize{#1}{#2pt}%
  \fontfamily{#3}\fontseries{#4}\fontshape{#5}%
  \selectfont}%
\fi\endgroup%
{\renewcommand{\dashlinestretch}{30}
\begin{picture}(407,939)(0,-10)
\path(395,462)(395,12)
\path(20,912)(20,462)
\path(234,462)(235,462)
\blacken\path(160.000,432.000)(235.000,462.000)(160.000,492.000)(182.500,462.000)(160.000,432.000)
\dottedline{60}(20,462)(395,462)
\put(0,347){\makebox(0,0)[lb]{\smash{{{\SetFigFont{5}{6.0}{\rmdefault}{\mddefault}{\updefault}1}}}}}
\put(355,507){\makebox(0,0)[lb]{\smash{{{\SetFigFont{5}{6.0}{\rmdefault}{\mddefault}{\updefault}2}}}}}
\end{picture}
}
\raisebox{4.0ex}{\, ,} 
\end{equation}
and the two terms from Eq.~(\ref{secondf3}) are given by
\begin{subequations}
\begin{eqnarray} 
\label{secondsb} 
\!\!\!\!
\raisebox{-4ex}{
\setlength{\unitlength}{0.00062500in}
\begingroup\makeatletter\ifx\SetFigFont\undefined%
\gdef\SetFigFont#1#2#3#4#5{%
  \reset@font\fontsize{#1}{#2pt}%
  \fontfamily{#3}\fontseries{#4}\fontshape{#5}%
  \selectfont}%
\fi\endgroup%
{\renewcommand{\dashlinestretch}{30}
\begin{picture}(815,1135)(0,-10)
\put(83,535){\ellipse{150}{300}}
\path(758,490)(758,489)
\blacken\path(713.000,624.000)(758.000,489.000)(803.000,624.000)(758.000,583.500)(713.000,624.000)
\path(83,640)(83,641)
\blacken\path(128.000,506.000)(83.000,641.000)(38.000,506.000)(83.000,546.500)(128.000,506.000)
\path(358,115)(357,115)
\blacken\path(458.250,148.750)(357.000,115.000)(458.250,81.250)(427.875,115.000)(458.250,148.750)
\path(483,1015)(484,1015)
\blacken\path(399.620,981.250)(484.000,1015.000)(399.620,1048.750)(424.934,1015.000)(399.620,981.250)
\put(8.000,565.000){\arc{1500.000}{5.6397}{6.9267}}
\put(833.000,565.000){\arc{1500.000}{2.4981}{3.7851}}
\dottedline{60}(233,1015)(608,1015)
\dottedline{60}(233,115)(608,115)
\put(578,0){\makebox(0,0)[lb]{\smash{{{\SetFigFont{5}{6.0}{\rmdefault}{\mddefault}{\updefault}2}}}}}
\put(578,1060){\makebox(0,0)[lb]{\smash{{{\SetFigFont{5}{6.0}{\rmdefault}{\mddefault}{\updefault}3}}}}}
\put(233,1060){\makebox(0,0)[lb]{\smash{{{\SetFigFont{5}{6.0}{\rmdefault}{\mddefault}{\updefault}4}}}}}
\put(233,0){\makebox(0,0)[lb]{\smash{{{\SetFigFont{5}{6.0}{\rmdefault}{\mddefault}{\updefault}1}}}}}
\end{picture}
}
} \hspace{-4ex} &&\raisebox{0.5ex}{$ \displaystyle  
\;\;\; = \; 
-
\varepsilon_\gamma^{-1} 
F_\gamma(\mathbf{x}_1,\mathbf{x}_2)
\gamma({\mathbf x}_2,{\mathbf x}_3)
F_\gamma(\mathbf{x}_3,\mathbf{x}_4)
\gamma({\mathbf x}_4,{\mathbf x}_1),$} 
\nonumber \\ \\ \nonumber
\!\!\!\! 
\label{secondsda} 
\raisebox{-4ex}{
\setlength{\unitlength}{0.00062500in}
\begingroup\makeatletter\ifx\SetFigFont\undefined%
\gdef\SetFigFont#1#2#3#4#5{%
  \reset@font\fontsize{#1}{#2pt}%
  \fontfamily{#3}\fontseries{#4}\fontshape{#5}%
  \selectfont}%
\fi\endgroup%
{\renewcommand{\dashlinestretch}{30}
\begin{picture}(887,1135)(0,-10)
\path(555,1015)(556,1015)
\blacken\path(471.620,981.250)(556.000,1015.000)(471.620,1048.750)(496.934,1015.000)(471.620,981.250)
\path(830,490)(830,489)
\blacken\path(785.000,624.000)(830.000,489.000)(875.000,624.000)(830.000,583.500)(785.000,624.000)
\path(155,640)(155,641)
\blacken\path(200.000,506.000)(155.000,641.000)(110.000,506.000)(155.000,546.500)(200.000,506.000)
\path(430,115)(429,115)
\blacken\path(530.250,148.750)(429.000,115.000)(530.250,81.250)(499.875,115.000)(530.250,148.750)
\put(80.000,565.000){\arc{1500.000}{5.6397}{6.9267}}
\put(905.000,565.000){\arc{1500.000}{2.4981}{3.7851}}
\dottedline{60}(305,1015)(680,1015)
\dottedline{60}(305,115)(680,115)
\put(650,1060){\makebox(0,0)[lb]{\smash{{{\SetFigFont{5}{6.0}{\rmdefault}{\mddefault}{\updefault}3}}}}}
\put(305,1060){\makebox(0,0)[lb]{\smash{{{\SetFigFont{5}{6.0}{\rmdefault}{\mddefault}{\updefault}4}}}}}
\put(650,0){\makebox(0,0)[lb]{\smash{{{\SetFigFont{5}{6.0}{\rmdefault}{\mddefault}{\updefault}2}}}}}
\put(305,0){\makebox(0,0)[lb]{\smash{{{\SetFigFont{5}{6.0}{\rmdefault}{\mddefault}{\updefault}1}}}}}
\put(0,520){\makebox(0,0)[lb]{\smash{{{\SetFigFont{5}{6.0}{\rmdefault}{\mddefault}{\updefault}\mbox{\scriptsize $\delta$}}}}}}
\end{picture}
}
} \hspace{-4ex} &&\raisebox{0.5ex}{$ \displaystyle  
\;\;\; = \; 
\varepsilon_\gamma^{-1} 
F_\gamma(\mathbf{x}_1,\mathbf{x}_2)
\gamma({\mathbf x}_2,{\mathbf x}_3)
F_\gamma(\mathbf{x}_3,\mathbf{x}_4)\delta(\mathbf{x}_4,\mathbf{x}_1).$} \\
\end{eqnarray}
\end{subequations}
These diagrams are evaluated by following the arrows in a backwards direction,
where the circled arrow in Eq.~(\ref{secondsb}) indicates that this line is {\em
not} a particle line, but a hole one with an additional factor of $-1$. This
convention conforms to the one used for folded diagrams in valence-universal
multireference perturbation theory
\cite{Brandow:67,Brandow:77,Sanders:69,Lindgren:74,Lindgren:86}. The
$\delta$--line in Eq.~(\ref{secondsda}) yields a
$\delta(\mathbf{x}_4,\mathbf{x}_1)$ factor; the integration over the
$\mathbf{x}_4$ can be performed, as in Eq.~(\ref{f.delta}), and the resulting
diagram can be represented by
\begin{eqnarray} 
\label{secondsa} 
\!\!\!\! \raisebox{-4ex}{ 
\setlength{\unitlength}{0.00062500in}
\begingroup\makeatletter\ifx\SetFigFont\undefined%
\gdef\SetFigFont#1#2#3#4#5{%
  \reset@font\fontsize{#1}{#2pt}%
  \fontfamily{#3}\fontseries{#4}\fontshape{#5}%
  \selectfont}%
\fi\endgroup%
{\renewcommand{\dashlinestretch}{30}
\begin{picture}(741,1150)(0,-10)
\path(684,490)(684,489)
\blacken\path(639.000,624.000)(684.000,489.000)(729.000,624.000)(684.000,583.500)(639.000,624.000)
\path(284,115)(283,115)
\blacken\path(384.250,148.750)(283.000,115.000)(384.250,81.250)(353.875,115.000)(384.250,148.750)
\path(409,1015)(410,1015)
\blacken\path(325.620,981.250)(410.000,1015.000)(325.620,1048.750)(350.934,1015.000)(325.620,981.250)
\put(-66.000,565.000){\arc{1500.000}{5.6397}{6.9267}}
\put(759.000,565.000){\arc{1500.000}{2.4981}{3.7851}}
\dottedline{60}(159,115)(534,115)
\dottedline{60}(159,1015)(534,1015)
\put(504,0){\makebox(0,0)[lb]{\smash{{{\SetFigFont{5}{6.0}{\rmdefault}{\mddefault}{\updefault}2}}}}}
\put(159,0){\makebox(0,0)[lb]{\smash{{{\SetFigFont{5}{6.0}{\rmdefault}{\mddefault}{\updefault}1}}}}}
\put(504,1060){\makebox(0,0)[lb]{\smash{{{\SetFigFont{5}{6.0}{\rmdefault}{\mddefault}{\updefault}3}}}}}
\put(159,1060){\makebox(0,0)[lb]{\smash{{{\SetFigFont{5}{6.0}{\rmdefault}{\mddefault}{\updefault}1}}}}}
\end{picture}
}
 
} \hspace{-4ex} &&\raisebox{0.5ex}{$ \displaystyle  
\;\;\; = \; 
\varepsilon_\gamma^{-1} 
F_\gamma(\mathbf{x}_1,\mathbf{x}_2)
\gamma({\mathbf x}_2,{\mathbf x}_3)
F_\gamma(\mathbf{x}_3,\mathbf{x}_1),$} \;\;\;\; 
\end{eqnarray}
where the sole purpose of the above internal line {\em without} an arrow is to
preserve the dummy index -- in this case $\mathbf{x}_1$ -- for the two vertices it
connects.

The diagrams from Eqs.~(\ref{secondsa}) and (\ref{secondsb}) correspond to the
first two terms on the right side of Eq.~(\ref{seconds}); the other terms are
given by the diagrams from Fig.~\ref{gamdiag}. The mirror-image diagrams (b) and
(c) are equivalent, or non-distinct. Therefore, one of them can be omitted if a
factor of $2$ is included when evaluating the other one. Similarly, diagrams (f)
and (g) are also non-distinct, so one can be omitted. Higher-order,
correlation-energy ${\mathcal E}_{\mathrm{co}}$ diagrams are obtained in a similar
way.
\begin{figure*}
\setlength{\unitlength}{0.00057220in}
\begingroup\makeatletter\ifx\SetFigFont\undefined%
\gdef\SetFigFont#1#2#3#4#5{%
  \reset@font\fontsize{#1}{#2pt}%
  \fontfamily{#3}\fontseries{#4}\fontshape{#5}%
  \selectfont}%
\fi\endgroup%
{\renewcommand{\dashlinestretch}{30}
\begin{picture}(11274,1764)(0,-10)
\put(537,277){\makebox(0,0)[lb]{\smash{{{\SetFigFont{8}{9.6}{\rmdefault}{\mddefault}{\updefault}(a)}}}}}
\put(1887,277){\makebox(0,0)[lb]{\smash{{{\SetFigFont{8}{9.6}{\rmdefault}{\mddefault}{\updefault}(b)}}}}}
\put(3237,277){\makebox(0,0)[lb]{\smash{{{\SetFigFont{8}{9.6}{\rmdefault}{\mddefault}{\updefault}(c)}}}}}
\put(4602,277){\makebox(0,0)[lb]{\smash{{{\SetFigFont{8}{9.6}{\rmdefault}{\mddefault}{\updefault}(d)}}}}}
\put(6012,277){\makebox(0,0)[lb]{\smash{{{\SetFigFont{8}{9.6}{\rmdefault}{\mddefault}{\updefault}(e)}}}}}
\put(7512,277){\makebox(0,0)[lb]{\smash{{{\SetFigFont{8}{9.6}{\rmdefault}{\mddefault}{\updefault}(f)}}}}}
\put(9012,277){\makebox(0,0)[lb]{\smash{{{\SetFigFont{8}{9.6}{\rmdefault}{\mddefault}{\updefault}(g)}}}}}
\put(10512,277){\makebox(0,0)[lb]{\smash{{{\SetFigFont{8}{9.6}{\rmdefault}{\mddefault}{\updefault}(h)}}}}}
\put(4887,957){\ellipse{150}{300}}
\path(7465,842)(7464,849)(7462,856)
	(7459,863)(7455,870)(7451,875)
	(7446,880)(7440,883)(7432,886)
	(7424,888)(7416,889)(7408,890)
	(7402,890)(7394,889)(7387,887)
	(7381,885)(7374,882)(7369,878)
	(7364,875)(7358,871)(7352,867)
	(7346,862)(7340,857)(7335,853)
	(7329,848)(7323,842)(7316,836)
	(7310,829)(7304,822)(7298,815)
	(7292,808)(7285,800)(7279,790)
	(7273,781)(7267,771)(7263,762)
	(7259,753)(7255,743)(7252,732)
	(7250,722)(7248,713)(7248,705)
	(7247,697)(7248,690)(7249,683)
	(7251,676)(7252,671)(7254,667)
	(7256,663)(7259,660)(7261,657)
	(7265,655)(7268,653)(7272,652)
	(7278,651)(7284,649)(7290,649)
	(7296,649)(7301,649)(7307,650)
	(7313,651)(7319,653)(7326,655)
	(7332,658)(7338,661)(7345,665)
	(7353,669)(7361,674)(7368,678)
	(7374,683)(7380,688)(7386,693)
	(7392,699)(7398,705)(7404,711)
	(7409,717)(7414,724)(7420,731)
	(7426,739)(7431,747)(7436,754)
	(7440,761)(7445,768)(7449,776)
	(7452,783)(7455,788)(7457,792)
	(7458,795)(7459,797)(7460,799)
	(7461,802)(7462,805)(7462,809)
	(7463,815)(7464,821)(7465,828)
	(7465,835)(7465,842)
\path(9195,842)(9196,849)(9198,856)
	(9201,863)(9205,870)(9209,875)
	(9214,880)(9220,883)(9228,886)
	(9236,888)(9244,889)(9252,890)
	(9258,890)(9266,889)(9273,887)
	(9279,885)(9286,882)(9291,878)
	(9296,875)(9302,871)(9308,867)
	(9314,862)(9320,857)(9325,853)
	(9331,848)(9337,842)(9344,836)
	(9350,829)(9356,822)(9362,815)
	(9368,808)(9375,800)(9381,790)
	(9387,781)(9393,771)(9397,762)
	(9401,753)(9405,743)(9408,732)
	(9410,722)(9412,713)(9413,705)
	(9413,697)(9412,690)(9411,683)
	(9409,676)(9408,671)(9406,667)
	(9404,663)(9401,660)(9399,657)
	(9395,655)(9392,653)(9388,652)
	(9382,651)(9376,649)(9370,649)
	(9364,649)(9359,649)(9353,650)
	(9347,651)(9341,653)(9334,655)
	(9328,658)(9322,661)(9315,665)
	(9307,669)(9299,674)(9292,678)
	(9286,683)(9280,688)(9274,693)
	(9268,699)(9262,705)(9256,711)
	(9251,717)(9246,724)(9240,731)
	(9234,739)(9229,747)(9224,754)
	(9220,761)(9215,768)(9211,776)
	(9208,783)(9205,788)(9203,792)
	(9202,795)(9201,797)(9200,799)
	(9199,802)(9199,805)(9198,809)
	(9197,815)(9196,821)(9195,828)
	(9195,835)(9195,842)
\path(10463,842)(10462,849)(10460,856)
	(10457,863)(10453,870)(10449,875)
	(10444,880)(10438,883)(10430,886)
	(10422,888)(10414,889)(10406,890)
	(10400,890)(10392,889)(10385,887)
	(10379,885)(10372,882)(10367,878)
	(10362,875)(10356,871)(10350,867)
	(10344,862)(10338,857)(10333,853)
	(10327,848)(10321,842)(10314,836)
	(10308,829)(10302,822)(10296,815)
	(10290,808)(10283,800)(10277,790)
	(10271,781)(10265,771)(10261,762)
	(10257,753)(10253,743)(10250,732)
	(10248,722)(10246,713)(10246,705)
	(10245,697)(10246,690)(10247,683)
	(10249,676)(10250,671)(10252,667)
	(10254,663)(10257,660)(10259,657)
	(10263,655)(10266,653)(10270,652)
	(10276,651)(10282,649)(10288,649)
	(10294,649)(10299,649)(10305,650)
	(10311,651)(10317,653)(10324,655)
	(10330,658)(10336,661)(10343,665)
	(10351,669)(10359,674)(10366,678)
	(10372,683)(10378,688)(10384,693)
	(10390,699)(10396,705)(10402,711)
	(10407,717)(10412,724)(10418,731)
	(10424,739)(10429,747)(10434,754)
	(10438,761)(10443,768)(10447,776)
	(10450,783)(10453,788)(10455,792)
	(10456,795)(10457,797)(10458,799)
	(10459,802)(10460,805)(10460,809)
	(10461,815)(10462,821)(10463,828)
	(10463,835)(10463,842)
\path(10695,842)(10696,849)(10698,856)
	(10701,863)(10705,870)(10709,875)
	(10714,880)(10720,883)(10728,886)
	(10736,888)(10744,889)(10752,890)
	(10758,890)(10766,889)(10773,887)
	(10779,885)(10786,882)(10791,878)
	(10796,875)(10802,871)(10808,867)
	(10814,862)(10820,857)(10825,853)
	(10831,848)(10837,842)(10844,836)
	(10850,829)(10856,822)(10862,815)
	(10868,808)(10875,800)(10881,790)
	(10887,781)(10893,771)(10897,762)
	(10901,753)(10905,743)(10908,732)
	(10910,722)(10912,713)(10913,705)
	(10913,697)(10912,690)(10911,683)
	(10909,676)(10908,671)(10906,667)
	(10904,663)(10901,660)(10899,657)
	(10895,655)(10892,653)(10888,652)
	(10882,651)(10876,649)(10870,649)
	(10864,649)(10859,649)(10853,650)
	(10847,651)(10841,653)(10834,655)
	(10828,658)(10822,661)(10815,665)
	(10807,669)(10799,674)(10792,678)
	(10786,683)(10780,688)(10774,693)
	(10768,699)(10762,705)(10756,711)
	(10751,717)(10746,724)(10740,731)
	(10734,739)(10729,747)(10724,754)
	(10720,761)(10715,768)(10711,776)
	(10708,783)(10705,788)(10703,792)
	(10702,795)(10701,797)(10700,799)
	(10699,802)(10699,805)(10698,809)
	(10697,815)(10696,821)(10695,828)
	(10695,835)(10695,842)
\put(2187,957){\ellipse{150}{300}}
\put(2787,957){\ellipse{150}{300}}
\put(4137,957){\ellipse{150}{300}}
\path(12,1737)(11262,1737)(11262,12)
	(12,12)(12,1737)
\path(4887,1062)(4887,1063)
\blacken\path(4932.000,928.000)(4887.000,1063.000)(4842.000,928.000)(4887.000,968.500)(4932.000,928.000)
\path(7427,833)(7428,834)
\blacken\path(7364.360,706.721)(7428.000,834.000)(7300.721,770.360)(7361.178,767.178)(7364.360,706.721)
\path(9233,833)(9232,834)
\blacken\path(9359.279,770.360)(9232.000,834.000)(9295.640,706.721)(9298.822,767.178)(9359.279,770.360)
\path(10425,833)(10426,834)
\blacken\path(10362.360,706.721)(10426.000,834.000)(10298.721,770.360)(10359.178,767.178)(10362.360,706.721)
\path(10733,833)(10732,834)
\blacken\path(10859.279,770.360)(10732.000,834.000)(10795.640,706.721)(10798.822,767.178)(10859.279,770.360)
\path(1737,912)(1737,911)
\blacken\path(1692.000,1046.000)(1737.000,911.000)(1782.000,1046.000)(1737.000,1005.500)(1692.000,1046.000)
\path(2487,912)(2487,911)
\blacken\path(2442.000,1046.000)(2487.000,911.000)(2532.000,1046.000)(2487.000,1005.500)(2442.000,1046.000)
\path(387,912)(387,911)
\blacken\path(342.000,1046.000)(387.000,911.000)(432.000,1046.000)(387.000,1005.500)(342.000,1046.000)
\path(1137,912)(1137,911)
\blacken\path(1092.000,1046.000)(1137.000,911.000)(1182.000,1046.000)(1137.000,1005.500)(1092.000,1046.000)
\path(3087,912)(3087,911)
\blacken\path(3042.000,1046.000)(3087.000,911.000)(3132.000,1046.000)(3087.000,1005.500)(3042.000,1046.000)
\path(3837,912)(3837,911)
\blacken\path(3792.000,1046.000)(3837.000,911.000)(3882.000,1046.000)(3837.000,1005.500)(3792.000,1046.000)
\path(4437,912)(4437,911)
\blacken\path(4392.000,1046.000)(4437.000,911.000)(4482.000,1046.000)(4437.000,1005.500)(4392.000,1046.000)
\path(5187,912)(5187,911)
\blacken\path(5142.000,1046.000)(5187.000,911.000)(5232.000,1046.000)(5187.000,1005.500)(5142.000,1046.000)
\path(2187,1062)(2187,1063)
\blacken\path(2232.000,928.000)(2187.000,1063.000)(2142.000,928.000)(2187.000,968.500)(2232.000,928.000)
\path(2787,1062)(2787,1063)
\blacken\path(2832.000,928.000)(2787.000,1063.000)(2742.000,928.000)(2787.000,968.500)(2832.000,928.000)
\path(4137,1062)(4137,1063)
\blacken\path(4182.000,928.000)(4137.000,1063.000)(4092.000,928.000)(4137.000,968.500)(4182.000,928.000)
\path(5487,912)(5487,911)
\blacken\path(5442.000,1046.000)(5487.000,911.000)(5532.000,1046.000)(5487.000,1005.500)(5442.000,1046.000)
\path(6687,912)(6687,911)
\blacken\path(6642.000,1046.000)(6687.000,911.000)(6732.000,1046.000)(6687.000,1005.500)(6642.000,1046.000)
\path(6987,912)(6987,911)
\blacken\path(6942.000,1046.000)(6987.000,911.000)(7032.000,1046.000)(6987.000,1005.500)(6942.000,1046.000)
\path(8187,912)(8187,911)
\blacken\path(8142.000,1046.000)(8187.000,911.000)(8232.000,1046.000)(8187.000,1005.500)(8142.000,1046.000)
\path(9687,912)(9687,911)
\blacken\path(9642.000,1046.000)(9687.000,911.000)(9732.000,1046.000)(9687.000,1005.500)(9642.000,1046.000)
\path(8487,912)(8487,911)
\blacken\path(8442.000,1046.000)(8487.000,911.000)(8532.000,1046.000)(8487.000,1005.500)(8442.000,1046.000)
\path(11187,912)(11187,911)
\blacken\path(11142.000,1046.000)(11187.000,911.000)(11232.000,1046.000)(11187.000,1005.500)(11142.000,1046.000)
\path(9987,912)(9987,911)
\blacken\path(9942.000,1046.000)(9987.000,911.000)(10032.000,1046.000)(9987.000,1005.500)(9942.000,1046.000)
\put(987.000,987.000){\arc{1500.000}{5.6397}{6.9267}}
\put(1737.000,987.000){\arc{1500.000}{5.6397}{6.9267}}
\put(-363.000,987.000){\arc{1500.000}{5.6397}{6.9267}}
\put(387.000,987.000){\arc{1500.000}{5.6397}{6.9267}}
\put(2337.000,987.000){\arc{1500.000}{5.6397}{6.9267}}
\put(3087.000,987.000){\arc{1500.000}{5.6397}{6.9267}}
\put(3687.000,987.000){\arc{1500.000}{5.6397}{6.9267}}
\put(4437.000,987.000){\arc{1500.000}{5.6397}{6.9267}}
\put(837.000,987.000){\arc{1500.000}{2.4981}{3.7851}}
\put(1587.000,987.000){\arc{1500.000}{2.4981}{3.7851}}
\put(2187.000,987.000){\arc{1500.000}{2.4981}{3.7851}}
\put(2937.000,987.000){\arc{1500.000}{2.4981}{3.7851}}
\put(3537.000,987.000){\arc{1500.000}{2.4981}{3.7851}}
\put(4287.000,987.000){\arc{1500.000}{2.4981}{3.7851}}
\put(4887.000,987.000){\arc{1500.000}{2.4981}{3.7851}}
\put(5637.000,987.000){\arc{1500.000}{2.4981}{3.7851}}
\put(6237.000,987.000){\arc{1500.000}{2.4981}{3.7851}}
\put(5937.000,987.000){\arc{1500.000}{5.6397}{6.9267}}
\put(7737.000,987.000){\arc{1500.000}{2.4981}{3.7851}}
\put(7437.000,987.000){\arc{1500.000}{5.6397}{6.9267}}
\put(9237.000,987.000){\arc{1500.000}{2.4981}{3.7851}}
\put(8937.000,987.000){\arc{1500.000}{5.6397}{6.9267}}
\put(10737.000,987.000){\arc{1500.000}{2.4981}{3.7851}}
\put(10437.000,987.000){\arc{1500.000}{5.6397}{6.9267}}
\path(5637,1437)(6539,535)
\path(5637,537)(6539,1439)
\path(7137,556)(8039,1458)
\path(7137,1437)(8039,535)
\path(8637,1437)(9539,535)
\path(8637,537)(9539,1439)
\path(10137,537)(11039,1439)
\path(10137,1437)(11039,535)
\dottedline{60}(1587,1437)(2337,1437)
\dottedline{60}(1587,537)(2337,537)
\dottedline{60}(237,537)(987,537)
\dottedline{60}(2937,1437)(3687,1437)
\dottedline{60}(2937,537)(3687,537)
\dottedline{60}(4287,1437)(5037,1437)
\dottedline{60}(4287,537)(5037,537)
\dottedline{60}(5637,1437)(6537,1437)
\dottedline{60}(7137,1437)(8037,1437)
\dottedline{60}(7137,537)(8037,537)
\dottedline{60}(237,1437)(987,1437)
\dottedline{60}(5637,537)(6537,537)
\dottedline{60}(8637,1437)(9537,1437)
\dottedline{60}(8637,537)(9537,537)
\dottedline{60}(10137,1437)(11037,1437)
\dottedline{60}(10137,537)(11037,537)
\put(207,1482){\makebox(0,0)[lb]{\smash{{{\SetFigFont{5}{6.0}{\rmdefault}{\mddefault}{\updefault}1}}}}}
\put(207,422){\makebox(0,0)[lb]{\smash{{{\SetFigFont{5}{6.0}{\rmdefault}{\mddefault}{\updefault}1}}}}}
\put(957,1482){\makebox(0,0)[lb]{\smash{{{\SetFigFont{5}{6.0}{\rmdefault}{\mddefault}{\updefault}2	}}}}}
\put(957,422){\makebox(0,0)[lb]{\smash{{{\SetFigFont{5}{6.0}{\rmdefault}{\mddefault}{\updefault}2}}}}}
\put(1557,1482){\makebox(0,0)[lb]{\smash{{{\SetFigFont{5}{6.0}{\rmdefault}{\mddefault}{\updefault}1}}}}}
\put(2307,1482){\makebox(0,0)[lb]{\smash{{{\SetFigFont{5}{6.0}{\rmdefault}{\mddefault}{\updefault}3}}}}}
\put(2307,422){\makebox(0,0)[lb]{\smash{{{\SetFigFont{5}{6.0}{\rmdefault}{\mddefault}{\updefault}2}}}}}
\put(2907,1482){\makebox(0,0)[lb]{\smash{{{\SetFigFont{5}{6.0}{\rmdefault}{\mddefault}{\updefault}3}}}}}
\put(2907,422){\makebox(0,0)[lb]{\smash{{{\SetFigFont{5}{6.0}{\rmdefault}{\mddefault}{\updefault}2}}}}}
\put(3657,1482){\makebox(0,0)[lb]{\smash{{{\SetFigFont{5}{6.0}{\rmdefault}{\mddefault}{\updefault}1}}}}}
\put(3657,422){\makebox(0,0)[lb]{\smash{{{\SetFigFont{5}{6.0}{\rmdefault}{\mddefault}{\updefault}1}}}}}
\put(4257,1482){\makebox(0,0)[lb]{\smash{{{\SetFigFont{5}{6.0}{\rmdefault}{\mddefault}{\updefault}3}}}}}
\put(4257,422){\makebox(0,0)[lb]{\smash{{{\SetFigFont{5}{6.0}{\rmdefault}{\mddefault}{\updefault}1}}}}}
\put(5007,1482){\makebox(0,0)[lb]{\smash{{{\SetFigFont{5}{6.0}{\rmdefault}{\mddefault}{\updefault}4}}}}}
\put(5007,422){\makebox(0,0)[lb]{\smash{{{\SetFigFont{5}{6.0}{\rmdefault}{\mddefault}{\updefault}2}}}}}
\put(5607,1482){\makebox(0,0)[lb]{\smash{{{\SetFigFont{5}{6.0}{\rmdefault}{\mddefault}{\updefault}2}}}}}
\put(5607,422){\makebox(0,0)[lb]{\smash{{{\SetFigFont{5}{6.0}{\rmdefault}{\mddefault}{\updefault}1}}}}}
\put(1557,422){\makebox(0,0)[lb]{\smash{{{\SetFigFont{5}{6.0}{\rmdefault}{\mddefault}{\updefault}1}}}}}
\put(6507,1482){\makebox(0,0)[lb]{\smash{{{\SetFigFont{5}{6.0}{\rmdefault}{\mddefault}{\updefault}1}}}}}
\put(6507,422){\makebox(0,0)[lb]{\smash{{{\SetFigFont{5}{6.0}{\rmdefault}{\mddefault}{\updefault}2}}}}}
\put(7107,1482){\makebox(0,0)[lb]{\smash{{{\SetFigFont{5}{6.0}{\rmdefault}{\mddefault}{\updefault}1}}}}}
\put(7107,422){\makebox(0,0)[lb]{\smash{{{\SetFigFont{5}{6.0}{\rmdefault}{\mddefault}{\updefault}3}}}}}
\put(8007,1482){\makebox(0,0)[lb]{\smash{{{\SetFigFont{5}{6.0}{\rmdefault}{\mddefault}{\updefault}2}}}}}
\put(8007,422){\makebox(0,0)[lb]{\smash{{{\SetFigFont{5}{6.0}{\rmdefault}{\mddefault}{\updefault}1}}}}}
\put(8607,1482){\makebox(0,0)[lb]{\smash{{{\SetFigFont{5}{6.0}{\rmdefault}{\mddefault}{\updefault}2}}}}}
\put(8607,422){\makebox(0,0)[lb]{\smash{{{\SetFigFont{5}{6.0}{\rmdefault}{\mddefault}{\updefault}1}}}}}
\put(9507,1482){\makebox(0,0)[lb]{\smash{{{\SetFigFont{5}{6.0}{\rmdefault}{\mddefault}{\updefault}1}}}}}
\put(9507,422){\makebox(0,0)[lb]{\smash{{{\SetFigFont{5}{6.0}{\rmdefault}{\mddefault}{\updefault}3}}}}}
\put(10107,1482){\makebox(0,0)[lb]{\smash{{{\SetFigFont{5}{6.0}{\rmdefault}{\mddefault}{\updefault}2}}}}}
\put(10107,422){\makebox(0,0)[lb]{\smash{{{\SetFigFont{5}{6.0}{\rmdefault}{\mddefault}{\updefault}1}}}}}
\put(11007,1482){\makebox(0,0)[lb]{\smash{{{\SetFigFont{5}{6.0}{\rmdefault}{\mddefault}{\updefault}4}}}}}
\put(11007,422){\makebox(0,0)[lb]{\smash{{{\SetFigFont{5}{6.0}{\rmdefault}{\mddefault}{\updefault}3}}}}}
\end{picture}
}
\caption{\label{gamdiag} Second-order correlation-energy ${\mathcal
E}_{\mathrm{co}}$ diagrams.
}
\end{figure*}
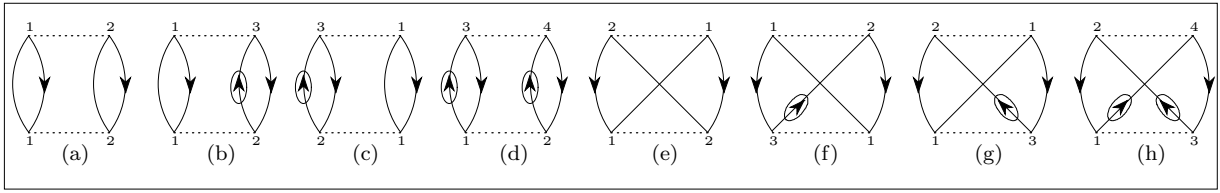

It is well known that ${\mathcal E}_{\mathrm{co}}$ is given by the set of
connected diagrams, when (at least some of) the exclusion principle violating
(EPV) diagrams are included
\cite{Goldstone:57,Hugenholtz:57,Sanders:69,Raimes:72,Paldus:75,Lindgren:74,Wilson:85,Lindgren:86,Harris:92}.
Diagrams (a) through (f) in Fig.~\ref{examples} present examples of diagrams that
contribute to ${\mathcal E}_{\mathrm{co}}$. (Diagram (g) is disconnected and does
not contribute.)  When orbital degeneracy is {\em not} imposed, the hole and
particle lines in these diagrams correspond to $\gamma_w$ and $\gamma_r$,
respectively. However, by imposing orbital degeneracy, as indicated by
Eq.~(\ref{H0.gammab}), and using Eq.~(\ref{delta}), these diagrams can be
converted into ones that explicitly depend on $\gamma$, where the arrows
representing the particle lines are either deleted or circled, in all unique ways.
Additional diagrams containing the $-\varepsilon_\gamma\{\hat{\gamma}\}$ (or
$-\{\hat{\mbox{\sc \Large $o$}}_\gamma\}$) interactions are generated by adding
unfilled arrows, as represented by Eqs.~(\ref{H01o.diag}) and (\ref{ident.oc}),
and including a factor of $\varepsilon_\gamma$ for each one appearing in a
diagram, as well as the appropriate energy-denominators. For example, the
following sixth-order diagram is generated from diagram (a) from
Fig.~\ref{examples}:
\begin{widetext}
\begin{equation}
\label{highord} 
\raisebox{-4ex}{
\setlength{\unitlength}{0.00062500in}
\begingroup\makeatletter\ifx\SetFigFont\undefined%
\gdef\SetFigFont#1#2#3#4#5{%
  \reset@font\fontsize{#1}{#2pt}%
  \fontfamily{#3}\fontseries{#4}\fontshape{#5}%
  \selectfont}%
\fi\endgroup%
{\renewcommand{\dashlinestretch}{30}
\begin{picture}(1437,1571)(0,-10)
\put(83,1107){\ellipse{150}{300}}
\path(333,1512)(334,1512)
\blacken\path(249.620,1478.250)(334.000,1512.000)(249.620,1545.750)(274.934,1512.000)(249.620,1478.250)
\path(83,1212)(83,1213)
\blacken\path(128.000,1078.000)(83.000,1213.000)(38.000,1078.000)(83.000,1118.500)(128.000,1078.000)
\path(561,812)(616,672)
\blacken\path(555.169,744.798)(616.000,672.000)(611.014,766.737)(592.964,730.637)(555.169,744.798)
\path(458,1362)(458,1237)
\blacken\path(428.000,1327.000)(458.000,1237.000)(488.000,1327.000)(458.000,1300.000)(428.000,1327.000)
\path(1358,537)(1358,412)
\whiten\path(1290.500,577.000)(1358.000,412.000)(1425.500,577.000)(1358.000,527.500)(1290.500,577.000)
\path(1358,777)(1358,652)
\whiten\path(1290.500,817.000)(1358.000,652.000)(1425.500,817.000)(1358.000,767.500)(1290.500,817.000)
\path(331,212)(212,119)
\blacken\path(264.440,198.057)(212.000,119.000)(301.386,150.782)(261.639,157.794)(264.440,198.057)
\path(458,1489)(458,1062)(758,315)
	(1358,1062)(1358,12)(458,315)
	(83,12)(83,1512)
\dottedline{60}(83,1512)(458,1512)
\dottedline{60}(461,312)(761,312)
\dottedline{60}(461,1062)(1358,1062)
\dottedline{60}(83,12)(1358,12)
\end{picture}
}

} \hspace{-4ex} \raisebox{0.5ex}{$ \displaystyle  
\;\;\; \;\;\; = \; \
\frac{1}{16} 
\varepsilon_\gamma^{-3}
\gamma(\mathbf{x}_1,\mathbf{x}_2)
\gamma({\mathbf x}_2,{\mathbf x}_3)
\gamma(\mathbf{x}_3,\mathbf{x}_4)
\gamma({\mathbf x}_4,{\mathbf x}_5)
F_\gamma({\mathbf x}_5,{\mathbf x}_6)
\gamma({\mathbf x}_6,{\mathbf x}_1)
,$} 
\end{equation}
\end{widetext} 
where the dummy index from the left-bottom vertex is ${\mathbf x}_1$.  The above
diagram contains four hole-lines, one loop, and a circled particle-line, yielding,
overall, a factor of $+1$.  In addition, each one of the two unfilled-arrow
interactions give a factor of $\varepsilon_\gamma$, and an energy-denominator
factor of $2\varepsilon_\gamma$ -- since both interactions appears within a double
excitation -- so overall a factor of $(2\varepsilon_\gamma)^{-1}
\varepsilon_\gamma$, or $\frac12$, is obtained for each of these interactions; in
general, an unfilled-arrow interaction gives a factor of $\frac1n$ when appearing
on a $n$--body excitation. In the above diagram, an additional factor of
$\frac{1}{4}\varepsilon_\gamma^{-3}$ also appears from the remaining three
energy-denominators.
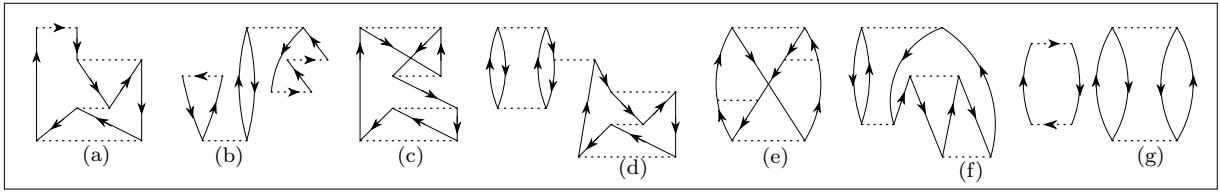
\begin{figure*}
\setlength{\unitlength}{0.00057220in}
\begingroup\makeatletter\ifx\SetFigFont\undefined%
\gdef\SetFigFont#1#2#3#4#5{%
  \reset@font\fontsize{#1}{#2pt}%
  \fontfamily{#3}\fontseries{#4}\fontshape{#5}%
  \selectfont}%
\fi\endgroup%
{\renewcommand{\dashlinestretch}{30}
\begin{picture}(11274,1764)(0,-10)
\put(3657,267){\makebox(0,0)[lb]{\smash{{{\SetFigFont{8}{9.6}{\rmdefault}{\mddefault}{\updefault}(c)}}}}}
\put(1959,262){\makebox(0,0)[lb]{\smash{{{\SetFigFont{8}{9.6}{\rmdefault}{\mddefault}{\updefault}(b)}}}}}
\put(5707,147){\makebox(0,0)[lb]{\smash{{{\SetFigFont{8}{9.6}{\rmdefault}{\mddefault}{\updefault}(d)}}}}}
\put(7037,257){\makebox(0,0)[lb]{\smash{{{\SetFigFont{8}{9.6}{\rmdefault}{\mddefault}{\updefault}(e)}}}}}
\put(10512,267){\makebox(0,0)[lb]{\smash{{{\SetFigFont{8}{9.6}{\rmdefault}{\mddefault}{\updefault}(g)}}}}}
\put(8857,123){\makebox(0,0)[lb]{\smash{{{\SetFigFont{8}{9.6}{\rmdefault}{\mddefault}{\updefault}(f)}}}}}
\path(3312,462)(3312,1512)(4062,1062)
	(4062,1512)(3612,1062)(4212,762)
	(4212,462)(3612,762)(3312,462)
\put(5937.000,1137.000){\arc{1950.000}{2.7468}{3.5364}}
\put(4137.000,1137.000){\arc{1950.000}{5.8884}{6.6780}}
\put(5487.000,1137.000){\arc{1950.000}{2.7468}{3.5364}}
\put(3687.000,1137.000){\arc{1950.000}{5.8884}{6.6780}}
\put(9237.000,1062.000){\arc{2704.163}{2.8023}{3.4809}}
\put(6687.000,1062.000){\arc{2704.163}{5.9439}{6.6225}}
\put(8949.500,830.750){\arc{1442.925}{2.8335}{4.3769}}
\put(8137.000,612.000){\arc{2136.001}{5.2809}{6.5679}}
\path(8262,612)(8412,1062)(8712,312)
	(8862,1062)(9162,312)
\path(12,1737)(11262,1737)(11262,12)
	(12,12)(12,1737)
\put(6592.250,987.000){\arc{1987.500}{5.7266}{6.8398}}
\put(7604.750,987.000){\arc{1987.500}{2.5850}{3.6982}}
\put(9012.000,987.000){\arc{1950.000}{5.8884}{6.6780}}
\put(4062.000,987.000){\arc{3750.000}{2.8578}{3.4254}}
\put(462.000,987.000){\arc{3750.000}{5.9994}{6.5670}}
\put(-648.447,1573.987){\arc{5467.634}{0.1884}{0.4189}}
\put(4347.447,1573.987){\arc{5467.634}{2.7227}{2.9532}}
\put(3349.500,855.750){\arc{1728.665}{3.2067}{4.0038}}
\put(10437.000,987.000){\arc{1950.000}{2.7468}{3.5364}}
\put(11130.750,987.000){\arc{1987.500}{2.5850}{3.6982}}
\put(9443.250,987.000){\arc{1987.500}{5.7266}{6.8398}}
\put(11730.750,987.000){\arc{1987.500}{2.5850}{3.6982}}
\put(10043.250,987.000){\arc{1987.500}{5.7266}{6.8398}}
\path(6761,1512)(7436,462)
\path(7436,1512)(6761,462)
\path(3012,1212)(2787,1512)
\path(2862,912)(2637,1212)
\path(3312,1212)(3312,1362)
\blacken\path(3342.000,1272.000)(3312.000,1362.000)(3282.000,1272.000)(3312.000,1299.000)(3342.000,1272.000)
\path(4212,687)(4212,537)
\blacken\path(4182.000,627.000)(4212.000,537.000)(4242.000,627.000)(4212.000,600.000)(4182.000,627.000)
\path(3887,929)(4020,857)
\blacken\path(3926.571,873.464)(4020.000,857.000)(3955.135,926.229)(3964.597,886.992)(3926.571,873.464)
\path(3499,1403)(3606,1338)
\blacken\path(3513.505,1359.087)(3606.000,1338.000)(3544.656,1410.367)(3552.156,1370.709)(3513.505,1359.087)
\path(4062,1206)(4062,1356)
\blacken\path(4092.000,1266.000)(4062.000,1356.000)(4032.000,1266.000)(4062.000,1293.000)(4092.000,1266.000)
\path(3912,615)(3799,668)
\blacken\path(3893.222,656.943)(3799.000,668.000)(3867.743,602.622)(3856.038,641.248)(3893.222,656.943)
\path(3924,1371)(3833,1285)
\blacken\path(3877.806,1368.621)(3833.000,1285.000)(3919.017,1325.013)(3878.788,1328.272)(3877.806,1368.621)
\path(4515,1119)(4515,1219)
\blacken\path(4545.000,1129.000)(4515.000,1219.000)(4485.000,1129.000)(4515.000,1156.000)(4545.000,1129.000)
\path(4662,1167)(4662,1067)
\blacken\path(4632.000,1157.000)(4662.000,1067.000)(4692.000,1157.000)(4662.000,1130.000)(4632.000,1157.000)
\path(5103,1023)(5085,924)
\blacken\path(5071.584,1017.915)(5085.000,924.000)(5130.616,1007.182)(5096.270,985.984)(5071.584,1017.915)
\path(5085,1377)(5106,1279)
\blacken\path(5057.808,1360.716)(5106.000,1279.000)(5116.476,1373.288)(5092.800,1340.602)(5057.808,1360.716)
\path(4959,1125)(4959,1225)
\blacken\path(4989.000,1135.000)(4959.000,1225.000)(4929.000,1135.000)(4959.000,1162.000)(4989.000,1135.000)
\path(7879,1076)(7882,976)
\blacken\path(7849.315,1065.060)(7882.000,976.000)(7909.288,1066.859)(7880.111,1038.972)(7849.315,1065.060)
\path(5522,497)(5418,388)
\blacken\path(5458.423,473.825)(5418.000,388.000)(5501.834,432.406)(5461.490,433.581)(5458.423,473.825)
\path(5739,807)(5847,702)
\blacken\path(5761.558,743.227)(5847.000,702.000)(5803.383,786.247)(5801.829,745.916)(5761.558,743.227)
\path(5955,448)(5822,520)
\blacken\path(5915.429,503.536)(5822.000,520.000)(5886.865,450.771)(5877.403,490.008)(5915.429,503.536)
\path(6037,707)(6137,820)
\blacken\path(6099.822,732.720)(6137.000,820.000)(6054.889,772.483)(6095.249,772.821)(6099.822,732.720)
\path(6237,687)(6237,537)
\blacken\path(6207.000,627.000)(6237.000,537.000)(6267.000,627.000)(6237.000,600.000)(6207.000,627.000)
\path(8034,1073)(8037,1173)
\blacken\path(8064.288,1082.141)(8037.000,1173.000)(8004.315,1083.940)(8035.111,1110.028)(8064.288,1082.141)
\path(8794,698)(8809,798)
\blacken\path(8825.317,704.546)(8809.000,798.000)(8765.981,713.446)(8799.655,735.697)(8825.317,704.546)
\path(9102,1061)(9063,1153)
\blacken\path(9125.747,1081.847)(9063.000,1153.000)(9070.506,1058.429)(9087.588,1094.996)(9125.747,1081.847)
\path(8326,809)(8359,903)
\blacken\path(8357.494,808.144)(8359.000,903.000)(8300.882,828.018)(8338.132,843.557)(8357.494,808.144)
\path(8406,1293)(8343,1216)
\blacken\path(8376.773,1304.653)(8343.000,1216.000)(8423.210,1266.659)(8382.894,1264.759)(8376.773,1304.653)
\path(8512,814)(8554,723)
\blacken\path(8489.046,792.145)(8554.000,723.000)(8543.524,817.288)(8527.599,780.201)(8489.046,792.145)
\path(8977,769)(9019,678)
\blacken\path(8954.046,747.145)(9019.000,678.000)(9008.524,772.288)(8992.599,735.201)(8954.046,747.145)
\path(2737,912)(2738,912)
\blacken\path(2653.620,878.250)(2738.000,912.000)(2653.620,945.750)(2678.934,912.000)(2653.620,878.250)
\path(2887,1212)(2888,1212)
\blacken\path(2803.620,1178.250)(2888.000,1212.000)(2803.620,1245.750)(2828.934,1212.000)(2803.620,1178.250)
\path(9662,612)(9661,612)
\blacken\path(9762.250,645.750)(9661.000,612.000)(9762.250,578.250)(9731.875,612.000)(9762.250,645.750)
\path(9787,1362)(9788,1362)
\blacken\path(9703.620,1328.250)(9788.000,1362.000)(9703.620,1395.750)(9728.934,1362.000)(9703.620,1328.250)
\path(562,1512)(563,1512)
\blacken\path(478.620,1478.250)(563.000,1512.000)(478.620,1545.750)(503.934,1512.000)(478.620,1478.250)
\path(1787,1062)(1786,1062)
\blacken\path(1887.250,1095.750)(1786.000,1062.000)(1887.250,1028.250)(1856.875,1062.000)(1887.250,1095.750)
\path(1005,598)(872,670)
\blacken\path(965.429,653.536)(872.000,670.000)(936.865,600.771)(927.403,640.008)(965.429,653.536)
\path(1287,837)(1287,712)
\blacken\path(1257.000,802.000)(1287.000,712.000)(1317.000,802.000)(1287.000,775.000)(1257.000,802.000)
\path(806,1042)(886,914)
\blacken\path(812.860,974.420)(886.000,914.000)(863.740,1006.220)(852.610,967.424)(812.860,974.420)
\path(1090,926)(1178,1048)
\blacken\path(1149.680,957.457)(1178.000,1048.000)(1101.019,992.557)(1141.145,996.905)(1149.680,957.457)
\path(7531,1302)(7501,1397)
\blacken\path(7556.709,1320.212)(7501.000,1397.000)(7499.494,1302.144)(7519.971,1336.924)(7556.709,1320.212)
\path(6627,1160)(6651,1257)
\blacken\path(6658.506,1162.429)(6651.000,1257.000)(6600.262,1176.840)(6635.869,1195.844)(6658.506,1162.429)
\path(7549,724)(7573,821)
\blacken\path(7580.506,726.429)(7573.000,821.000)(7522.262,740.840)(7557.869,759.844)(7580.506,726.429)
\path(6671,643)(6641,738)
\blacken\path(6696.709,661.212)(6641.000,738.000)(6639.494,643.144)(6659.971,677.924)(6696.709,661.212)
\path(6896,663)(6839,582)
\blacken\path(6866.260,672.867)(6839.000,582.000)(6915.329,638.338)(6875.256,633.522)(6866.260,672.867)
\path(7338,1346)(7281,1265)
\blacken\path(7308.260,1355.867)(7281.000,1265.000)(7357.329,1321.338)(7317.256,1316.522)(7308.260,1355.867)
\path(7188,1121)(7131,1040)
\blacken\path(7158.260,1130.867)(7131.000,1040.000)(7207.329,1096.338)(7167.256,1091.522)(7158.260,1130.867)
\path(9987,987)(9987,887)
\blacken\path(9957.000,977.000)(9987.000,887.000)(10017.000,977.000)(9987.000,950.000)(9957.000,977.000)
\path(2185,960)(2185,1060)
\blacken\path(2215.000,970.000)(2185.000,1060.000)(2155.000,970.000)(2185.000,997.000)(2215.000,970.000)
\path(2335,1012)(2335,912)
\blacken\path(2305.000,1002.000)(2335.000,912.000)(2365.000,1002.000)(2335.000,975.000)(2305.000,1002.000)
\path(312,1162)(312,1287)
\blacken\path(342.000,1197.000)(312.000,1287.000)(282.000,1197.000)(312.000,1224.000)(342.000,1197.000)
\path(566,665)(448,572)
\blacken\path(500.116,651.272)(448.000,572.000)(537.255,604.148)(497.480,610.997)(500.116,651.272)
\path(1731,776)(1758,680)
\blacken\path(1704.753,758.516)(1758.000,680.000)(1762.512,774.761)(1740.943,740.647)(1704.753,758.516)
\path(1946,716)(1973,812)
\blacken\path(1977.512,717.239)(1973.000,812.000)(1919.753,733.484)(1955.943,751.353)(1977.512,717.239)
\path(2766,1043)(2703,1120)
\blacken\path(2783.210,1069.341)(2703.000,1120.000)(2736.773,1031.347)(2742.894,1071.241)(2783.210,1069.341)
\path(2916,1348)(2853,1425)
\blacken\path(2933.210,1374.341)(2853.000,1425.000)(2886.773,1336.347)(2892.894,1376.241)(2933.210,1374.341)
\path(2636,1353)(2585,1268)
\blacken\path(2605.580,1360.609)(2585.000,1268.000)(2657.029,1329.739)(2617.413,1322.022)(2605.580,1360.609)
\path(9462,962)(9462,1062)
\blacken\path(9492.000,972.000)(9462.000,1062.000)(9432.000,972.000)(9462.000,999.000)(9492.000,972.000)
\path(687,1408)(687,1283)
\blacken\path(657.000,1373.000)(687.000,1283.000)(717.000,1373.000)(687.000,1346.000)(657.000,1373.000)
\path(3487,640)(3418,567)
\blacken\path(3458.020,653.014)(3418.000,567.000)(3501.625,611.799)(3461.276,612.784)(3458.020,653.014)
\path(5407,714)(5419,814)
\blacken\path(5438.063,721.067)(5419.000,814.000)(5378.491,728.215)(5411.494,751.449)(5438.063,721.067)
\path(5563,1069)(5602,977)
\blacken\path(5539.253,1048.153)(5602.000,977.000)(5594.494,1071.571)(5577.412,1035.004)(5539.253,1048.153)
\path(6913,1273)(6970,1192)
\blacken\path(6893.671,1248.338)(6970.000,1192.000)(6942.740,1282.867)(6933.744,1243.522)(6893.671,1248.338)
\path(10439,1012)(10439,912)
\blacken\path(10409.000,1002.000)(10439.000,912.000)(10469.000,1002.000)(10439.000,975.000)(10409.000,1002.000)
\path(11039,962)(11039,1062)
\blacken\path(11069.000,972.000)(11039.000,1062.000)(11009.000,972.000)(11039.000,999.000)(11069.000,972.000)
\path(10737,987)(10737,887)
\blacken\path(10707.000,977.000)(10737.000,887.000)(10767.000,977.000)(10737.000,950.000)(10707.000,977.000)
\path(10137,962)(10137,1062)
\blacken\path(10167.000,972.000)(10137.000,1062.000)(10107.000,972.000)(10137.000,999.000)(10167.000,972.000)
\path(5487,1212)(5637,912)(5937,612)
	(6237,912)(6237,312)(5637,612)
	(5337,312)(5487,1212)
\path(687,1512)(687,1212)(987,765)
	(1287,1212)(1287,465)(687,765)
	(312,462)(312,1512)
\dottedline{60}(3615,762)(4212,762)
\dottedline{60}(3315,462)(4215,462)
\dottedline{60}(3615,1062)(4062,1062)
\dottedline{60}(3312,1512)(4062,1512)
\dottedline{60}(4587,1512)(5037,1512)
\dottedline{60}(4587,762)(5037,762)
\dottedline{60}(5640,612)(5940,612)
\dottedline{60}(5640,912)(6240,912)
\dottedline{60}(5340,312)(6240,312)
\dottedline{60}(7962,1512)(8712,1512)
\dottedline{60}(7962,612)(8262,612)
\dottedline{60}(8412,1062)(8862,1062)
\dottedline{60}(8712,312)(9162,312)
\dottedline{60}(2487,912)(2862,912)
\dottedline{60}(2637,1212)(3012,1212)
\dottedline{60}(9537,612)(9912,612)
\dottedline{60}(9537,1362)(9912,1362)
\dottedline{60}(312,1512)(687,1512)
\dottedline{60}(1662,1062)(2037,1062)
\dottedline{60}(690,762)(990,762)
\dottedline{60}(690,1212)(1290,1212)
\dottedline{60}(10287,462)(10887,462)
\dottedline{60}(10287,1512)(10887,1512)
\dottedline{60}(6761,462)(7436,462)
\dottedline{60}(6761,1512)(7436,1512)
\dottedline{60}(6626,837)(7001,837)
\dottedline{60}(7248,1212)(7563,1212)
\dottedline{60}(5113,1212)(5487,1212)
\dottedline{60}(2262,1512)(2787,1512)
\dottedline{60}(312,462)(1290,462)
\dottedline{60}(1849,462)(2262,462)
\put(734,270){\makebox(0,0)[lb]{\smash{{{\SetFigFont{8}{9.6}{\rmdefault}{\mddefault}{\updefault}(a)}}}}}
\end{picture}
}
\caption{\label{examples} Examples of ${\mathcal E}_{\mathrm{co}}$ and
$E_{\mathrm{co}}^{\scscs (\eta)}$ diagrams.}
\end{figure*}

A discussion of perturbative convergence when imposing orbital degeneracies is
presented in Appendix \ref{ORBDEG}.

\subsection{The correlation-energy functionals $E_{\mathrm{co}}^{\scscs (\eta)}[\gamma]$} 
\label{EXPL2} 

We now briefly discuss the diagrammatic representation of the correlation-energy
functionals $E_{\mathrm{co}}^{\scscs (\eta)}$; the details are given elsewhere
\cite{tobe}. The diagrams for $E_{\mathrm{co}}^{\scscs (\eta)}$ are a subset of
the correlation-energy ${\mathcal E}_{\mathrm{co}}$ diagrams. Therefore, we
continue to use the diagrams from Fig.~\ref{examples} as examples, where, as
mentioned previously, these diagrams can be converted into ones that explicitly
depend on $\gamma$; additional diagrams containing
$-\varepsilon_\gamma\{\hat{\gamma}\}$ interactions are easily generated from the
ones in the figure.

Comparing Eqs.~(\ref{Ecorr.cc}) and (\ref{E.corrbb}), we see that the ${\mathcal
E}_{\mathrm{co}}$ diagrams that contribute to $({H}^{\gamma}_1
S_1^{\gamma})_{\text{cl}}$, or $(\{\hat{F}_\gamma\}S_1^{\gamma})_{\text{cl}}$, do
{\em not} contribute to $E_{\mathrm{co}}^{\scscs (\mathrm{I})}$, where these
diagrams have a final interaction, associated with ${H}^{\gamma}_1$, and a
single-excitation below this last interaction, associated with
$S_1^{\gamma}$. Diagram (a) from \mbox{Fig.\ \ref{examples}} is an example of a
$({H}^{\gamma}_1S_1^{\gamma})_{\text{cl}}$ diagram.

Eqs.~(\ref{E.corrbb}) and (\ref{Eco.II}) indicate that the
$E_{\mathrm{co}}^{\scscs (\mathrm{II})}$ diagrams are a subset of the
$E_{\mathrm{co}}^{\scscs (\mathrm{I})}$ diagrams, where diagrams contributing to
$(\frac12 {H}^{\gamma}_2 S_1^{\gamma}S_1^{\gamma})_{\text{cl}}$ do {\em not}
contribute to $E_{\mathrm{co}}^{\scscs (\mathrm{II})}$. Diagram (b) in \mbox{Fig.\
\ref{examples}} is an example of a $(\frac12 {H}^{\gamma}_2
S_1^{\gamma}S_1^{\gamma})_{\text{cl}}$ diagram \cite{Lindgren:74,Lindgren:86}; when
the top ${H}^{\gamma}_2$ interaction is removed, the resultant diagram is
disconnected -- it possessing two {\em pieces}, where, using the factorization
theorem, it is easily demonstrated that each of these {\em fragments} contribute
to $S_1^{\gamma}$. In general, any $E_{\mathrm{co}}^{\scscs (\mathrm{II})}$ diagrams
diagram that generates a disconnected diagram by removing its top interaction is
{\em not} a $E_{\mathrm{co}}^{\scscs (\mathrm{II})}$ diagram, and the resulting
disconnected diagram contributes to $\frac12 S_1^{\gamma}S_1^{\gamma}$.

It is easily shown that the $E_{\mathrm{co}}^{\scscs (\mathrm{III})}$ diagrams are
a subset of the $E_{\mathrm{co}}^{\scscs (\mathrm{II})}$ diagrams \cite{tobe}. In
general, any $E_{\mathrm{co}}^{\scscs (\mathrm{II})}$ diagram that has an
intermediate single-excitation does not contribute to $E_{\mathrm{co}}^{\scscs
(\mathrm{III})}$. An example is given by diagram (c) in the figure. This diagram
possesses a single-excitation that appears in the second order. In other words, if
we remove the top-two interactions, a second-order, single-excitation diagram is
produced.

Another type of $E_{\mathrm{co}}^{\scscs (\mathrm{II})}$ diagram that does not
contribute to $E_{\mathrm{co}}^{\scscs (\mathrm{III})}$ is diagram (d). If we
remove the top two interactions from this diagram we obtain a disconnected diagram
comprised of two fragments -- one fragment being a third-order {\em single-excitation}
diagram and the other one being a first-order double-excitation diagram. In general,
if we remove any of the top interaction of a $E_{\mathrm{co}}^{\scscs
(\mathrm{II})}$ diagram and obtain a disconnected diagram, in which one or more
fragments are single-excitation diagrams, the (parent) $E_{\mathrm{co}}^{\scscs
(\mathrm{II})}$ diagram does {\em not} contribute to $E_{\mathrm{co}}^{\scscs
(\mathrm{III})}$.

Diagrams (e) and (f) contribute to $E_{\mathrm{co}}^{\scscs (\mathrm{II})}$. Diagram
(e) is connected in each order; diagram (f) is disconnected in the second order, but
both fragments are double-excitation diagrams.

By including certain EPV diagrams, disconnected diagram can be excluded from the
set of diagrams that contribute to the first three correlation-energy functionals:
$E_{\mathrm{co}}^{\scscs (\mathrm{I})}$, $E_{\mathrm{co}}^{\scscs (\mathrm{II})}$,
and $E_{\mathrm{co}}^{\scscs (\mathrm{III})}$.  However, disconnected diagrams,
like diagram (g), {\em do} contribute to $E_{\mathrm{co}}^{\scscs (\mathrm{IV})}$;
the set of diagrams representing $E_{\mathrm{co}}^{\scscs (\mathrm{IV})}$ excludes
diagrams with intermediate single-excitations, and this omission removes certain
disconnected diagrams that are needed to invoke the factorization theorem and
obtain a linked diagram expansion.

\section{Treatment of the External Potential} \label{extpot} 

\subsection{ External potential dependence on ${\cal E}_{\mathrm{co}}$ and
$E_{\mathrm{co}}^{\scscs (\eta)}$} 

Density functional theory employs a universal exchange-correlation functional,
independent of the external potential, permitting approximations to be derived
from model systems, where, in the vicinity of the model systems, the general form
of the exchange-correlation functional is known. In contrasts, our
correlation-energy functionals depend on the external potential $v(\mathbf{r})$:
$E_{\mathrm{co}}^{\scscs (\eta)}[\gamma,v]$, and are, therefore, in this sense,
{\em not} universal. Nevertheless, by partitioning $E_{\mathrm{co}}^{\scscs
(\eta)}[\gamma,v]$ into individual terms that are -- to a varying degree --
universal, we can also obtain approximations from model systems.

In order to pursue this approach, and partition $E_{\mathrm{co}}^{\scscs
(\eta)}[\gamma,v]$, we first partition the Fock-operator, given by
Eq.~(\ref{fock.dm}), into two terms:
\begin{equation} \label{fock.part} 
\hat{F}_\gamma = \hat{F}^g_{\gamma} + v,
\end{equation}
where $\hat{F}^g_{\gamma}$ is the Fock operator for an electron gas: 
\begin{equation}
\hat{F}^g_\gamma=
{-}\mbox{\small$\frac{1}{2}$}\nabla^2 + J_\gamma - K_\gamma,
\end{equation}
and this operator is independent of the external potential $v$.  Using this
partitioning, the one-body part of the perturbation $V_1^{\gamma}$, given by
Eq.~(\ref{V1}), can be written as
\begin{equation}\label{V1.p} 
V_1^{\gamma}= \{\hat{F}^g_\gamma\} + \{v\} 
- \{\hat{\mbox{\sc \Large $o$}}_\gamma\}  
- \hat{\mbox{\sc \large $u$}}_\gamma,
\end{equation}
where the normal-ordered, uncontracted portions of $\hat{F}^g_\gamma$ and $v$ have
the following forms:
\begin{eqnarray}
\{\hat{F}_{\gamma}^{g}
\} &=& \sum_{ij} [i|\hat{F}_{\gamma}^{g}|j] \{a_i^\dagger a_j\}, \\
\{v\} &=& \sum_{ij} [i|v|j] \{a_i^\dagger a_j\}.
\end{eqnarray}

The correlation energy and our correlation-energy functionals, ${\cal
E}_{\mathrm{co}}[\gamma,v]$ and $E_{\mathrm{co}}^{\scscs (\eta)}[\gamma,v]$, have
a dependence on the external potential $v$ that arises, exclusively, from the
$\{v\}$ contribution to $V_1^{\gamma}$, as indicated in Eq.~(\ref{V1.p}). In the
diagrammatic treatment presented in Sec.~\ref{EXPL}, $V_1^{\gamma}$ is partitioned
into three terms, given by Eq.~(\ref{V1}). We now consider an approach where we
partitioned it into the four terms, given by Eq.~(\ref{V1.p}).

The diagrammatic representations of $-\{\hat{\mbox{\sc \Large $o$}}_\gamma\}$ and
$-\{\hat{\mbox{\sc \large $u$}}_\gamma\}$ are given by Eqs.~(\ref{H01.diag}); the
other two terms from Eq.~(\ref{V1.p}) use the following:
\begin{subequations}  
\label{fgvb.diag} 
\begin{eqnarray}  \label{fg.diag} 
\raisebox{3.6ex}{$\{\hat{F}^g_\gamma\}$} &\raisebox{3.6ex}{$\;=\;\;$}& 
\setlength{\unitlength}{0.00062500in}
\begingroup\makeatletter\ifx\SetFigFont\undefined%
\gdef\SetFigFont#1#2#3#4#5{%
  \reset@font\fontsize{#1}{#2pt}%
  \fontfamily{#3}\fontseries{#4}\fontshape{#5}%
  \selectfont}%
\fi\endgroup%
{\renewcommand{\dashlinestretch}{30}
\begin{picture}(474,939)(0,-10)
\path(12,462)(49,397)(124,397)
	(162,462)(124,527)(50,527)(12,462)
\path(462,912)(462,12)
\dottedline{45}(162,462)(432,462)
\end{picture}
}
\raisebox{4.0ex}{\, ,} \\ \label{v.diag} 
\raisebox{3.6ex}{$\{v\}$} &\raisebox{3.6ex}{$\;=\;\;$}& 
\setlength{\unitlength}{0.00062500in}
\begingroup\makeatletter\ifx\SetFigFont\undefined%
\gdef\SetFigFont#1#2#3#4#5{%
  \reset@font\fontsize{#1}{#2pt}%
  \fontfamily{#3}\fontseries{#4}\fontshape{#5}%
  \selectfont}%
\fi\endgroup%
{\renewcommand{\dashlinestretch}{30}
\begin{picture}(474,939)(0,-10)
\path(12,462)(87,387)(162,462)
	(87,537)(12,462)
\path(462,912)(462,12)
\dottedline{45}(162,462)(432,462)
\end{picture}
}
\raisebox{4.0ex}{\, .}
\end{eqnarray}
\end{subequations} 

For simplicity, in this section, we consider diagrammatic examples that do not
distinguish between the various functionals, do not contain $-\{\hat{\mbox{\sc
\Large $o$}}_\gamma\}$ and $- \{\hat{\mbox{\sc \large $u$}}_\gamma\}$ insert, and
do not explicitly depend on $\gamma$.  For example, diagrams (a) through (d) in
Fig.~\ref{fgv2} are simple examples of fourth-order diagrams that contain the
$\{\hat{F}^g_\gamma\}$ and $\{v\}$ operators.  These diagrams contribute to the
correlation energy ${\cal E}_{\mathrm{co}}[\gamma,v]$, and any of the
correlation-energy functionals $E_{\mathrm{co}}^{\scscs
(\eta)}[\gamma,v]$. Diagram (e) is the corresponding diagram that uses the
$\{\hat{F}_\gamma\}$ operator, given by Eq.~(\ref{h1.diag}). All four diagram, (a)
through (d), can be obtained from diagram (e) by replacing the
$\{\hat{F}_\gamma\}$ operators with the $\{\hat{F}^g_\gamma\}$ and $\{v\}$
operators in all unique ways; the sum of diagrams (a) through (d) is equal to
diagram~(e). In general, all diagrams containing $\{\hat{F}^g_\gamma\}$ and
$\{v\}$ can be obtained from the $\{\hat{F}_\gamma\}$ diagrams by using this
approach. A similar procedure can be used when adding an additional
perturbation~\cite{Lindgren:86}.
\begin{figure}
\setlength{\unitlength}{0.00062500in}
\begingroup\makeatletter\ifx\SetFigFont\undefined%
\gdef\SetFigFont#1#2#3#4#5{%
  \reset@font\fontsize{#1}{#2pt}%
  \fontfamily{#3}\fontseries{#4}\fontshape{#5}%
  \selectfont}%
\fi\endgroup%
{\renewcommand{\dashlinestretch}{30}
\begin{picture}(5199,1239)(0,-10)
\put(517,122){\makebox(0,0)[lb]{\smash{{{\SetFigFont{8}{9.6}{\rmdefault}{\mddefault}{\updefault}(a)}}}}}
\put(1587,122){\makebox(0,0)[lb]{\smash{{{\SetFigFont{8}{9.6}{\rmdefault}{\mddefault}{\updefault}(b)}}}}}
\put(2637,122){\makebox(0,0)[lb]{\smash{{{\SetFigFont{8}{9.6}{\rmdefault}{\mddefault}{\updefault}(c)}}}}}
\put(3687,122){\makebox(0,0)[lb]{\smash{{{\SetFigFont{8}{9.6}{\rmdefault}{\mddefault}{\updefault}(d)}}}}}
\put(4737,122){\makebox(0,0)[lb]{\smash{{{\SetFigFont{8}{9.6}{\rmdefault}{\mddefault}{\updefault}(e)}}}}}
\path(12,1212)(5187,1212)(5187,12)
	(12,12)(12,1212)
\put(5937.000,687.000){\arc{1950.000}{2.7468}{3.5364}}
\put(4137.000,687.000){\arc{1950.000}{5.8884}{6.6780}}
\put(5487.000,687.000){\arc{1950.000}{2.7468}{3.5364}}
\put(3687.000,687.000){\arc{1950.000}{5.8884}{6.6780}}
\put(1737.000,687.000){\arc{1950.000}{2.7468}{3.5364}}
\put(-63.000,687.000){\arc{1950.000}{5.8884}{6.6780}}
\put(2787.000,687.000){\arc{1950.000}{2.7468}{3.5364}}
\put(987.000,687.000){\arc{1950.000}{5.8884}{6.6780}}
\put(3837.000,687.000){\arc{1950.000}{2.7468}{3.5364}}
\put(2037.000,687.000){\arc{1950.000}{5.8884}{6.6780}}
\put(4887.000,687.000){\arc{1950.000}{2.7468}{3.5364}}
\put(3087.000,687.000){\arc{1950.000}{5.8884}{6.6780}}
\put(1287.000,687.000){\arc{1950.000}{2.7468}{3.5364}}
\put(-513.000,687.000){\arc{1950.000}{5.8884}{6.6780}}
\put(2337.000,687.000){\arc{1950.000}{2.7468}{3.5364}}
\put(537.000,687.000){\arc{1950.000}{5.8884}{6.6780}}
\put(3387.000,687.000){\arc{1950.000}{2.7468}{3.5364}}
\put(1587.000,687.000){\arc{1950.000}{5.8884}{6.6780}}
\put(4437.000,687.000){\arc{1950.000}{2.7468}{3.5364}}
\put(2637.000,687.000){\arc{1950.000}{5.8884}{6.6780}}
\path(62,837)(87,794)(137,794)
	(162,837)(137,880)(87,880)(62,837)
\path(62,537)(87,494)(137,494)
	(162,537)(137,580)(87,580)(62,537)
\path(1112,837)(1137,794)(1187,794)
	(1212,837)(1187,880)(1137,880)(1112,837)
\path(1114,537)(1163,488)(1212,537)
	(1163,586)(1114,537)
\path(2164,837)(2213,788)(2262,837)
	(2213,886)(2164,837)
\path(2162,537)(2187,494)(2237,494)
	(2262,537)(2237,580)(2187,580)(2162,537)
\path(3214,837)(3263,788)(3312,837)
	(3263,886)(3214,837)
\path(3214,537)(3263,488)(3312,537)
	(3263,586)(3214,537)
\path(4556,398)(4532,495)
\blacken\path(4564.612,441.260)(4532.000,495.000)(4528.210,432.253)(4542.088,454.229)(4564.612,441.260)
\path(4538,906)(4562,1003)
\blacken\path(4565.790,940.253)(4562.000,1003.000)(4529.388,949.260)(4551.912,962.229)(4565.790,940.253)
\path(4512,639)(4512,739)
\blacken\path(4530.750,679.000)(4512.000,739.000)(4493.250,679.000)(4512.000,697.000)(4530.750,679.000)
\path(4962,639)(4962,739)
\blacken\path(4980.750,679.000)(4962.000,739.000)(4943.250,679.000)(4962.000,697.000)(4980.750,679.000)
\path(4662,752)(4662,652)
\blacken\path(4643.250,712.000)(4662.000,652.000)(4680.750,712.000)(4662.000,694.000)(4643.250,712.000)
\path(5112,752)(5112,652)
\blacken\path(5093.250,712.000)(5112.000,652.000)(5130.750,712.000)(5112.000,694.000)(5093.250,712.000)
\path(356,398)(332,495)
\blacken\path(364.612,441.260)(332.000,495.000)(328.210,432.253)(342.088,454.229)(364.612,441.260)
\path(338,906)(362,1003)
\blacken\path(365.790,940.253)(362.000,1003.000)(329.388,949.260)(351.912,962.229)(365.790,940.253)
\path(312,639)(312,739)
\blacken\path(330.750,679.000)(312.000,739.000)(293.250,679.000)(312.000,697.000)(330.750,679.000)
\path(762,639)(762,739)
\blacken\path(780.750,679.000)(762.000,739.000)(743.250,679.000)(762.000,697.000)(780.750,679.000)
\path(462,752)(462,652)
\blacken\path(443.250,712.000)(462.000,652.000)(480.750,712.000)(462.000,694.000)(443.250,712.000)
\path(912,752)(912,652)
\blacken\path(893.250,712.000)(912.000,652.000)(930.750,712.000)(912.000,694.000)(893.250,712.000)
\path(1406,398)(1382,495)
\blacken\path(1414.612,441.260)(1382.000,495.000)(1378.210,432.253)(1392.088,454.229)(1414.612,441.260)
\path(1388,906)(1412,1003)
\blacken\path(1415.790,940.253)(1412.000,1003.000)(1379.388,949.260)(1401.912,962.229)(1415.790,940.253)
\path(1362,639)(1362,739)
\blacken\path(1380.750,679.000)(1362.000,739.000)(1343.250,679.000)(1362.000,697.000)(1380.750,679.000)
\path(1812,639)(1812,739)
\blacken\path(1830.750,679.000)(1812.000,739.000)(1793.250,679.000)(1812.000,697.000)(1830.750,679.000)
\path(1512,752)(1512,652)
\blacken\path(1493.250,712.000)(1512.000,652.000)(1530.750,712.000)(1512.000,694.000)(1493.250,712.000)
\path(1962,752)(1962,652)
\blacken\path(1943.250,712.000)(1962.000,652.000)(1980.750,712.000)(1962.000,694.000)(1943.250,712.000)
\path(2456,398)(2432,495)
\blacken\path(2464.612,441.260)(2432.000,495.000)(2428.210,432.253)(2442.088,454.229)(2464.612,441.260)
\path(2438,906)(2462,1003)
\blacken\path(2465.790,940.253)(2462.000,1003.000)(2429.388,949.260)(2451.912,962.229)(2465.790,940.253)
\path(2412,639)(2412,739)
\blacken\path(2430.750,679.000)(2412.000,739.000)(2393.250,679.000)(2412.000,697.000)(2430.750,679.000)
\path(2862,639)(2862,739)
\blacken\path(2880.750,679.000)(2862.000,739.000)(2843.250,679.000)(2862.000,697.000)(2880.750,679.000)
\path(2562,752)(2562,652)
\blacken\path(2543.250,712.000)(2562.000,652.000)(2580.750,712.000)(2562.000,694.000)(2543.250,712.000)
\path(3012,752)(3012,652)
\blacken\path(2993.250,712.000)(3012.000,652.000)(3030.750,712.000)(3012.000,694.000)(2993.250,712.000)
\path(3506,398)(3482,495)
\blacken\path(3514.612,441.260)(3482.000,495.000)(3478.210,432.253)(3492.088,454.229)(3514.612,441.260)
\path(3488,906)(3512,1003)
\blacken\path(3515.790,940.253)(3512.000,1003.000)(3479.388,949.260)(3501.912,962.229)(3515.790,940.253)
\path(3462,639)(3462,739)
\blacken\path(3480.750,679.000)(3462.000,739.000)(3443.250,679.000)(3462.000,697.000)(3480.750,679.000)
\path(3912,639)(3912,739)
\blacken\path(3930.750,679.000)(3912.000,739.000)(3893.250,679.000)(3912.000,697.000)(3930.750,679.000)
\path(3612,752)(3612,652)
\blacken\path(3593.250,712.000)(3612.000,652.000)(3630.750,712.000)(3612.000,694.000)(3593.250,712.000)
\path(4062,752)(4062,652)
\blacken\path(4043.250,712.000)(4062.000,652.000)(4080.750,712.000)(4062.000,694.000)(4043.250,712.000)
\put(4312,837){\blacken\ellipse{100}{52}}
\put(4312,837){\ellipse{100}{52}}
\put(4312,537){\blacken\ellipse{100}{52}}
\put(4312,537){\ellipse{100}{52}}
\dottedline{60}(4587,1062)(5037,1062)
\dottedline{60}(4587,312)(5037,312)
\dottedline{45}(4362,837)(4522,837)
\dottedline{45}(4362,537)(4522,537)
\dottedline{60}(387,1062)(837,1062)
\dottedline{60}(387,312)(837,312)
\dottedline{45}(162,837)(322,837)
\dottedline{45}(162,537)(322,537)
\dottedline{60}(1437,1062)(1887,1062)
\dottedline{60}(1437,312)(1887,312)
\dottedline{45}(1212,837)(1372,837)
\dottedline{45}(1212,537)(1372,537)
\dottedline{60}(2487,1062)(2937,1062)
\dottedline{60}(2487,312)(2937,312)
\dottedline{45}(2262,837)(2422,837)
\dottedline{45}(2262,537)(2422,537)
\dottedline{60}(3537,1062)(3987,1062)
\dottedline{60}(3537,312)(3987,312)
\dottedline{45}(3312,837)(3472,837)
\dottedline{45}(3312,537)(3472,537)
\end{picture}
}
\caption{\label{fgv2} Fourth-order ${\mathcal E}_{\mathrm{co}}$
and $E_{\mathrm{co}}^{\scscs (\eta)}$ diagrams.
}
\end{figure}
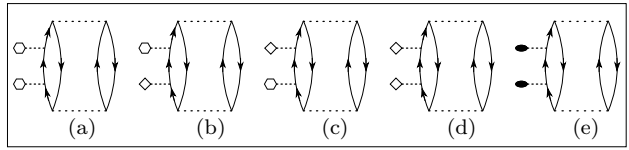

\subsection{Electron gas terms}

Note that the diagrams from Fig.~\ref{fgv2} can be partitioned into one that does
{\em not} depend on the external potential $v$, diagram (a) in the figure, and the
remaining ones that do depend on $v$. This is a general result when using the
partitioned Fock operator $\hat{F}_{\gamma}$, given by Eq.~(\ref{fock.part});
therefore, the correlation energy ${\cal E}_{\mathrm{co}}[\gamma,v]$, and the
correlation-energy functionals $E_{\mathrm{co}}^{\scscs (\eta)}[\gamma,v]$, can be
divided in the following manner:
\begin{subequations}  
\label{E.part} 
\begin{eqnarray}  \label{Eco.part} 
{\cal E}_{\mathrm{co}}[\gamma,v]&=& 
{\cal E}_{\mathrm{co}}^{\scscs (0)}[\gamma] +
{\cal E}_{\mathrm{co}}^{\scscs (1+)}[\gamma,v],  \\
E_{\mathrm{co}}^{\scscs (\eta)}[\gamma,v] &=&
E_{\mathrm{co}}^{\scscs (\eta,0)}[\gamma] +
E_{\mathrm{co}}^{\scscs (\eta,1+)}[\gamma,v],
\end{eqnarray}
\end{subequations} 
where ${\cal E}_{\mathrm{co}}^{\scscs (0)}[\gamma]$ and $E_{\mathrm{co}}^{\scscs
(\eta,0)}[\gamma]$ are universal functionals of $\gamma$, independent of the
external potential; they are given by the sum of diagrams that do {\em not}
contain $\{v\}$ inserts; including diagram (a) in the figure, and the second-order
Coulomb and exchange diagrams, given by Eqs.~(\ref{secondj.d}) and
(\ref{secondk.d}). The remaining diagrams of ${\cal E}_{\mathrm{co}}[\gamma,v]$
and $E_{\mathrm{co}}^{\scscs (\eta)}[\gamma,v]$ contribute to ${\cal
E}_{\mathrm{co}}^{\scscs (1+)}[\gamma,v]$ and $E_{\mathrm{co}}^{\scscs
(\eta,1+)}[\gamma,v]$, respectively, including diagrams (b) through (d) in the
figure.

We refer to ${\cal E}_{\mathrm{co}}^{\scscs (0)}[\gamma]$ and
$E_{\mathrm{co}}^{\scscs (\eta,0)}[\gamma]$ as {\em electron-gas} terms, since,
for an electron gas, $v$ is constant, so $\{v\}$ is zero, and we have the
following:
\begin{subequations} \label{E.gas} 
\label{E.part.gas} 
\begin{eqnarray}  \label{Eco.part.gas} 
\begin{picture}(0,0)(0,0)
\put(90,-12){$\left. \makebox(0,26){} \right\}$ (electron gas)}
\end{picture}
{\cal E}_{\mathrm{co}}[\gamma]&=& 
{\cal E}_{\mathrm{co}}^{\scscs (0)}[\gamma],  \\
\label{Eco.part.gas.a} 
E_{\mathrm{co}}^{\scscs (\eta)}[\gamma] &=&
E_{\mathrm{co}}^{\scscs (\eta,0)}[\gamma]. \hspace{20ex}
\end{eqnarray}
\end{subequations}
From Eq.~(\ref{Eco.ident}), the terms on the left sides of Eqs.~(\ref{E.gas}) are
equal when $\gamma$ is the one-particle density-matrix for the Brueckner state;
hence, we have
\begin{equation} \label{E.part.tau} 
E_{\mathrm{co}}^{\scscs (\eta,0)}[\tau_g] =
{\cal E}_{\mathrm{co}}^{\scscs (0)}[\tau_g],
\end{equation}
where $\tau_g$ is the Brueckner-state one-particle density-matrix for an electron
gas. For future use, we denote the correlation energy of an electron gas by ${\cal
E}_{\mathrm{co}}^{\scscs (\text{gas})}[\gamma]$; explicitly, we have
\begin{equation} \label{Ecorr.gas.def} 
{\cal E}_{\mathrm{co}}^{\scscs (\text{gas})}[\gamma] = 
{\cal E}_{\mathrm{co}}[\gamma,v],\;\;\; \mbox{for $v=\text{constant}$},
\end{equation}
so, using Eq.~(\ref{Eco.part.gas}), Eq.~(\ref{E.part.tau}) can be written as
\begin{equation} \label{E.part.taub} 
E_{\mathrm{co}}^{\scscs (\eta,0)}[\tau_g] =
{\cal E}_{\mathrm{co}}^{\scscs (\text{gas})}[\tau_g].
\end{equation}

\subsection{Atomic and molecular terms}

In order to further generalize Eqs.~(\ref{E.part}), we partition the external
potential into its individual components:
\begin{equation} \label{vi.pot} 
v = \sum_m v_m,
\end{equation}
where $v_m(\mathbf{r})$ is the external potential from the $m$th nuclear
point-charge at $\mathbf{R}_m$:
\begin{equation} \label{vm} 
v_m(\mathbf{r}) = -\frac{Z_m}{|\mathbf{R}_m-\mathbf{r}|},
\end{equation}
and we choose the following diagrammatic representation for this operator:
\begin{equation}  \label{vm.diag} 
\raisebox{3.6ex}{$\{v_m\}$} \raisebox{3.6ex}{$\;=\;\;$} 
\setlength{\unitlength}{0.00062500in}
\begingroup\makeatletter\ifx\SetFigFont\undefined%
\gdef\SetFigFont#1#2#3#4#5{%
  \reset@font\fontsize{#1}{#2pt}%
  \fontfamily{#3}\fontseries{#4}\fontshape{#5}%
  \selectfont}%
\fi\endgroup%
{\renewcommand{\dashlinestretch}{30}
\begin{picture}(492,939)(0,-10)
\put(0,417){\makebox(0,0)[lb]{\smash{{{\SetFigFont{9}{10.8}{\rmdefault}{\mddefault}{\itdefault}m}}}}}
\path(480,912)(480,12)
\dottedline{45}(180,462)(450,462)
\end{picture}
}
\raisebox{4.0ex}{\, .}
\end{equation}

Using the decomposition of $v$, given by Eq.~(\ref{vi.pot}), the one-body part of
the perturbation $V_1^{\gamma}$, Eq.~(\ref{V1.p}), becomes
\begin{equation}\label{V1b.p} 
V_1^{\gamma}= \{\hat{F}^g_\gamma\} + \sum_m \{v_m\} 
- \{\hat{\mbox{\sc \Large $o$}}_\gamma\} - \{\hat{\mbox{\sc \large $u$}}_\gamma\},
\end{equation}
permitting us to generalize Eqs.~(\ref{E.part}) in the following manner:
\begin{subequations}  
\label{Eat.part} 
\begin{eqnarray}  \label{Eatco.part} 
{\cal E}_{\mathrm{co}}[\gamma,v]&=& 
{\cal E}_{\mathrm{co}}^{\scscs (0)}[\gamma] +
\sum_m{\cal E}_{\mathrm{co}}^{\scscs (1)}[\gamma,v_m] +
{\cal E}_{\mathrm{co}}^{\scscs (2+)}[\gamma,v],  \\
E_{\mathrm{co}}^{\scscs (\eta)}[\gamma,v] &=&
E_{\mathrm{co}}^{\scscs (\eta,0)}[\gamma] +
\sum_m E_{\mathrm{co}}^{\scscs (\eta,1)}[\gamma,v_m] +
E_{\mathrm{co}}^{\scscs (\eta,2+)}[\gamma,v],
\end{eqnarray}
\end{subequations}  
where the {\em atomic} terms, ${\cal E}_{\mathrm{co}}^{\scscs (1)}[\gamma,v_m]$
and $E_{\mathrm{co}}^{\scscs (\eta,1)}[\gamma,v_m]$, depend on $v_m$; they are the
sum of all diagrams that contain one or more $\{v_m\}$ inserts, but have no
inserts from other external-potential components, say $\{v_n\}$, where ($n\ne
m$). For example, diagrams (b), (c), and (d) from Fig.~\ref{fgv3} contribute to
${\cal E}_{\mathrm{co}}^{\scscs (1)}[\gamma,v_m]$ and $E_{\mathrm{co}}^{\scscs
(\eta,1)}[\gamma,v_m]$. These diagrams are generated from the corresponding
diagrams in Fig.~\ref{fgv2} by replacing all the external-potential inserts,
$\{v\}$, given by Eq.~(\ref{v.diag}), with the individual components, $\{v_m\}$,
given by Eq.~(\ref{vm.diag}).

The remaining diagrams contribute to ${\cal E}_{\mathrm{co}}^{\scscs
(2+)}[\gamma,v]$ and $E_{\mathrm{co}}^{\scscs (\eta,2+)}[\gamma,v]$. These
diagrams have two or more inserts from {\em different} external-potential
components, e.g., diagrams with {\em both} $\{v_m\}$ and $\{v_n\}$ inserts, where
($m\ne n$). Diagram (a) in the figure is an example of a ${\cal
E}_{\mathrm{co}}^{\scscs (2+)}[\gamma,v]$ or $E_{\mathrm{co}}^{\scscs
(\eta,2+)}[\gamma,v]$ diagram. Diagram (e) is also a $E_{\mathrm{co}}^{\scscs
(\eta,2+)}[\gamma,v]$ diagram, unless ($m=n=p$), where it then contributes to ${\cal
E}_{\mathrm{co}}^{\scscs (1)}[\gamma,v_m]$ and $E_{\mathrm{co}}^{\scscs
(\eta,1)}[\gamma,v_m]$.
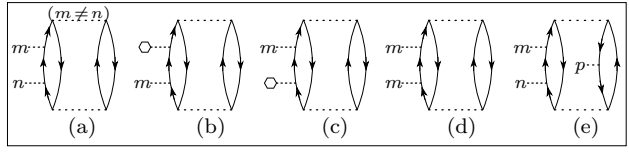
\begin{figure}
\setlength{\unitlength}{0.00062500in}
\begingroup\makeatletter\ifx\SetFigFont\undefined%
\gdef\SetFigFont#1#2#3#4#5{%
  \reset@font\fontsize{#1}{#2pt}%
  \fontfamily{#3}\fontseries{#4}\fontshape{#5}%
  \selectfont}%
\fi\endgroup%
{\renewcommand{\dashlinestretch}{30}
\begin{picture}(5199,1239)(0,-10)
\put(342,1085){\mbox{\tiny ($m\!\ne \!n$)}}
\put(517,122){\makebox(0,0)[lb]{\smash{{{\SetFigFont{8}{9.6}{\rmdefault}{\mddefault}{\updefault}(a)}}}}}
\put(1587,122){\makebox(0,0)[lb]{\smash{{{\SetFigFont{8}{9.6}{\rmdefault}{\mddefault}{\updefault}(b)}}}}}
\put(2637,122){\makebox(0,0)[lb]{\smash{{{\SetFigFont{8}{9.6}{\rmdefault}{\mddefault}{\updefault}(c)}}}}}
\put(3687,122){\makebox(0,0)[lb]{\smash{{{\SetFigFont{8}{9.6}{\rmdefault}{\mddefault}{\updefault}(d)}}}}}
\put(4737,122){\makebox(0,0)[lb]{\smash{{{\SetFigFont{8}{9.6}{\rmdefault}{\mddefault}{\updefault}(e)}}}}}
\path(12,1212)(5187,1212)(5187,12)
	(12,12)(12,1212)
\put(1737.000,687.000){\arc{1950.000}{2.7468}{3.5364}}
\put(-63.000,687.000){\arc{1950.000}{5.8884}{6.6780}}
\put(2787.000,687.000){\arc{1950.000}{2.7468}{3.5364}}
\put(987.000,687.000){\arc{1950.000}{5.8884}{6.6780}}
\put(3837.000,687.000){\arc{1950.000}{2.7468}{3.5364}}
\put(2037.000,687.000){\arc{1950.000}{5.8884}{6.6780}}
\put(4887.000,687.000){\arc{1950.000}{2.7468}{3.5364}}
\put(3087.000,687.000){\arc{1950.000}{5.8884}{6.6780}}
\put(5937.000,687.000){\arc{1950.000}{2.7468}{3.5364}}
\put(4137.000,687.000){\arc{1950.000}{5.8884}{6.6780}}
\put(1287.000,687.000){\arc{1950.000}{2.7468}{3.5364}}
\put(-513.000,687.000){\arc{1950.000}{5.8884}{6.6780}}
\put(2337.000,687.000){\arc{1950.000}{2.7468}{3.5364}}
\put(537.000,687.000){\arc{1950.000}{5.8884}{6.6780}}
\put(3387.000,687.000){\arc{1950.000}{2.7468}{3.5364}}
\put(1587.000,687.000){\arc{1950.000}{5.8884}{6.6780}}
\put(4437.000,687.000){\arc{1950.000}{2.7468}{3.5364}}
\put(2637.000,687.000){\arc{1950.000}{5.8884}{6.6780}}
\put(5487.000,687.000){\arc{1950.000}{2.7468}{3.5364}}
\put(3687.000,687.000){\arc{1950.000}{5.8884}{6.6780}}
\path(1112,837)(1137,794)(1187,794)
	(1212,837)(1187,880)(1137,880)(1112,837)
\path(2162,537)(2187,494)(2237,494)
	(2262,537)(2237,580)(2187,580)(2162,537)
\put(2112,792){\makebox(0,0)[lb]{\smash{{{\SetFigFont{7}{8.4}{\rmdefault}{\mddefault}{\itdefault}m}}}}}
\put(1062,492){\makebox(0,0)[lb]{\smash{{{\SetFigFont{7}{8.4}{\rmdefault}{\mddefault}{\itdefault}m}}}}}
\put(3162,492){\makebox(0,0)[lb]{\smash{{{\SetFigFont{7}{8.4}{\rmdefault}{\mddefault}{\itdefault}m}}}}}
\put(3162,792){\makebox(0,0)[lb]{\smash{{{\SetFigFont{7}{8.4}{\rmdefault}{\mddefault}{\itdefault}m}}}}}
\put(47,492){\makebox(0,0)[lb]{\smash{{{\SetFigFont{7}{8.4}{\rmdefault}{\mddefault}{\itdefault}n}}}}}
\put(37,792){\makebox(0,0)[lb]{\smash{{{\SetFigFont{7}{8.4}{\rmdefault}{\mddefault}{\itdefault}m}}}}}
\put(4237,792){\makebox(0,0)[lb]{\smash{{{\SetFigFont{7}{8.4}{\rmdefault}{\mddefault}{\itdefault}m}}}}}
\put(4752,642){\makebox(0,0)[lb]{\smash{{{\SetFigFont{7}{8.4}{\rmdefault}{\mddefault}{\itdefault}p}}}}}
\put(4247,492){\makebox(0,0)[lb]{\smash{{{\SetFigFont{7}{8.4}{\rmdefault}{\mddefault}{\itdefault}n}}}}}
\path(356,398)(332,495)
\blacken\path(364.612,441.260)(332.000,495.000)(328.210,432.253)(342.088,454.229)(364.612,441.260)
\path(338,906)(362,1003)
\blacken\path(365.790,940.253)(362.000,1003.000)(329.388,949.260)(351.912,962.229)(365.790,940.253)
\path(312,639)(312,739)
\blacken\path(330.750,679.000)(312.000,739.000)(293.250,679.000)(312.000,697.000)(330.750,679.000)
\path(762,639)(762,739)
\blacken\path(780.750,679.000)(762.000,739.000)(743.250,679.000)(762.000,697.000)(780.750,679.000)
\path(462,752)(462,652)
\blacken\path(443.250,712.000)(462.000,652.000)(480.750,712.000)(462.000,694.000)(443.250,712.000)
\path(912,752)(912,652)
\blacken\path(893.250,712.000)(912.000,652.000)(930.750,712.000)(912.000,694.000)(893.250,712.000)
\path(1406,398)(1382,495)
\blacken\path(1414.612,441.260)(1382.000,495.000)(1378.210,432.253)(1392.088,454.229)(1414.612,441.260)
\path(1388,906)(1412,1003)
\blacken\path(1415.790,940.253)(1412.000,1003.000)(1379.388,949.260)(1401.912,962.229)(1415.790,940.253)
\path(1362,639)(1362,739)
\blacken\path(1380.750,679.000)(1362.000,739.000)(1343.250,679.000)(1362.000,697.000)(1380.750,679.000)
\path(1812,639)(1812,739)
\blacken\path(1830.750,679.000)(1812.000,739.000)(1793.250,679.000)(1812.000,697.000)(1830.750,679.000)
\path(1512,752)(1512,652)
\blacken\path(1493.250,712.000)(1512.000,652.000)(1530.750,712.000)(1512.000,694.000)(1493.250,712.000)
\path(1962,752)(1962,652)
\blacken\path(1943.250,712.000)(1962.000,652.000)(1980.750,712.000)(1962.000,694.000)(1943.250,712.000)
\path(2456,398)(2432,495)
\blacken\path(2464.612,441.260)(2432.000,495.000)(2428.210,432.253)(2442.088,454.229)(2464.612,441.260)
\path(2438,906)(2462,1003)
\blacken\path(2465.790,940.253)(2462.000,1003.000)(2429.388,949.260)(2451.912,962.229)(2465.790,940.253)
\path(2412,639)(2412,739)
\blacken\path(2430.750,679.000)(2412.000,739.000)(2393.250,679.000)(2412.000,697.000)(2430.750,679.000)
\path(2862,639)(2862,739)
\blacken\path(2880.750,679.000)(2862.000,739.000)(2843.250,679.000)(2862.000,697.000)(2880.750,679.000)
\path(2562,752)(2562,652)
\blacken\path(2543.250,712.000)(2562.000,652.000)(2580.750,712.000)(2562.000,694.000)(2543.250,712.000)
\path(3012,752)(3012,652)
\blacken\path(2993.250,712.000)(3012.000,652.000)(3030.750,712.000)(3012.000,694.000)(2993.250,712.000)
\path(3506,398)(3482,495)
\blacken\path(3514.612,441.260)(3482.000,495.000)(3478.210,432.253)(3492.088,454.229)(3514.612,441.260)
\path(3488,906)(3512,1003)
\blacken\path(3515.790,940.253)(3512.000,1003.000)(3479.388,949.260)(3501.912,962.229)(3515.790,940.253)
\path(3462,639)(3462,739)
\blacken\path(3480.750,679.000)(3462.000,739.000)(3443.250,679.000)(3462.000,697.000)(3480.750,679.000)
\path(3912,639)(3912,739)
\blacken\path(3930.750,679.000)(3912.000,739.000)(3893.250,679.000)(3912.000,697.000)(3930.750,679.000)
\path(3612,752)(3612,652)
\blacken\path(3593.250,712.000)(3612.000,652.000)(3630.750,712.000)(3612.000,694.000)(3593.250,712.000)
\path(4062,752)(4062,652)
\blacken\path(4043.250,712.000)(4062.000,652.000)(4080.750,712.000)(4062.000,694.000)(4043.250,712.000)
\path(4556,398)(4532,495)
\blacken\path(4564.612,441.260)(4532.000,495.000)(4528.210,432.253)(4542.088,454.229)(4564.612,441.260)
\path(4538,906)(4562,1003)
\blacken\path(4565.790,940.253)(4562.000,1003.000)(4529.388,949.260)(4551.912,962.229)(4565.790,940.253)
\path(4512,639)(4512,739)
\blacken\path(4530.750,679.000)(4512.000,739.000)(4493.250,679.000)(4512.000,697.000)(4530.750,679.000)
\path(4662,752)(4662,652)
\blacken\path(4643.250,712.000)(4662.000,652.000)(4680.750,712.000)(4662.000,694.000)(4643.250,712.000)
\path(5112,639)(5112,739)
\blacken\path(5130.750,679.000)(5112.000,739.000)(5093.250,679.000)(5112.000,697.000)(5130.750,679.000)
\path(4985,908)(4969,808)
\blacken\path(4959.965,870.209)(4969.000,808.000)(4996.994,864.284)(4975.636,849.473)(4959.965,870.209)
\path(4970,554)(4989,454)
\blacken\path(4959.380,509.446)(4989.000,454.000)(4996.221,516.445)(4981.160,495.262)(4959.380,509.446)
\dottedline{45}(2262,837)(2422,837)
\dottedline{45}(1212,537)(1372,537)
\dottedline{45}(3312,537)(3472,537)
\dottedline{45}(3312,837)(3472,837)
\dottedline{60}(387,1062)(837,1062)
\dottedline{60}(387,312)(837,312)
\dottedline{60}(1437,1062)(1887,1062)
\dottedline{60}(1437,312)(1887,312)
\dottedline{45}(1212,837)(1372,837)
\dottedline{60}(2487,1062)(2937,1062)
\dottedline{60}(2487,312)(2937,312)
\dottedline{45}(2262,537)(2422,537)
\dottedline{60}(3537,1062)(3987,1062)
\dottedline{60}(3537,312)(3987,312)
\dottedline{60}(4587,1062)(5037,1062)
\dottedline{60}(4587,312)(5037,312)
\dottedline{45}(162,537)(322,537)
\dottedline{45}(187,837)(322,837)
\dottedline{45}(4362,537)(4522,537)
\dottedline{45}(4387,837)(4522,837)
\dottedline{45}(4862,687)(4962,687)
\end{picture}
}
\caption{\label{fgv3} ${\mathcal E}_{\mathrm{co}}$
and $E_{\mathrm{co}}^{\scscs (\eta)}$ diagrams with $v$ components.
}
\end{figure}

For an atomic system we only have a single nucleus, and we get 
\begin{subequations}
\begin{eqnarray}  \label{Eco.part.atsh} 
\begin{picture}(0,0)(0,0)
\put(180,-12){$\left. \makebox(0,26){} \right\}$ (atomic systems)}
\end{picture}
{\cal E}_{\mathrm{co}}[\gamma,v]&=& 
{\cal E}_{\mathrm{co}}^{\scscs (0)}[\gamma]  + 
{\cal E}_{\mathrm{co}}^{\scscs(1)}[\gamma,v_m],\\
E_{\mathrm{co}}^{\scscs (\eta)}[\gamma,v] &=&
E_{\mathrm{co}}^{\scscs (\eta,0)}[\gamma] + 
E_{\mathrm{co}}^{\scscs (\eta,1)}[\gamma,v_m].
\hspace{15ex}
\end{eqnarray}
\end{subequations}
The atomic terms, ${\cal E}_{\mathrm{co}}^{\scscs(1)}[\gamma,v_m]$ and
$E_{\mathrm{co}}^{\scscs (\eta,1)}[\gamma,v_m]$, are -- for a range of systems --
universal functionals: They contribute to ${\cal E}_{\mathrm{co}}[\gamma,v]$ and
$E_{\mathrm{co}}^{\scscs (\eta)}[\gamma,v]$ for any system with a nuclei with a
charge of $Z_m$. Note that two atomic tems, say ${\cal
E}_{\mathrm{co}}^{\scscs(1)}[\gamma,v_m]$ and ${\cal
E}_{\mathrm{co}}^{\scscs(1)}[\gamma,v_{m^\prime}]$, are the same functionals when
they are from the same nuclei ($Z_m=Z_{m^\prime}$), but they are expressed from
different nuclear locations, since we have ($\mathbf{R}_m\ne\mathbf{R}_{m^\prime}$).

Factoring $Z_m$ from the diagrams, we get the following expansions:
\begin{subequations} \label{factor.z} 
\begin{eqnarray} 
{\cal E}_{\mathrm{co}}^{\scscs(1)}[\gamma,v_m]
&=& Z_m{\cal E}_{\mathrm{co}\mbox{\tiny $(1)$}}^{\scscs(1)}[\gamma] +
Z_m^2{\cal E}_{\mathrm{co}\mbox{\tiny $(2)$}}^{\scscs(1)}[\gamma] +
Z_m^3{\cal E}_{\mathrm{co}\mbox{\tiny $(3)$}}^{\scscs(1)}[\gamma] + \ldots,
\\
\label{factor.z.co} 
E_{\mathrm{co}}^{\scscs (\eta,1)}[\gamma,v_m]
&=&
Z_m E_{\mathrm{co}\mbox{\tiny $(1)$}}^{\scscs (\eta,1)}[\gamma] +
Z_m^2 E_{\mathrm{co}\mbox{\tiny $(2)$}}^{\scscs (\eta,1)}[\gamma] +
Z_m^3 E_{\mathrm{co}\mbox{\tiny $(3)$}}^{\scscs (\eta,1)}[\gamma] + \ldots,
\hspace{15ex}
\end{eqnarray}
\end{subequations}
where ${\cal E}_{\mathrm{co}\mbox{\tiny $(i)$}}^{\scscs(1)}[\gamma]$ and
$E_{\mathrm{co}\mbox{\tiny $(i)$}}^{\scscs (\eta,1)}[\gamma]$ are universal
functionals, independent of $Z_m$; they are given by the sum of all diagrams with
$i$ inserts of $\{v_m\}$, with ($Z_m=1$). For example, diagrams diagrams (b) and
(c) from Fig.~\ref{fgv3} contribute to ${\cal E}_{\mathrm{co}\mbox{\tiny
$(1)$}}^{\scscs(1)}[\gamma]$ and $E_{\mathrm{co}\mbox{\tiny $(1)$}}^{\scscs
(\eta,1)}[\gamma]$, for ($Z_m=1$), since they have a single insert; diagram (d)
contributes to ${\cal E}_{\mathrm{co}\mbox{\tiny $(2)$}}^{\scscs(1)}[\gamma]$ and
$E_{\mathrm{co}\mbox{\tiny $(2)$}}^{\scscs (\eta,1)}[\gamma]$; diagram (e)
contributes to ${\cal E}_{\mathrm{co}\mbox{\tiny $(3)$}}^{\scscs(1)}[\gamma]$ and
$E_{\mathrm{co}\mbox{\tiny $(3)$}}^{\scscs (\eta,1)}[\gamma]$, if ($n=p=m$).

Further generalizing Eqs.~(\ref{Eat.part}), we have the following {\em
external-potential expansions}:
\begin{subequations} 
\label{Egen.part} 
\begin{eqnarray}
\label{Egenco.partA} 
{\cal E}_{\mathrm{co}}[\gamma,v]&=& 
{\cal E}_{\mathrm{co}}^{\scscs (0)}[\gamma] +
\sum_m{\cal E}_{\mathrm{co}}^{\scscs (1)}[\gamma,v_m] +
\sum_{m>n}{\cal E}_{\mathrm{co}}^{\scscs (2)}[\gamma,v_m,v_n] + \ldots,  \\
\label{Egenco.part} 
E_{\mathrm{co}}^{\scscs (\eta)}[\gamma,v] &=&
E_{\mathrm{co}}^{\scscs (\eta,0)}[\gamma] +
\sum_m E_{\mathrm{co}}^{\scscs (\eta,1)}[\gamma,v_m] +
\sum_{m>n} E_{\mathrm{co}}^{\scscs (\eta,2)}[\gamma,v_m,v_n] + \ldots,
\end{eqnarray}
\end{subequations}  
where the {\em diatomic} terms, ${\cal
E}_{\mathrm{co}}^{\scscs(2)}[\gamma,v_m,v_n]$ and $E_{\mathrm{co}}^{\scscs
(\eta,2)}[\gamma,v_m,v_n]$, depend on $v_m$ and $v_n;$ they are the sum of all
diagrams that contain one or more $\{v_m\}$ and $\{v_n\}$ inserts, but have no
inserts from other external-potential components, say $\{v_p\}$, where ($p\ne m$)
and ($p\ne n$).  They are universal functionals for any system with one or more
nuclei with charges of $Z_m$ and $Z_n$, but, in addition, they also depend on the
molecular geometry -- the distance between the charges. Diagrams (a) from
Fig.~\ref{fgv3} is one example.  By factoring the $Z_m$ and $Z_n$ terms from the
diagrams, analogous expansions, as in Eqs.~(\ref{factor.z}), can also be
obtained. {\em Triatomic} and {\em molecular} terms are defined in a similar
manner.

\section{Approximations} \label{APPROX}

The diagrammatic methods presented in Sec.~\ref{BDMT}, and in the previous
section, gives explicit expansions for the functionals ${\cal
E}_{\mathrm{co}}[\gamma,v]$ and $E_{\mathrm{co}}^{\scscs(\eta)}[\gamma,v]$, and
for the terms on the right side of Eq.~(\ref{Egen.part}). However, in the
approximations we consider below, we often assume that these functionals can be
expressed in a more simplified form, for example, as integrals involving the
coordinates of only two electrons:
\begin{subequations}
\begin{eqnarray}
{\cal E}_{\mathrm{co}}[\gamma,v]
&=&
\frac12 \int\!\!\int 
{\cal G}_{\mathrm{co}}(\mathbf{x}_1,\mathbf{x}_2)\,d\mathbf{x}_1\,d\mathbf{x}_2, \\
E_{\mathrm{co}}^{\scscs(\eta)}[\gamma,v]&=&
\frac12 \int\!\!\int 
G_{\mathrm{co}}^{\scscs(\eta)}(\mathbf{x}_1,\mathbf{x}_2)\,d\mathbf{x}_1\,d\mathbf{x}_2,
\end{eqnarray}
\end{subequations}
where the integrands, ${\cal G}_{\mathrm{co}}(\mathbf{x}_1,\mathbf{x}_2)$ and
$G_{\mathrm{co}}^{\scscs(\eta)}(\mathbf{x}_1,\mathbf{x}_2)$, explicitly depends
upon $\gamma(\mathbf{x}_1,\mathbf{x}_2)$, $\gamma(\mathbf{x}_2,\mathbf{x}_1)$,
$v(\mathbf{r}_1)$, and $v(\mathbf{r}_2)$, and can include gradients or
higher-order derivatives, e.g.,
$\nabla_1^2\gamma(\mathbf{x}_1,\mathbf{x}_2)$.

If ${\cal E}_{\mathrm{co}}^{\scscs (0)}$ and $E_{\mathrm{co}}^{\scscs (\eta,0)}$
are known for a model, one-particle, density-matrix, say $\gamma_m$, then possible
approximations are the following:
\begin{eqnarray}
{\cal E}_{\mathrm{co}}^{\scscs (0)}[\gamma] 
&=&
{\cal E}_{\mathrm{co}}^{\scscs (0)} [\gamma_m]_{(\gamma_{\mbox{\tiny $m$}}=\gamma)},\\
\label{gamma.mod.app} 
E_{\mathrm{co}}^{\scscs (\eta,0)}[\gamma]
&=& 
E_{\mathrm{co}}^{\scscs (\eta,0)}[\gamma_m]_{(\gamma_{\mbox{\tiny $m$}}=\gamma)}.
\end{eqnarray}
If $\gamma_m$ is equal to, or similar to, the Brueckner-state one-particle
density-matrix for an electron gas $\tau_g$, then, from Eqs.~(\ref{E.part.taub})
and (\ref{gamma.mod.app}), we get the following approximation:
\begin{equation}\label{Egas.approx2a} 
E_{\mathrm{co}}^{\scscs (\eta,0)}[\gamma] \approx
{\cal E}_{\mathrm{co}}^{\scscs (\text{gas})} [\tau_g]_{(\tau_{\mbox{\tiny $g$}}=\gamma)}.
\end{equation}

If periodic boundary conditions are used, the Brueckner orbitals are known to be
plane waves \cite{Gellmann:57,March:67,Raimes:72,Mattuck:76,Parr:89}, so $\tau_g$
is known.  Because of conservation of momentum, there are many diagrams that are
absent in the ${\cal E}_{\mathrm{co}}^{\scscs (\text{gas})} [\tau_g]$ expansion
\cite{Gellmann:57,March:67,Raimes:72,Mattuck:76}, but, in general, these diagrams
appear in the expansion of ${\cal E}_{\mathrm{co}}^{\scscs (\text{gas})}
[\gamma]$.  Since, apparently, {\em all} of these excluded-diagrams are also
excluded in the $E_{\mathrm{co}}^{\scscs (\mathrm{III},0)}[\gamma]$ expansion,
the above approximation, Eq.~(\ref{Egas.approx2a}), is probably most appropriate
for ($\eta = \mbox{\small III}$):
\begin{equation}\label{Egas.approx2} 
E_{\mathrm{co}}^{\scscs (\mathrm{III},0)}[\gamma] \approx
{\cal E}_{\mathrm{co}}^{\scscs (\text{gas})} [\tau_g]_{(\tau_{\mbox{\tiny $g$}}=\gamma)}.
\end{equation}

For systems with a non-constant external potential, $E_{\mathrm{co}}^{\scscs
(\eta,0)}[\gamma]$ is, in many instances, the dominant portion of
$E_{\mathrm{co}}^{\scscs (\eta)}[\gamma,v]$. In that case, we can neglect all
terms except the electron gas one from Eq.~(\ref{Egenco.part}), and using
Eq~(\ref{Egas.approx2}), we get the following {\em electron-gas approximation}:
\begin{equation}\label{Egas.approxB} 
E_{\mathrm{co}}^{\scscs (\mathrm{III})}[\gamma] \approx
{\cal E}_{\mathrm{co}}^{\scscs (\text{gas})} [\tau_g]_{(\tau_{\mbox{\tiny $g$}}=\gamma)},
\end{equation}
where, in this approximation, the dependence on $v$ is neglected.

Eq.~(\ref{Egas.approxB}) shares many similarities with the LDA of
density-functional theory \cite{Kohn:65,Parr:89,Dreizler:90}, where this approach
constructs approximate energy-functionals from expressions derived from a {\em
uniform} electron gas.  If periodic boundary conditions are used, the Brueckner
orbitals (and the Hartree--Fock ones) are known to be plane waves
\cite{March:67,Raimes:72}, so $\tau_g$ is known.  If, in addition to requiring the
volume to be infinitely large, the number of particles becomes infinite,
a {\em uniform electron gas} is obtained. In this limiting case, the density of
the Brueckner reference-state $|\Theta\rangle$, say $\rho_{\text{ug}}$, is
identical to the density of the target state $|\Psi\rangle$, both being a
constant; the correlation energy of a uniform electron gas, say ${\cal
E}_{\mathrm{co}}^{\scscs (\text{gas})} (\rho_{\text{ug}})$, is a function of this
density, not a functional \cite{Gellmann:57,March:67,Raimes:72,Mattuck:76}.  In
the LDA a functional is constructed using the function ${\cal
E}_{\mathrm{co}}^{\scscs (\text{gas})} (\rho_{\text{ug}})$ divided by the number
of electrons -- the correlation energy per particle. An analogous approach may be
necessary when constructing the functional ${\cal E}_{\mathrm{co}}^{\scscs
(\text{gas})} [\tau_g]_{(\tau_{\mbox{\tiny $g$}}=\gamma)}$, although the
one-particle density matrix for an electron gas is {\em not} a constant
\cite{March:67,Raimes:72,Parr:89}. Furthermore, when evaluating the diagrams for
${\cal E}_{\mathrm{co}}^{\scscs (\text{gas})} (\rho_{\text{ug}})$, as in the
random phase approximation (RPA) \cite{Gellmann:57,March:67,Raimes:72,Mattuck:76},
the summations over the occupied, plane-wave states are replaced by integrals. For
an exact treatment of ${\cal E}_{\mathrm{co}}^{\scscs (\text{gas})} [\tau_g]$,
this approach cannot be used, and, mathematically speaking, this is the
difference between ${\cal E}_{\mathrm{co}}^{\scscs (\text{gas})} [\tau_g]$ and
${\cal E}_{\mathrm{co}}^{\scscs (\text{gas})} (\rho_{\text{ug}})$.

In using the above electron gas approximation, Eq.~(\ref{Egas.approxB}), or other
similar approximations, especially for atoms or molecules, one must take into
consideration the boundary conditions imposed on the wavefunction. For example, if
the wavefunction for an electron gas is required to vanish at the end points of a
cube -- and the cube has a finite size -- this may lead to an electron gas
functional ${\cal E}_{\mathrm{co}}^{\scscs (\text{gas})}
[\tau_g]_{(\tau_{\mbox{\tiny $g$}}=\gamma)}$ that explicitly depends on the
boundary conditions. However, probably in many cases, this explicit dependents can
be transformed, or contained within $\tau_g$, since $\tau_g$ also depends on the
boundary conditions. 

We have, so far, used the external-potential expansions, Eqs.~(\ref{Egen.part}),
to obtain approximate functionals.  An alternative approach is obtained, if
$E_{\mathrm{co}}^{\scscs (\eta)}[\gamma,v]$ is known for some system, say the
helium atom, in the vicinity of some one-particle density-matrix, say the
Brueckner one. In that case, the following prescription yields an approximate
correlation-energy functional:
\begin{equation} \label{He.approx} 
E_{\mathrm{co}}^{\scscs (\eta)}[\gamma,v] \approx
E_{\mathrm{co}}^{\scscs (\eta)}
[\tau_{\mbox{\tiny\textsc{h}e}},v_{\mbox{\tiny\textsc{h}e}}
]_{(\tau_{\mbox{\tiny\textsc{h}e}} = \gamma,\,v_{\mbox{\tiny\textsc{h}e}} = v)},
\end{equation}
where $\tau_{\mbox{\tiny\textsc{h}e}}$ is the Brueckner, one-particle,
density-matrix for the helium atom, and $v_{\mbox{\tiny\textsc{h}e}}$ is the
external potential for this system, where using Eq.~(\ref{Eco.ident}), we have
\begin{equation} \label{He.approxB} 
E_{\mathrm{co}}^{\scscs (\eta)}[\gamma,v] \approx
{\cal E}_{\mathrm{co}}
[\tau_{\mbox{\tiny\textsc{h}e}},v_{\mbox{\tiny\textsc{h}e}}]
_{(\tau_{\mbox{\tiny\textsc{h}e}} = \gamma,\,v_{\mbox{\tiny\textsc{h}e}} = v)}.
\end{equation}
In the limit of $\rho \longrightarrow \tau_x$, necessarily, many terms from ${\cal
E}_{\mathrm{co}} [\tau_x,v_x]$ must vanish, where $v_x$ is the external potential
associated with the Brueckner one-particle density matrix, $\tau_x$. Since,
apparently, {\em all} of these vanishing-diagrams are also excluded in the
$E_{\mathrm{co}}^{\scscs (\mathrm{III},0)}[\gamma,v_x]$ expansion, most probably, the
above approximation, Eq.~(\ref{He.approxB}), is most appropriate for ($\eta =
\mbox{\small III}$). Furthermore, since the Brueckner density matrix
$\tau_{\mbox{\tiny\textsc{h}e}}$ is approximately equal to the Hartree--Fock one,
say $\tilde{\tau}_{\mbox{\tiny\textsc{h}e}}$, we can write
\begin{equation} \label{He.approxC} 
E_{\mathrm{co}}^{\scscs (\mathrm{III})}[\gamma,v] \approx
{\cal E}_{\mathrm{co}}
[\tilde{\tau}_{\mbox{\tiny\textsc{h}e}},v_{\mbox{\tiny\textsc{h}e}}]
_{(\tilde{\tau}_{\mbox{\tiny\textsc{h}e}} = \gamma,\,v_{\mbox{\tiny\textsc{h}e}} = v)}.
\end{equation}
Using the expansion for ${\cal E}_{\mathrm{co}}$, given by
Eq.~(\ref{Egenco.partA}), and neglecting all terms except the electron gas one, we
get
\begin{equation} \label{He.approxD} 
E_{\mathrm{co}}^{\scscs (\mathrm{III})}[\gamma,v] \approx
{\cal E}_{\mathrm{co}}^{\scscs (0)}[\tilde{\tau}_{\mbox{\tiny\textsc{h}e}}]
_{(\tilde{\tau}_{\mbox{\tiny\textsc{h}e}} = \gamma)}.
\end{equation}
The above approximation assumes that the terms arising from the helium potential
$v_{\mbox{\tiny\textsc{h}e}}$ are small; therefore, including them in the
following way should yield only a small error:
\begin{equation} \label{He.approxE} 
E_{\mathrm{co}}^{\scscs (\mathrm{III})}[\gamma] \approx
{\cal E}_{\mathrm{co}}
[\tilde{\tau}_{\mbox{\tiny\textsc{h}e}},v_{\mbox{\tiny\textsc{h}e}}]
_{(\tilde{\tau}_{\mbox{\tiny\textsc{h}e}} = \gamma)}.
\end{equation}

A well known approximation for ${\cal E}_{\mathrm{co}}
[\tilde{\tau}_{\mbox{\tiny\textsc{h}e}},v_{\mbox{\tiny\textsc{h}e}}]$ is given by
Colle and Salvetti functional \cite{Colle:75,Lee:88}, say ${\cal
E}_{\mathrm{co}}^{\mathrm{cs}} [\tilde{\tau}_{\mbox{\tiny\textsc{h}e}}]$;
so we have
\begin{equation} \label{CS} 
E_{\mathrm{co}}^{\scscs (\mathrm{III})}[\gamma] \approx
{\cal E}_{\mathrm{co}}^{\mathrm{cs}} [\tilde{\tau}_{\mbox{\tiny\textsc{h}e}}]
_{(\tilde{\tau}_{\mbox{\tiny\textsc{h}e}} = \gamma)},
\end{equation}
where we have suppressed any mention of $v_{\mbox{\tiny\textsc{h}e}}$,
since this functional has no explicit dependence on the external
potential. However, this functional can still possess an implicit dependence on
$v_{\mbox{\tiny\textsc{h}e}}$, since, for example, its four empirical parameters
are determined by using data from the helium atom. An improved functional,
perhaps, can be obtained by appending a potential dependence on these parameters
and using the approximation given by Eq.~(\ref{He.approxC}).

We also mention that the correlation potentials $v_{\mathrm{co}}^{\scscs \tau}$
can be treated in a similar manner as the correlation-energy functionals
$E_{\mathrm{co}}^{\scscs (\eta)}[\gamma,v]$, since they also depend on the
external potential: $v_{\mathrm{co}}^{\scscs \tau}[v]$; Equations, or
approximations, analogous to Eqs.~(\ref{Egenco.part}), (\ref{factor.z.co}),
(\ref{Egas.approxB}), and (\ref{He.approx}), are easily defined. However, we now
pursue a different approach, permitting correlation potentials to be obtained as
functional derivatives of variation correlation-energy functionals.

\section{Variational formalism} \label{VARI} 

We now introduce variational energy-functionals $\bar{E}_\eta[\gamma]$. By
functional differentiating these functionals with respect to the one-particle
density-matrix $\gamma$, generalized Fock operators are obtained.  These operators
-- denoted by $\mathcal{\hat{\zeta}}_{\tau}^{\scscs (\eta)}$ -- satisfy the same
Brillouin-Brueckner and commutation relations, Eqs.\ (\ref{FBB-cond.dm}) and
(\ref{commute.dm}), as the corresponding non-variational ones, ${\cal
\hat{F}}_{\tau}^{\scscs (\eta)}$.

Using our trial wavefunctions from Sec.\ (\ref{TRIAL}), we can construct
variational energy-functionals:
\begin{equation} \label{Efuncts.var} 
\bar{E}_\eta[\gamma] = 
\frac{\langle\Psi_\gamma^{\scscs (\eta)}|
H|\Psi_\gamma^{\scscs (\eta)}\rangle}
{\langle\Psi_\gamma^{\scscs (\eta)}|\Psi_\gamma^{\scscs (\eta)}\rangle}
= E_1[\gamma] + \bar{E}_{\mathrm{co}}^{\scscs (\eta)}[\gamma],
\end{equation}
where the first-order energy $E_1[\gamma]$ is given by Eq.\ (\ref{first.eb}), our
variational correlation-energy functionals are
\begin{equation} 
\bar{E}_{\mathrm{co}}^{\scscs (\eta)}[\gamma]
=
\frac{\langle\Psi_\gamma^{\scscs (\eta)}|
(\{\hat{F}_\gamma\} + \{r_{12}^{-1}\}_\gamma)
|\Psi_\gamma^{\scscs (\eta)}\rangle}
{\langle\Psi_\gamma^{\scscs (\eta)}|\Psi_\gamma^{\scscs (\eta)}\rangle},
\end{equation}
and we have used Eqs.\ (\ref{H.norm}), (\ref{Hc}) and (\ref{H.F}).  We also define
exchange-correlation (xc) energy-functionals:
\begin{equation} \label{Excorr.v} 
\bar{E}_{\mathrm{xc}}^{\scscs (\eta)}[\gamma] = \bar{E}_{\mathrm{co}}^{\scscs (\eta)}[\gamma] - 
E_{\mathrm{x}}[\gamma],
\end{equation}
where the exchange energy $E_{\mathrm{x}}[\gamma]$ is given by Eq.\ (\ref{exch.dm}).

The exact energy is also given by
\begin{equation} \label{Eex.v} 
{\cal E} = \frac{\langle\Psi|H|\Psi\rangle}{\langle\Psi|\Psi\rangle}
= E_1[\gamma] + 
\frac{\langle\Psi|
(\{\hat{F}_\gamma\} + \{r_{12}^{-1}\}_\gamma)
|\Psi\rangle}{\langle\Psi|\Psi\rangle}.
\end{equation}
Comparing this Eq.\ to Eq.\ (\ref{E0.a}) gives
\begin{equation} \label{Eco.v} 
{\cal E}_{\mathrm{co}}[\gamma]=
\frac{\langle\Psi|
(\{\hat{F}_\gamma\} + \{r_{12}^{-1}\}_\gamma)
|\Psi\rangle}{\langle\Psi|\Psi\rangle},
\end{equation}

From the variational theorem, the fourth-trial wavefunctions are equal:
\begin{equation} 
\bar{E}_{\scscs \mathrm{IV}}[\gamma]=E_{\scscs \mathrm{IV}}[\gamma],
\end{equation}
and we have
\begin{equation} 
\bar{E}_{\mathrm{co}}^{\scscs (\mathrm{IV})}[\gamma]=E_{\mathrm{co}}^{\scscs
(\mathrm{IV})}[\gamma].
\end{equation}

From the variational theorem, we also have 
\begin{equation} \label{EvarX} 
\bar{E}_\eta[\gamma] \ge {\cal E},
\end{equation}
where Eqs.\ (\ref{trwf=exact}), (\ref{Eex.v}), (\ref{Eco.v}), (\ref{Efuncts.var}),
(\ref{Excorr.v}), and (\ref{EXC}) give the following equalities that appear for
the Brueckner one-particle density-matrix $\tau$:
\begin{eqnarray} \label{fy.exact.v} 
{\cal E}&=&\bar{E}_\eta[\tau]\;\,=E_\eta[\tau],\\ 
\label{Eco.ident.v} 
{\cal E}_{\mathrm{co}}[\tau]&=&\bar{E}_{\mathrm{co}}^{\scscs (\eta)}[\tau]
=E_{\mathrm{co}}^{\scscs (\eta)}[\tau],\\
\label{Exc.ident.v} 
{\cal E}_{\mathrm{xc}}[\tau]&=&\bar{E}_{\mathrm{xc}}^{\scscs (\eta)}[\tau]
=E_{\mathrm{xc}}^{\scscs (\eta)}[\tau],
\end{eqnarray}
where the latter relations use Eqs.\ (\ref{fy.exact}), (\ref{Eco.ident}), and
(\ref{Exc.ident}); Eqs.~(\ref{EvarX}) and (\ref{fy.exact.v}) indicate that the
minimization of $\bar{E}_\eta[\gamma]$ occurs at $\tau$.

We now pursue, in a formal way, the minimization of $\bar{E}_\eta[\gamma]$, using
an approach that is similar to the procedure used by Parr and Yang \cite{Parr:89}
in their treatment of Hartree--Fock theory, where $\bar{E}_\eta[\gamma]$ is
subject to the constraint that the one-particle density-matrix comes from a
single-determinantal state $\mbox{\large $\gamma$}\mbox{\footnotesize
$(|\Phi\rangle)$}$. This condition is imposed by requiring the one-particle
density-matrix $\gamma$ to have a trace equal to the number of electrons
$N_\gamma$, and that it is also idempotent \cite{Lowdin:55b,Blaizot:86}:
\begin{eqnarray} \label{trace}  
\int \!\! \int \gamma(\mathbf{x}_3,\mathbf{x}_4) 
\delta(\mathbf{x}_4-\mathbf{x}_3) \,d\mathbf{x}_3 \,d\mathbf{x}_4 = N_\gamma, \\
\label{indep} 
\int 
\gamma(\mathbf{x}_3,\mathbf{x}_5) 
\gamma(\mathbf{x}_5,\mathbf{x}_4) 
\,d\mathbf{x}_5=
\gamma(\mathbf{x}_3,\mathbf{x}_4).
\end{eqnarray}
The normalization constraint, Eq.\ (\ref{trace}), insures that $\gamma$ can be
constructed from $N_\gamma$ orbitals; Eq.\ (\ref{indep}) insures that the
density-matrix operator $\hat{\gamma}$ -- when acting within the one-particle
Hilbert space -- is a projector into the occupied subspace, as indicated by
Eq.~(\ref{dm}), where $\gamma$ serves as the kernel of the one-particle
density-matrix operator $\hat{\gamma}$, Eq.~(\ref{denmat.op}).

Using the above constraints, the variational problem is expressed by
\begin{equation} \label{deltaL} 
\left.\delta \mathcal{L}(\gamma)\right|_{\tau}=0,
\end{equation}
where
\begin{widetext}
\begin{eqnarray} \label{L} 
\mathcal{L}(\gamma) =
\bar{E}_\eta[\gamma] 
-  \beta \left[ 
\int \!\! \int \gamma(\mathbf{x}_3,\mathbf{x}_4) 
\delta(\mathbf{x}_4-\mathbf{x}_3) \,d\mathbf{x}_3 \,d\mathbf{x}_4 - N_\gamma
\right] \hspace{29ex} 
\\  \nonumber  \hspace{24ex}
\mbox{}-
\int \!\! \int d\mathbf{x}_3 \, d\mathbf{x}_4 \, \alpha(\mathbf{x}_4,\mathbf{x}_3)
\left[
\int 
\gamma(\mathbf{x}_3,\mathbf{x}_5) 
\gamma(\mathbf{x}_5,\mathbf{x}_4) 
\,d\mathbf{x}_5-
\gamma(\mathbf{x}_3,\mathbf{x}_4)
\right],
\end{eqnarray}
\end{widetext}
\noindent
and where $\alpha$ and $\beta$ are the Lagrangian multipliers. Eq.\
(\ref{deltaL}) is satisfied when the functional derivative of $\mathcal{L}$ vanishes:
\begin{equation} \label{delL} 
\left.\frac{\delta \mathcal{L}(\gamma)}{\delta \gamma(\mathbf{x}_2,\mathbf{x}_1)} 
\right|_{\tau}
= 0,
\end{equation}
where the definition of the functional derivative is 
\begin{equation} 
\delta \mathcal{L}(\gamma)=\int \!\! \int 
\frac{\delta \mathcal{L}(\gamma)}{\delta \gamma(\mathbf{x}_2,\mathbf{x}_1)}
\delta \gamma(\mathbf{x}_2,\mathbf{x}_1) 
d\mathbf{x}_1 \, d\mathbf{x}_2.
\end{equation}
\begin{widetext}
Substituting Eq.\ (\ref{L}) into (\ref{delL}), yields
{
\begin{equation} \label{delL2} 
{\cal \zeta}_{\tau}^{\scscs (\eta)}(\mathbf{x}_1,\mathbf{x}_2) - 
\int d\mathbf{x}_3 \, \left[
\tau(\mathbf{x}_1,\mathbf{x}_3)
\alpha(\mathbf{x}_3,\mathbf{x}_2)
+
\alpha(\mathbf{x}_1,\mathbf{x}_3)
\tau(\mathbf{x}_3,\mathbf{x}_2)\right]
+ \alpha(\mathbf{x}_1,\mathbf{x}_2) 
- \beta \delta(\mathbf{x}_1-\mathbf{x}_2)
= 0,
\end{equation}}
\end{widetext}
where 
\begin{equation} \label{extF.v} 
{\cal \zeta}_{\tau}^{\scscs (\eta)}(\mathbf{x}_1,\mathbf{x}_2) =
\left. \frac{\delta \bar{E}_\eta[\gamma]}
{\delta \gamma(\mathbf{x}_2,\mathbf{x}_1)}\right|_\tau .
\end{equation}

Let the two-body functions, $\alpha(\mathbf{x}_1,\mathbf{x}_2)$ and ${\cal
\zeta}_{\tau}^{\scscs (\eta)}(\mathbf{x}_1,\mathbf{x}_2)$, serve as kernels of
operators, $\hat{\alpha}$ and $\mathcal{\hat{\zeta}}_{\tau}^{\scscs
(\eta)}$; explicitly, we have
\begin{eqnarray} \label{alph.op} 
\hat{\alpha} \phi(\mathbf{x}_1) 
&=& 
\int 
\alpha(\mathbf{x}_1,\mathbf{x}_2)
\phi(\mathbf{x}_2)
\,d\mathbf{x}_2, \\
\label{Fex.op} 
\mathcal{\hat{\zeta}}_{\tau}^{\scscs (\eta)}
\phi(\mathbf{x}_1) 
&=& 
\int 
{\cal \zeta}_{\tau}^{\scscs (\eta)}(\mathbf{x}_1,\mathbf{x}_2)
\phi(\mathbf{x}_2)
\,d\mathbf{x}_2.
\end{eqnarray}
Using this notation, the operator form of Eq.\ (\ref{delL2}) is given by
\begin{equation}\label{delL3} 
\mathcal{\hat{\zeta}}_{\tau}^{\scscs (\eta)} - \hat{\tau}\hat{\alpha}
- \hat{\alpha}\hat{\tau} + \hat{\alpha} - \beta = 0,
\end{equation}
where $\hat{\tau}$ is defined by Eq.\ (\ref{denmat.op.bru}).

The identity operator $\hat{I}$, defined by Eq.\ (\ref{ident.op}), or expressed by
\begin{equation}\label{ident.op2} 
\hat{\mbox{\small $I$}} = \hat{\tau} + \hat{\kappa}_\tau,
\end{equation}
can be used to obtain the following relation:
\begin{equation}
\hat{\alpha} - \hat{\tau}\hat{\alpha}
- \hat{\alpha}\hat{\tau} =
\hat{\mbox{\small $I$}}\left(\hat{\alpha} - \hat{\tau}\hat{\alpha}
- \hat{\alpha}\hat{\tau}\right)\hat{\mbox{\small $I$}} = 
\hat{\kappa}_\tau\hat{\alpha}\hat{\kappa}_\tau -
\hat{\tau}\hat{\alpha}\hat{\tau},
\end{equation}
which we substitute into Eq.\ (\ref{delL3}); this procedure gives
\begin{equation}\label{delL4} 
\mathcal{\hat{\zeta}}_{\tau}^{\scscs (\eta)} - 
\hat{\tau}\hat{\alpha}\hat{\tau} + 
\hat{\kappa}_\tau\hat{\alpha}\hat{\kappa}_\tau - \beta = 0,
\end{equation}
and yields the following requirements:
\begin{subequations}
\label{FBB-conds.dm.v} 
\begin{eqnarray}\label{FBB-cond.dm.v} 
\mbox{\large $\hat{\kappa}$}_\tau
\mathcal{\hat{\zeta}}_{\tau}
\mbox{\large $\hat{\tau}$}
&=&0, \\
\label{Herm.F.dm.v} 
\mbox{\large $\hat{\tau}$}
\mathcal{\hat{\zeta}}_{\tau}
\mbox{\large $\hat{\kappa}$}_\tau
&=&0,
\end{eqnarray}
\end{subequations}
where we have dropped the $\eta$ superscript since, for ($\gamma=\tau$), all
operators are equal within these blocks:
\begin{subequations}
\label{ident.z} 
\begin{eqnarray} 
\mbox{\large $\hat{\kappa}$}_\tau
\mathcal{\hat{\zeta}}_{\tau}^{\scscs (\eta)}
\mbox{\large $\hat{\tau}$}
=
\mbox{\large $\hat{\kappa}$}_\tau
\mathcal{\hat{\zeta}}_{\tau}^{\scscs (\eta^\prime)}
\mbox{\large $\hat{\tau}$}
=
\mbox{\large $\hat{\kappa}$}_\tau
\mathcal{\hat{\zeta}}_{\tau}
\mbox{\large $\hat{\tau}$}, \\
\mbox{\large $\hat{\tau}$}
\mathcal{\hat{\zeta}}_{\tau}^{\scscs (\eta)}
\mbox{\large $\hat{\kappa}$}_\tau
=
\mbox{\large $\hat{\tau}$}
\mathcal{\hat{\zeta}}_{\tau}^{\scscs (\eta^\prime)}
\mbox{\large $\hat{\kappa}$}_\tau
=
\mbox{\large $\hat{\tau}$}
\mathcal{\hat{\zeta}}_{\tau}
\mbox{\large $\hat{\kappa}$}_\tau.
\end{eqnarray}
\end{subequations}
Eq.\ (\ref{FBB-cond.dm.v}) is yet another representation of the
Brillouin--Brueckner condition; comparing Eqs.\ (\ref{FBB-conds.dm}) and
(\ref{FBB-conds.dm.v}) give
\begin{subequations}
\label{exFock.idents} 
\begin{eqnarray} 
\mbox{\large $\hat{\kappa}$}_\tau
\mathcal{\hat{\zeta}}_{\tau}
\mbox{\large $\hat{\tau}$}&=&
\mbox{\large $\hat{\kappa}$}_\tau
\mathcal{\hat{F}}_{\tau}
\mbox{\large $\hat{\tau}$}, \\
\mbox{\large $\hat{\tau}$}
\mathcal{\hat{\zeta}}_{\tau}
\mbox{\large $\hat{\kappa}$}_\tau.
&=&
\mbox{\large $\hat{\tau}$}
\mathcal{\hat{F}}_{\tau}
\mbox{\large $\hat{\kappa}$}_\tau,
\end{eqnarray}
\end{subequations}
and it is easily verified that the commutation relation, Eq.\ (\ref{commute.dm}),
also holds for the variational one-body operators
$\mathcal{\hat{\zeta}}_{\tau}$:
\begin{equation} \label{commute.dm.v} 
\left[\mathcal{\hat{\zeta}}_{\tau},\hat{\mbox{\large $\tau$}}\right] = 0.
\end{equation}
An alternative to the exact Hartree--Fock Eq. (\ref{eF.eqs.dm}) is
\begin{equation} \label{eF.eqs.v} 
\mathcal{\hat{\zeta}}_{\tau}^{\scscs (\eta)}
|\psi_w\rangle =
\sum_{x\in \{\psi_o\rightarrow\tau\}} 
\xi_{xw}^{\mbox{\tiny $\tau\eta$}}
|\psi_x\rangle,
\end{equation}
where the appended $\eta$ superscripts appear since the occupied-block
matrix-elements $\xi_{xw}^{\mbox{\tiny $\tau\eta$}}$, perhaps, depend on
$\eta$. However, we can redefine the variational operators
$\mathcal{\hat{\zeta}}_{\tau}^{\scscs (\eta)}$ to remove this dependence, since
Eqs.\ (\ref{FBB-conds.dm.v}) still holds. In any event, we assume that 
$\mathcal{\hat{\zeta}}_{\tau}^{\scscs (\eta)}$ is independent of $\eta$ and
choose orbitals that diagonalize $\xi_{xw}^{\mbox{\tiny $\tau$}}$, giving a
generalized Hartree-Fock Eq.\ that is an alternative to Eq.~(\ref{F.eigen}):
\begin{equation} \label{eF.can} 
\mathcal{\hat{\zeta}}_{\tau}
\bar{\psi}_i^\tau(\mathbf{x}) =
\xi_{i}^{\mbox{\tiny $\tau$}}
\bar{\psi}_i^\tau(\mathbf{x}).
\end{equation}

Substituting Eq.\ (\ref{Efuncts.var}) into (\ref{extF.v}) for ($\tau=\gamma$) gives
\begin{equation} \label{extF.v2} 
{\cal \zeta}_{\gamma}^{\scscs (\eta)}(\mathbf{x}_1,\mathbf{x}_2) =
F_\gamma(\mathbf{x}_1,\mathbf{x}_2) +
\nu_{\mathrm{co}}^{\scscs \gamma\eta}(\mathbf{x}_1,\mathbf{x}_2),
\end{equation}
where $F_\gamma(\mathbf{x}_1,\mathbf{x}_2)$ and $\nu_{\mathrm{co}}^{\scscs
\gamma\eta}(\mathbf{x}_1,\mathbf{x}_2)$ are the kernel of the Fock operator
$\hat{F}_\gamma$ -- defined by Eq.~(\ref{kernfock}) -- and variational
correlation-potentials $\hat{\nu}_{\mathrm{co}}^{\scscs \gamma\eta}$:
\begin{eqnarray} \label{kernfock.def} 
F_\gamma(\mathbf{x}_1,\mathbf{x}_2)
&=&
\frac{\delta E_1[\gamma]} 
{\delta \gamma(\mathbf{x}_2,\mathbf{x}_1)}, \\
\label{kerncorpot.def} 
\nu_{\mathrm{co}}^{\scscs \gamma\eta}(\mathbf{x}_1,\mathbf{x}_2)
&=&
\frac{\delta \bar{E}_{\mathrm{co}}^{\scscs (\eta)}[\gamma]} 
{\delta \gamma(\mathbf{x}_2,\mathbf{x}_1)}.
\end{eqnarray}

The operator form of Eq.\ (\ref{extF.v2}) is
\begin{equation} \label{extF.vop} 
{\cal \hat{\zeta}}_{\gamma}^{\scscs (\eta)} = 
\hat{F}_\gamma + \hat{\nu}_{\mathrm{co}}^{\scscs \gamma\eta}.
\end{equation}
Substituting this Eq.\ into Eq.\ (\ref{ident.z}) indicates that we have
\begin{subequations}
\label{copot.identBB} 
\begin{eqnarray}
\mbox{\large $\hat{\kappa}$}_\tau
\hat{\nu}_{\mathrm{co}}^{\scscs \tau\eta}
\mbox{\large $\hat{\tau}$}
=
\mbox{\large $\hat{\kappa}$}_\tau
\hat{\nu}_{\mathrm{co}}^{\scscs \tau\eta^\prime}
\mbox{\large $\hat{\tau}$}
=
\mbox{\large $\hat{\kappa}$}_\tau
\hat{\nu}_{\mathrm{co}}^{\scscs \tau}
\mbox{\large $\hat{\tau}$}
, \\
\mbox{\large $\hat{\tau}$}
\hat{\nu}_{\mathrm{co}}^{\scscs \tau\eta}
\mbox{\large $\hat{\kappa}$}_\tau
=
\mbox{\large $\hat{\tau}$}
\hat{\nu}_{\mathrm{co}}^{\scscs \tau\eta^\prime}
\mbox{\large $\hat{\kappa}$}_\tau
=
\mbox{\large $\hat{\tau}$}
\hat{\nu}_{\mathrm{co}}^{\scscs \tau}
\mbox{\large $\hat{\kappa}$}_\tau.
\end{eqnarray}
\end{subequations}
Substituting Eqs.\ (\ref{extF.vop}) and (\ref{exactF.dm}) into (\ref{exFock.idents})
and using the two above definitions, yields
\begin{subequations} \label{copot.ident} 
\begin{eqnarray} 
\mbox{\large $\hat{\kappa}$}_\tau
\hat{\nu}_{\mathrm{co}}^{\scscs \tau}
\mbox{\large $\hat{\tau}$}&=&
\mbox{\large $\hat{\kappa}$}_\tau
v_{\mathrm{co}}^{\scscs \tau}
\mbox{\large $\hat{\tau}$}, \\
\mbox{\large $\hat{\tau}$}
\hat{\nu}_{\mathrm{co}}^{\scscs \tau}
\mbox{\large $\hat{\kappa}$}_\tau.
&=&
\mbox{\large $\hat{\tau}$}
v_{\mathrm{co}}^{\scscs \tau}
\mbox{\large $\hat{\kappa}$}_\tau.
\end{eqnarray}
\end{subequations}

In order acquire to the kernels of the generalized Fock operators ${\cal
\zeta}_{\tau}^{\scscs (\eta)}(\mathbf{x}_1,\mathbf{x}_2)$, given by
Eq.~(\ref{extF.v2}), it is necessary obtain the functional derivatives of $E_1[\gamma]$
and $\bar{E}_{\mathrm{co}}^{\scscs (\eta)}[\gamma]$, as indicated by
Eq.~(\ref{kernfock.def}) and (\ref{kerncorpot.def}).  The functional derivative for
$E_1[\gamma]$ can be evaluated using Eqs.~(\ref{first.eb}), (\ref{coul.dm}), and
(\ref{exch.dm}), yielding Eq.~(\ref{kernfock}). The functional derivative of the
diagrammatic terms of $\bar{E}_{\mathrm{co}}^{\scscs (\eta)}[\gamma]$ can also be
obtained; the details are presented elsewhere \cite{tobe}.  Here we only mention
that by imposing the same occupied and virtual orbital degeneracies as in Sec.\
\ref{EXPL}, each diagram in the expansion is given by a product of one-particle
density-matrices, and can, therefore, be differentiated in the same manner as in
the treatment of $E_1[\gamma]$. After the functional derivative is taken, the
nondegeneracy of the orbitals can be restored, since the entire expansion is
invariant to the choice of orbital energies, but this removes the explicit
dependence on $\gamma$ for each term.

It is easily demonstrated that an analogous external-potential expansion, as given
in Eq.~(\ref{Egenco.part}), also holds:
\begin{equation}
\label{Egenco.part.v} 
\bar{E}_{\mathrm{co}}^{\scscs (\eta)}[\gamma,v] =
\bar{E}_{\mathrm{co}}^{\scscs (\eta,0)}[\gamma] +
\sum_m \bar{E}_{\mathrm{co}}^{\scscs (\eta,1)}[\gamma,v_m] +
\sum_{m>n} \bar{E}_{\mathrm{co}}^{\scscs (\eta,2)}[\gamma,v_m,v_n] + \ldots.
\end{equation}
Furthermore -- by using, Eqs.~(\ref{fy.exact.v}), (\ref{Eco.ident.v}) and
(\ref{Exc.ident.v}) -- all approximations presented in Sec.~\ref{APPROX} are valid
when $E_{\mathrm{co}}^{\scscs (\eta)}$ is replaced by
$\bar{E}_{\mathrm{co}}^{\scscs (\eta)}$. For example, consider the electron-gas
approximation, Eq.~(\ref{Egas.approxB}), and the Colle--Salvetti functional,
Eq~(\ref{CS}):
\begin{eqnarray}\label{Egas.approxB.v} 
\bar{E}_{\mathrm{co}}^{\scscs (\mathrm{III})}[\gamma]&\approx& {\cal
E}_{\mathrm{co}}^{\scscs (\text{gas})} [\tau_g]_{(\tau_{\mbox{\tiny$g$}}=\gamma)},\\
\label{CS.v} 
\bar{E}_{\mathrm{co}}^{\scscs (\mathrm{III})}[\gamma]&\approx&
{\cal E}_{\mathrm{co}}^{\mathrm{cs}} [\tilde{\tau}_{\mbox{\tiny\textsc{h}e}}]
_{(\tilde{\tau}_{\mbox{\tiny\textsc{h}e}} = \gamma)}.
\end{eqnarray}
Assuming both approximations are reasonable ones, we can use a linear combination
of the two:
\begin{equation}
\bar{E}_{\mathrm{co}}^{\scscs (\mathrm{III})}[\gamma]\approx
a_c{\cal E}_{\mathrm{co}}^{\mathrm{cs}} [\tilde{\tau}_{\mbox{\tiny\textsc{h}e}}]
_{(\tilde{\tau}_{\mbox{\tiny\textsc{h}e}} = \gamma)}
+
(1-a_c) \hspace{0.1ex} {\cal
E}_{\mathrm{co}}^{\scscs (\text{gas})} [\tau_g]_{(\tau_{\mbox{\tiny$g$}}=\gamma)},
\end{equation}
where $a_c$ is an empirical parameter. This Eq.\ is an alternative to the B3LYP
functional \cite{Becke:93,Stephens:94} which uses analogous correlation-energy
functionals: They use the LYP correlation-energy functional \cite{Lee:88},
derived from Colle--Salvetti one ${\cal E}_{\mathrm{co}}^{\mathrm{cs}}$, and a
uniform-electron-gas functional, derived from the RPA \cite{Vosko:80}; they set
($a_c=8.1$).
\section{Acknowledgments}
The author thanks Peter Pulay, Kimihiko Hirao, and Karl Freed for useful
discussions.  This work was initiated while at the University of Lund, Department
of Theoretical Chemsitry, Sweden; the author thanks Bj\"orn~Roos for useful
discussions and suggesting this area of research.  This work was supported by the
Air Force Office of Scientific Research under grant No.\ F49620-00-1-0281, the
National Science Foundation under grant No.\ CHE0111101, the Japanese Society for
the Promotion of Science (JSPS), Swedish Natural Science Research Council (NFR),
and an Internal Research Grant at Eastern New Mexico University.  
\appendix 
\section{Variational Brueckner-orbital formalisms} \label{ESCF} 

In the exact SCF theory by L\"owdin \cite{Lowdin:62}, an orbital variation of an
energy functional is used to derive the Brillouin--Brueckner condition. By using a
slight modification of Kobe's formulation \cite{Kobe:71}, L\"owdin's
energy-functional can be written in the following manner:
\begin{equation}
E_\eta[\Theta,\Phi] =  \langle \Phi | \hat{H}_{\mathrm{eff}}^\Theta | \Phi \rangle =  
\langle \Phi |(H_0^\Theta + t^\Theta)| \Phi \rangle,
\end{equation}
where $t^\Theta$, or $V_\Theta\Omega_\Theta$, is the reaction operator, which we
also define above by the effective Hamiltonian $\hat{H}_{\mathrm{eff}}^\Phi$; by
definition, this operator satisfies the following relation:
\begin{equation} \label{E.Heff} 
{\cal E}=\langle \Phi | \hat{H}_{\mathrm{eff}}^\Phi | \Phi \rangle
= \langle \Phi |H\Omega_\Phi | \Phi \rangle,
= \langle \Phi |H\left(1 + \chi_\Phi\right)| \Phi \rangle,
\end{equation}
and it acts only within the one-dimensional reference-space, with projector
$P_\Phi$: ($\hat{H}_{\mathrm{eff}}^\Phi=P_\Phi\hat{H}_{\mathrm{eff}}^\Phi
P_\Phi$); the wave operator $\Omega_\Phi$ and correlation operator $\chi_\Phi$ are
given by Eqs.~(\ref{WO}) and (\ref{corr.op}).  As in valence-universal
multireference perturbation theory
\cite{Wilson:85,Ellis:77,Brandow:67,Brandow:77,Lindgren:86,Lindgren:74,Freed:89},
$\hat{H}_{\mathrm{eff}}^\Phi$ can be written as a sum of constant, one-, two- and
higher-body excitations, where the vacuum state is at our disposal.

By varying the occupied orbitals from the reference state $|\Phi\rangle$, i.e.,
$\{\psi_o\mbox{\small $\rightarrow\Phi$}\}$, L\"owdin and Kobe derived a form of
the Brillouin--Brueckner condition, indicating that the above functional
$E_\eta[\Theta,\Phi]$ -- when $\hat{H}_{\mathrm{eff}}^\Theta$ is held constant --
satisfies a variational condition that yields the exact energy at the extremum:
(${\cal E}=E_\eta[\Theta,\Theta]$), where the reference state is the Brueckner one,
$|\Theta\rangle$. A modified version of L\"owdin and Kobe's formulation involving
a one-particle density matrix approach, instead of an orbital one, is easily
derived \cite{tobe}.

A generalization of L\"owdin and Kobe's theory, by Brueckner and Goldman
\cite{Brueckner:59}, minimizes the following functional: $E_\eta[\Phi,\Phi]$,
given by the right side of Eq.~(\ref{E.Heff}). Such an approach has been
criticized, since this functional is invariant to the reference state
$|\Phi\rangle$, and its a constant -- the exact energy ${\cal E}$; also, if an
approximate $\hat{H}_{\mathrm{eff}}^\Phi$ is used, the energy functional does not
satisfy the Rayleigh-Ritz principle, as pointed out by Brandow
\cite{Brandow:67}. This method leads to the so-called rearrangement potential
\cite{Brueckner:59,Kumar:62} that arises from the variation of the term involving
the reaction operator $t^\Phi$.  (For further references regarding the
rearrangement potential and an historical account of Brueckner orbital theory, see
the bibliography notes within Kobe's article \cite{Kobe:71}.)

Another variant of the energy functional $E_\eta[\Phi,\Phi]$, replaces the
correlation operator $\chi_\Phi$ with $\chi_\Phi^{\scscs \eta}$ in the right side
of Eq.~(\ref{E.Heff}); this energy functional is simply the non-variational one
$E_\eta[\Phi]$, as indicated by Eq.~(\ref{Efunct.b}). L\"owdin and Kobe's exact
SCF theory is obtained by using $\chi_\Theta^{\scscs \eta}$ and not permitting it
to vary, where from Eq.~(\ref{chi.ident}) we have ($\chi_\Theta^{\scscs
\eta}=\chi_\Theta$).

\section{Partitioning of second quantized operators} \label{orb.part} 

Any second quantized operator, say $\hat{O}$, can be partitioned into open (op) and
closed (cl) portions with respect to a single- or multi-reference space
\cite{Lindgren:85,Lindgren:86,Lindgren:87}.  In our case, where the reference
space is only spanned by a single-determinantal state, the closed portion of
$\hat{O}$, say $\hat{O}_{\mathrm{cl}}$, is simply a constant -- as in Eqs.\
(\ref{Efirst.cl}) and (\ref{Ecorr.cl}) -- and is given by the fully contracted
part of $\hat{O}$, where the operator is written in normal-ordered form with
respect to the reference state~$|\Phi\rangle$
\cite{Cizek:66,Cizek:69,Lindgren:86,Paldus:75}. Explicitly, we have
\begin{eqnarray} \label{closed.op} 
\hat{O}_{\mathrm{cl}} = \hat{O}_0 =
\langle \Phi| \hat{O} |\Phi\rangle ,
\end{eqnarray}
where the $0$ subscript indicates the zero-body term.

The open portion of $\hat{O}$, say $\hat{O}_{\mathrm{op}}$, is usually defined as
the remaining portion; it is given by the one-, two- and higher-body terms, where,
again, the operator is written in normal-ordered form. However, for our purposes,
we use a more restrictive definition for $\hat{O}_{\mathrm{op}}$, and define it by
the following conditions:
\begin{eqnarray} \label{open.op} 
\hat{O}_{\mathrm{op}}|\Phi \rangle &=&Q_\Phi \hat{O}|\Phi \rangle,\\
\hat{O}_{\mathrm{op}}Q_\Phi  &=& 0.
\end{eqnarray}

We define the remaining portion, $\hat{O}_{\mathrm{re}}$, by the following:
\begin{equation} \label{ext.op} 
\hat{O} = \hat{O}_{\mathrm{cl}} + \hat{O}_{\mathrm{op}} + \hat{O}_{\mathrm{re}},
\end{equation}
so the following identities are satisfied:
\begin{eqnarray} 
(\hat{O}_{\mathrm{cl}} + \hat{O}_{\mathrm{op}})|\Phi\rangle&=&\hat{O}|\Phi\rangle, \\
\label{extb.op} 
\hat{O}_{\mathrm{re}}|\Phi \rangle  &=& 0.
\end{eqnarray}
The remaining portion $\hat{O}_{\mathrm{re}}$ has at least one hole or particle
annihilation-operator. $\hat{O}_{\mathrm{op}}$ has at least one pair of
hole-particle creation-operators and no hole or particle
annihilation-operators. In terms of diagrams, $\hat{O}_{\mathrm{re}}$ has at least one
external line below the vertex; $\hat{O}_{\mathrm{op}}$ has no lines below the
vertex and at least one pair of externals lines above it; $\hat{O}_{\mathrm{cl}}$
has no external free-lines.

As an alternative to the above normal-ordered partitioning, we find it convenient
to partition one-body operators, say $\hat{h}$, 
\begin{equation} \label{h.one} 
\hat{h} = \sum_{ij} \hat{h}_{ij},
\end{equation}
into the following four
components:
\begin{subequations} \label{one.part} 
\begin{equation} \label{h.one.part} 
\hat{h} =
\hat{h}_{\text{ex}} +
\hat{h}_{\text{de}} +
\hat{h}_{\text{oc}} +
\hat{h}_{\text{un}},
\end{equation}
where the excitation (ex), de-excitation (de), occupied (oc), and unoccupied (un)
parts are given by the following expressions:
\begin{eqnarray}  \label{h.one.comp} 
\hat{h}_{\text{ex}}&=& 
\sum_{wr} h_{rw} a_r^\dagger a_w,
\\
\hat{h}_{\text{de}}&=& 
\sum_{wr} h_{wr} a_w^\dagger a_r,
\\
\hat{h}_{\text{oc}}&=& 
\sum_{wx} h_{wx} a_w^\dagger a_x,
\\
\hat{h}_{\text{un}}&=& 
\sum_{rs} h_{rs} a_r^\dagger a_s.
\end{eqnarray}
\end{subequations}
and the orbitals are defined with respect to a reference state $|\Phi\rangle$,
as indicated by Eqs.~(\ref{phi.orbs.index}). Note that the open and excited portions are
identical:
\begin{equation} \label{ex=opa} 
\hat{h}_{\text{ex}} = \hat{h}_{\text{op}}.
\end{equation}

\section{Discussion of perturbative convergence when using orbital degeneracies} \label{ORBDEG}

Our primary reason for choosing orbital degeneracies within the occupied and
virtual subspaces is that it provides a means of obtaining an explicit dependence
on the one-particle density-matrix $\gamma$ for the individual perturbative terms
(the diagrams) that represent the correlation energy ${\mathcal E}_c[\gamma]$ and
correlation-energy functionals $E_{\mathrm{co}}^{\scscs (\eta)}[\gamma]$. While
this choice greatly restricts the zeroth-order Hamiltonian, it can still yield
convergent series for ground states, as long as the parameter \mbox{\footnotesize
$\varepsilon_\gamma$} is chosen to be sufficiently large, and negative, so that the dominant
configurations do not become intruder-states from the presence of small-energy
denominators or incorrect energy ordering, as in the case for multireference
perturbation theory
\cite{Finley:N2,Finley:H4,Finley:Be,Schucan:72,Schucan:73,Ellis:77}. However,
because of incorrect energy-ordering \cite{Finley:Be} it is probably not possible
to generate convergent expansions for excited states. However, infinite-order
summation methods \cite{Ellis:77,Harris:92,Bartlett:76,Brueckner:57,Gellmann:57b}
can be attempted, or the series can be asymptotically convergent.

Consider a suitable partitioning method, like, for example, M{\o}ller--Plesset
\cite{Moller:34,Bartlett:75,Pople:76,Szabo:82}, Epstein-Nesbet
\cite{Claverie:67,Epstein:26,Nesbet:55}, or maximum radius of convergence ($R_c$)
perturbation theory \cite{Finley:maxrc,Finley:maxrc2}. Any of these approaches
generate a separate energy-denominator for each orthogonal-space state
$|\mbox{\large $q$}\rangle$, say ($\Delta_q=E_0^\Phi - E_0^{\mbox{\footnotesize
$q$}}$), where $E_0^\Phi$ and $E_0^{\mbox{\footnotesize $q$}}$ are the,
respective, zeroth-order energies.  Since small energy-denominators can yield
convergence problems, a reasonable choice for \mbox{\footnotesize
$2\varepsilon_\gamma$} is given by the maximum $\Delta_q$ within the
doubly-excited subspace:
\begin{equation}
\mbox{\footnotesize $\varepsilon_\gamma$} \; = 
\frac12 \;\;\; \mbox{Max} \hspace{-6ex} 
\raisebox{-2ex}{\scriptsize $|q\rangle\in\{|\Phi_{wx}^{rs}\rangle\}$}
\Delta_q,
\end{equation}
where the search is over all orthogonal-space states $\{|\mbox{\large
$q$}\rangle\}$ from within the set of double excited-states
$\{|\Phi_{wx}^{rs}\rangle\}$ that arise from a reasonable set of orbitals, e.g.,
from the $\{\varphi_o \mbox{\footnotesize $\leftarrow\Phi,\hat{f}_o$}\}$ and
$\{\varphi_u \mbox{\footnotesize $\leftarrow\Phi,\hat{f}_u$}\}$ sets. (Some
modification is necessary for M{\o}ller--Plesset, since this method is restricted
to Hartree--Fock orbitals, e.g., choose $H_0$ to be given by Eqs.~(\ref{H0.nd}),
(\ref{H0.wx}) and (\ref{H0.rs}), where ($\hat{f}_o^{\scscs \Phi} =
\hat{f}_u^{\scscs \Phi} = \hat{F}_\Phi$).)

At least for ground states, we anticipate that perturbative expansions using
orbital degeneracies can often converge, but, perhaps, at a slow rate.  On the
other hand, this deficiency is partially compensated by a greater computational
efficiency, since, for example, costly two-electron integral-transformations can be
avoided. Furthermore, approximations that involve infinite-order summations are
often invariant to the choice of $H_0$ -- and hence \mbox{\footnotesize
$\varepsilon_\gamma$} -- including the coupled cluster method that can be viewed
as an infinite-order partial-summation method \cite{Lindgren:86,Lindgren:78}.

\bibliography{ref}
\end{document}